\documentclass[hidelinks,aps,pra,superscriptaddress,10pt]{revtex4-2} 
\usepackage{amsmath}
\usepackage{amssymb}
\usepackage{graphicx}
\usepackage{color}
\usepackage{url}
\newcommand{\be}[1]{\begin{equation}\label{#1}}
\newcommand{\ee}{\end{equation}}
\newcommand{\bea}[1]{\begin{eqnarray}\label{#1}}
\newcommand{\eea}{\end{eqnarray}}
\newcommand{\no}{\nonumber \\}
\newcommand{\Fig}[1]{Fig.(\ref{#1})}
\newcommand{\Tbl}[1]{Table \ref{#1}}
\newcommand{\Eq}[1]{Eq.(\ref{#1})}
\newcommand{\App}[1]{Appendix~\ref{#1}}
\newcommand{\Sec}[1]{Section~\ref{#1}}
\newcommand{\bsub}{\begin{subequations}}
\newcommand{\esub}{\end{subequations}}
\newcommand{\bwt}{\begin{widetext}}
\newcommand{\ewt}{\end{widetext}}

\usepackage{color}

\def\trm#1{\textrm{#1}}
\def\tit#1{\textit{#1}}
\def\tbf#1{\textbf{#1}}
\def\ttt#1{\texttt{#1}}
\def\defn{\overset{\textrm{def}}{=}}
\newcommand{\om}{\omega}
\newcommand{\Om}{\Omega}

\def\a0{{\alpha_0}}

\def\da0{{\dot{\alpha}_0}}

\def\myoverDefn#1#2{\hbox{\space \raise-2mm\hbox{$\textstyle{#1} \atop \scriptstyle{#2}$} }}

\def\mS{{\mathcal{S}}}

\def\a{{\alpha}}

\def\dag{\dagger}

\def\L{\Lambda}

\def\rp{r_{P}}
\def\rp2{r_{p}^{2}}

\def\Mod{\textrm{Mod}}

\def\Perm{\trm{Perm}}
\def\PermL{\trm{Perm}(\L)}

\def\ketnNN{\ket{\tfrac{n}{N}}^{\otimes N}}
\def\overSNto{\overset{S_N}{\to}}
\def\n{\mathbf{n}}
\def\m{\mathbf{m}}
\newcommand{\half}{\frac{1}{2}}

\newcommand{\ket}[1]{|#1\rangle}
\newcommand{\bra}[1]{\langle #1|}

\newcommand{\IP}[2]{\langle {#1} | {#2} \rangle}

\begin{document}

\title{Multiphoton Interference with a symmetric SU(N) beam splitter \\ and the generalization of the extended 
Hong-Ou-Mandel effect }
\author{Paul M. Alsing}\email{corresponding author: alsingpm@gmail.com/palsing@albany.edu}
\affiliation{University at Albany-SUNY, Albany, NY, 20222, USA}
\author{Richard J. Birrittella}
\affiliation{Booz Allen Hamilton, 8283 Greensboro Drive, McLean, Virginia 22102, USA}
\affiliation{Air Force Research Laboratory, Information Directorate, 525 Brooks Road, Rome, New York, 13411, USA}
\author{Peter L. Kaulfuss}
\affiliation{Booz Allen Hamilton, 8283 Greensboro Drive, McLean, Virginia 22102, USA}
\affiliation{Air Force Research Laboratory, Information Directorate, 525 Brooks Road, Rome, New York, 13411, USA}

\date{\today}

\begin{abstract}
We examine multiphoton interference with a symmetric $SU(N)$ beam splitter $S_N$, an  extension  of features of the $SU(2)$ 50/50 beam splitter extended Hong-Ou-Mandel (eHOM) effect, whereby one obtains a zero amplitude (probability) for the output coincidence state (defined by equal number of photons $n/N$ in each output port), when a total number $n$ of  photons impinges on the $N$-port device.
These are transitions of the form $\ket{n_1,n_2,\ldots,n_N}\overset{S_N}{\to}\ket{n/N}^{\otimes N}$, where
$n=\sum_{i=1}^N n_i$, which generalize the 
Hong-Ou-Mandel (HOM) effect $\ket{1,1}\overset{S_2}{\to}\ket{1,1}$,  
the eHOM effect $\ket{n_1,n_2}\overset{S_2}{\to}\ket{\tfrac{n_1+n_2}{2},\tfrac{n_1+n_2}{2}}$, and the 
generalized HOM effect (gHOM) $\ket{1}^{\otimes N}\overset{S_N}{\to}\ket{1}^{\otimes N}$, which have previously been studied in the literature.
The emphasis of this work is on illuminating how the over all destructive interference occurs in separate groups of destructive interferences of sub-amplitudes of the total zero amplitude.  
We also consider the more general case for zero-coincidences for the symmetric $SU(N)$
beam splitter transformations on multiphoton Fock input states 
$\ket{n_1,n_2,\ldots,n_N}\overset{S}{\to}\ket{m_1,m_2,\ldots,m_N}$ such that
$\sum_{i=1}^N n_i = \sum_{i=1}^N m_i \defn n$.
We relate these zero-coincidences  to the symmetry properties of $S_N$, 
beyond the well known condition that zero-coincidence implies \tit{zero-permanent} Perm$(S_N)$=0, which governs the transformation  $\ket{1}^{\otimes N}\overset{S_N}{\to}\ket{1}^{\otimes N}$ for arbitrary 
$N\in even$.
We extend these symmetry properties to the case of the generalized eHOM effect (geHOM)
$\ket{n_1,n_2,\ldots,n_N}\overset{S_N}{\to}\ket{n/N}^{\otimes N}$ involving a zero amplitude
governed by  $\PermL=0$, for an appropriately constructed matrix $\L(S_N)$ 
built from the matrix elements of $S_N$.
We develop an analytical constraint equation for $\PermL$  for arbitrary $N$ that allows us to determine when it is zero, implying complete destructive interference on the geHOM output coincident state $\ketnNN$. 
We generalize the SU(2) beam splitter feature of central nodal line (CNL),  which has a zero diagonal along the output probability distribution when one of the input states is of odd parity (containing only odd number of photons), to general case of $N = 2 * N'$ where $N'\in odd$.
\end{abstract}

\maketitle

\section{Introduction}\label{sec:Intro}
Multiphoton interference  is an important topic of active research with a myriad of applications including spectroscopy, sensing, quantum communications and networking, boson sampling, quantum computing, and atom-photon quantum memory, photonic-interfaces and quantum information processing (QIP).
Multiphoton interference effects have critical application across a variety of QIP tasks, involving numerous discussions of the $SU(3)$ beam splitter ``tritter" \cite{Spagnolo:2013}, and its applications for high-fidelity photonic quantum information processing \cite{Lu:2018}, and three-party quantum key distribution 
\cite{Suryadi:2025}. Higher order multiphoton effects at balanced beam splitters have also been discussed as quantum fourier transform (QFT) interferometers, with applications including quantum metrology 
\cite{Su:2017}.
\smallskip

The process that gives rise to  two-mode states of light via (passive) beam splitting is known 
as two-photon quantum interference  \cite{Ou:1996,Ou:2007,Ou_Book:2017},  
and serves as a critical element in  applications including quantum optical interferometry \cite{Pan:2012}, and quantum state engineering where beam splitters and conditional measurements are utilized to perform post-selection techniques such as photon subtraction \cite{Dakna:1997,Carranza:2012,Magana-Loaiza:2019}, photon addition  \cite{Dakna:1998}, and photon catalysis \cite{Lvovsky:2002, Bartley:2012, Birrittella:2018}.  
\smallskip

The quintessential example of two-photon quantum interference is the celebrated Hong-Ou-Mandel (HOM) two-photon interference effect \cite{HOM:1987} describing the transition $\ket{1,1}\overset{S_2}{\to}\ket{1,1}$ by which two single photons enter each of the two input ports of an ideal, lossless, balanced 50/50 beam splitter (BS), producing a zero amplitude (probability) for the output coincident state (for a recent extensive review, see
Bouchard \tit{et al}. \cite{Bouchard:2021}). 
Famously, this zero coincidence occurs because the amplitude for both photons to transmit $t$, or for both to reflect $r$  at the BS (such that $|t|^2 + |r|^2=1$), have equal magnitude  $|t|=|r| = \tfrac{1}{\sqrt{2}}$, yet opposite signs, and therefore cancel each other. Here,
$S_2=\tiny{\tfrac{1}{\sqrt{2}}\,\left(\begin{array}{cc}1 & 1 \\1 & -1\end{array}\right)}$ is the symmetric $SU(2)$ BS, which is unitarily equivalent to the more common forms found in the literature such as
$S_2=
\big\{ 
{
\tiny{\tfrac{1}{\sqrt{2}}\,\left(\begin{array}{cc}1 & i \\i & 1\end{array}\right)},
\tiny{\tfrac{1}{\sqrt{2}}\,\left(\begin{array}{cc}1 &1 \\-1 & 1\end{array}\right)}
}
\big\}$.

In a series of recent papers \cite{eHOM_PRA:2022,eHOM_PRA:2025}, two of the current authors generalized the HOM effect to the 50/50 BS $SU(2)$ muliphoton transitions 
$\ket{n_1,n_2}\overset{S_2}{\to}\ket{\tfrac{n_1+n_2}{2},\tfrac{n_1+n_2}{2}}$ for 
$(n_1,n_2)\in \{(odd,odd), (even,even)\}$, which they termed the extended Hong-Ou-Mandel (eHOM) effect. 
The key results of this work were that only for 
$(n_1,n_2)\in (odd,odd)$ does one obtain a zero amplitude $A=0$ for the eHOM output coincident state  
$\ket{\tfrac{n_1+n_2}{2},\tfrac{n_1+n_2}{2}}$, while one obtains $A\ne0$ for $(n_1,n_2)\in (even, even)$.
Further, for $(n_1,n_2)\in (odd,odd)$ the overall zero amplitude $A=0$ consisted of an even number of sub-amplitudes which canceled in pairs, having the same pair-dependent combinatorial amplitude, yet opposite signs. This pair cancellation of sub-amplitudes generalizes the single-pair cancelation in the original HOM effect.
A further consequence of that work was that for any arbitrary input state consisting only of odd number of photons entering port-1 of the symmetric 2-port BS, then regardless of the input into the second port, be it a pure or a mixed state, there will always be a central nodal line (CNL) of zeros $P(m,m)=0$ in the output probability $P(m_1,m_2)$ of the BS. This CNL will dramatically bifurcate the output probability distribution of the BS (as will be illustrated later).
\smallskip

With the intense interest in boson-sampling \cite{Aaronson:2010} in mid-2000s, much effort was turned to the study of arbitrary transitions $\ket{n_1,n_2,\ldots,n_N}\overset{U_N}{\to}\ket{m_1,m_2,\ldots,m_N}$ and its evaluation in terms of the permanent $\PermL$ of a matrix $\L(U_N)$ constructed from the matrix elements of an arbitrary unitary matrix $U_N$ \cite{Scheel:2004,Scheel:2008}. 
In terms of the HOM effect, Lim and Beige \cite{Lim:Beige:2005} considered the transition $\ket{1}^{\otimes N}\overSNto\ket{1}^{\otimes N}$ for arbitrary $N$ and showed that in this case $\L(U_N)\equiv S_N$, and that $A=\PermL=0$ for $N\in even$. They termed this the \tit{generalized HOM effect} (gHOM).
These authors also expanded their  investigations to  multiphoton entanglement in thes $SU(N)$ beam splitters \cite{Lim:Beige_PRA:2005}. 
Seminal work in this area was also carried out by Tichy and collaborators who developed an important \tit{zero-transmission law} \cite{Tichy:2010} for $SU(N)$ beam splitters, deriving strict transmission laws for most possible output events consistent with a generic bosonic behavior after a suitable coarse graining. These authors subsequently applied their results to the investigation of stringent and efficient assessment of boson-sampling devices, falsifying physically plausible alternatives to coherent many-boson propagation \cite{Tichy:2014}.

\smallskip

In this work, we consider the $SU(N)$ extension of the $SU(2)$ (50/50 BS) eHOM effect by considering transitions of the form
$\ket{n_1,n_2,\ldots,n_N}\overset{S_N}{\to}\ket{n/N}^{\otimes N}$ for arbitrary $N$, where 
$n\defn \sum_{i=1}^N n_i$. 
Here the symmetric $SU(N)$ BS is defined by the matrix elements \cite{Zukowski:Zeilinger:Horne:1997,Campos:2000,Lim:Beige:2005,Tichy:2010} as 
$(S_N)_{ij} = \tfrac{1}{\sqrt{N}}\,\om^{(i-1)(j-1)}$ with $\om = e^{i\,2\,\pi/N}$, the fundamental root of unity for dimension $N$. 
We call the output state  $\ketnNN$ the \tit{eHOM coincident output state}, since like the HOM and gHOM output states, it contains an equal number $\tfrac{n}{N}$ of photons in each of the output ports of the symmetric $S_N$ BS. We term the analytic determination of the zero amplitude $A=0$ for this output state the \tit{generalized eHOM effect} (geHOM).
\Tbl{Table:HOM:effects:terms:new} lists the terminology used for the various HOM effects discussed in this work, and the associated transitions and dimension $N$.
\begin{table}[!ht]
\begin{tabular}{|c|c|c|c|}\hline
\multicolumn{4}{|c|}{\bf Hong-Ou-Mandel (HOM) effect terms used in this work}\\ \hline
\multicolumn{1}{|c|}{\bf Term} & 
\multicolumn{1}{|c|}{\bf symmetric BS} & 
\multicolumn{1}{|c|}{\bf transition} &
\multicolumn{1}{|c|}{\bf \;authors/citation\;} \\ \hline \hline
HOM effect (HOM) & $SU(2)$  & $\ket{1,1}\overset{S_2}{\to}\ket{1,1}$ & HOM \cite{HOM:1987} \\ \hline
extended HOM effect (eHOM) & $SU(2)$ & $\ket{n_1,n_2}\overset{S_2}{\to}\ket{\tfrac{n_1+n_2}{2},\tfrac{n_1+n_2}{2}}$  & Alsing \tit{et al.} \cite{eHOM_PRA:2022, eHOM_PRA:2025}\\ \hline
generalized HOM effect (gHOM) & $SU(N)$ & $\ket{1}^{\otimes N}\overset{S_N}{\to}\ket{1}^{\otimes N}$ & Lim and Beige \cite{Lim:Beige:2005}\\ \hline
generalized eHOM effect (geHOM) & $SU(N)$ & $\ket{n_1,n_2,\ldots,n_N}\overset{S_N}{\to}\ket{\tfrac{n}{N}}^{\otimes N}$ & this work \\ \hline
\hline
\end{tabular}
\caption{HOM effects terminology used in this work.  $n\defn\sum_{i=1}^N n_i$ is the total input/output photon number of the $N\times N$ symmetric beam splitter (BS) $S_N\defn SU(N)$ with matrix elements
$(S_N)_{ij} = \tfrac{1}{\sqrt{N}}\,\om^{(i-1)(j-1)}$ with $\om = e^{i\,2\,\pi/N}$.}
\label{Table:HOM:effects:terms:new}
\end{table}
We can interpret $(S_N)_{ij}$ as the amplitude for a single photon entering input port-$i$ 
to scatter to output port-$j$, and  $(S_N)^k_{ij}$, as $k$-photons entering input port-$i$  and all scattering
to output port-$j$. We take the boson transformation of the input photons written in terms of the output photons as $a^{\dag(in)}_i = \sum_{j=1} (S_N)_{ij}\,a^{\dag(out)}_j$ for the symmetric $N\to N$ port device. In the future, we drop the $(in)$ and $(out)$ labels on the boson operators, and simply employ the rhs of the above equation for any input to output transformations.
\smallskip

In this work, we are primarily concerned with investigating which features of the $SU(2)$ eHOM effect generalize, or have analogues, in the $SU(N)$ extension. In particular, we wish to be able to determine under what conditions does one obtain a zero amplitude $A=0$ for the output eHOM coincident state. We achieve this by developing a symmetry constraint on the permanent $\PermL$ of the associated matrix $\L(S_N)$ constructed from the matrix elements of the symmetric BS $S_N$ which allows us to analytically determine when $A=\PermL=0$. Additionally, rather than just determining whether or not $A=0$, we also show how the overall amplitude becomes zero by the grouping of sub-amplitudes with different combinatorial coefficients, which separately sum to zero, thus generalizing the pairwise amplitude cancelations that arise in the HOM and eHOM effects.
Lastly, with our analytic constraint equation on $\PermL$, we show how to construct CNLs for various types of input states composed of superposition of Fock states with arbitrary quantum amplitude coefficients. 
The work reported here develops both analytical results, and symbolic/numerical calculations (in \tit{Mathematica}) illustrating, and explicitly verifying, these features from $N=3-16$.
\smallskip

The investigations presented in this work are mostly closely related to the above referenced papers by  Lim and Beige (2005) \cite{Lim:Beige:2005},  and by Tichy \tit{et al}. (2010) \cite{Tichy:2010}.
We generalize the symmetry constraint of Lim and Beige (2005) of all-single-photons input and output, to arbitrary photon number input to the $SU(N)$ BS, concentrating on the transitions to the eHOM output coincident state with equal  photon number in each output port.
Analogous to  Tichy \tit{et al}. (2010), we develop our own analytic zero-transmission constraints, though again we focus primarily on the eHOM output coincident state, while those authors investigate more general output states, as well as conditions and approximation for non-zero amplitudes and quantum enhancement (ratio of quantum to classical event probabilities).
Our work (which was completed prior of learning of Tichy \tit{et al}. (2010) \cite{Tichy:2010}) also differs from theirs in that we also present a more detailed investigation of how sub-amplitudes group together and sum separately to zero within a total zero amplitude, thus generalizing both 
the $SU(2)$ HOM \cite{HOM:1987}, and eHOM pairwise sub-amplitude cancellations found in \cite{eHOM_PRA:2022,eHOM_PRA:2025}. We also generalize the CNL effect, discussed above, from $SU(2)$ to $SU(N)$, and illustrate it on $N=4$.
\smallskip

The outline of this paper is a follows:
\smallskip

In \Sec{sec:SU2:eHOM}
we review the $SU(2) $ eHOM effect as an extension of the HOM effect. In particular we focus on the pair cancelations of sub-amplitudes leading to an overall zero amplitude on the eHOM output coincident state.
\smallskip

In \Sec{sec:SUN:BS}
we define and illustrate the $SU(N)$ BS. 
We discuss the \tit{fundamental summation relation} (FSR) $\sum_{i=1} \om^{i-1}=0$ for a given $N$ (i.e. the sum of the roots of unity equals zero), and how it governs in general, the ability of transitions to group together to form a zero amplitude.
\smallskip

In \Sec{sec:calc:of:A}
we detail  two methods to compute the 
amplitude $A$ for the general transition $\ket{\n}\overset{U_N}{\to}\ket{\m}$, 
where $\n\defn\{n_1,n_2,\ldots,n_N\}$ and $\m\defn\{m_1,m_2,\ldots,m_N\}$.
The first exhaustive search method involves $N\times N$ matrices which we call $K = \{k_{ij}\}$ 
whose $i$-th row-sum equals the photon number $n_i$ entering input port-$i$ $\sum_{j=1}^N k_{ij} = n_i$, and 
whose $j$-th column-sum equals the photon number $m_j$ exiting output port-$j$, $\sum_{i=1}^N k_{ij} = m_j$.
The total input/output photon number is given by 
$n=\sum_{i,j=1}^N k_{ij} = \sum_{i=1}^N n_i =  \sum_{j=1}^N m_j$.
While not the most computationally  efficient method to determine the amplitude $A$, its advantage is that we can interrogate the valid matrices $K$ (i.e. satisfying the row-sum and column-sum conditions) in order to determine the how and which sub-amplitudes group together to separately sum to zero within the total zero amplitude $A=0$. 
We present specific illustrative example transitions for $N=3,4$ with zero amplitudes, and the analysis of their sub-amplitude groupings summing separately to zero.

\smallskip
For the second method we review 
the more common and computationally efficient method to compute the amplitude $A$ of 
the general transition $\ket{\n}\overset{U_N}{\to}\ket{\m}$.
We outline the $SU(N)$ gHOM result of Lim and Beige \cite{Lim:Beige:2005} and its relationship to the permanent of $SU(N)$. We review the work of  Scheel \cite{Scheel:2004,Scheel:2008}, 
Aaronson and Arkhipov \cite{Aaronson:2010},  and Chabaud \tit{et al}. \cite{Chabaud:2022} on the construction of $A=\PermL$ from the matrix elements of $U_N$.
\smallskip

In \Sec{sec:Results:cancellation:in:groups}
we present results for the zero amplitudes for various illustrative cases within $N=3,4$, focusing on how and when groups of sub-amplitudes separately sum to zero within a total zero amplitude $A=0$. At the center of these results is how groups of sub-amplitudes, with equal coefficients, collect to yield expressions whose values are zero when evaluated on $\om = e^{i 2\pi /N}$.
\smallskip

In \Sec{sec:psym}
we present results for the geHOM effect for $N\in odd, \{3,5,\ldots,15\}$, and for 
$N\in even, \{4,6,\ldots,14\}$, for both the number of $A=0$ and $A\ne0$ amplitudes, 
and discuss trends seen in the results. 
The results were computed symbolically, i.e. with 
$\PermL$ as a function of $\om$, and subsequently numerically evaluated when the value of $\om$ was substituted into the analytic expression. 
Note that the required $\L$ for a given $N$ and input/output photon number $n$ is an $n\times n$ matrix 
(such that $\tfrac{n}{N}\in \mathbb{Z}_+$), and formally contains $n!$ terms, which limits the practical size of $N$ and $n$ that can be computed in a reasonable amount of time and/or memory.
\smallskip

In \Sec{sec:sym:constraint}
we develop a symmetry constraint on the $A=\PermL$ by generalizing the symmetry argument of Lim and Beige employed for the gHOM effect. With the use of two auxiliary matrices, we determine two different expressions for the value of  $\PermL$. When these two expression disagree, it implies that $\PermL=0$. On the other hand, when these expressions agree, we end up with a trivial identity $\PermL=\PermL$ (even for the gHOM effect). Though this does not necessarily imply that $A=\PermL\ne0$ (since one could have the case $0=0$), we find that most often it does. The two cases found where it does not, are interesting since they involve variants of the FSR, which we explore. We analyze this case analytically as well, to determine when this trivial identity actually implies instead $A=0$ in the special case when $n=N/2$ for $N\in even$.
Lastly, we show how using our analytic constraint equation for $A=\PermL$ we can construct $N$-dependent states that produces CNLs, generalizing those found in the $SU(2)$ eHOM case.
\smallskip

In \Sec{sec:Conclusion} we state our conclusions and discuss prospects for future research.
\smallskip
In \App{app:LChabaud} we present (\tit{Mathematica}) code that constructs the matrix $\PermL$ from the symmetric BS $S_N$, and then factorizes it. This code is readily translatable into other common programming languages such as \tit{Python}.

\section{A review of the $\mathbf{SU(2)}$ \lowercase{e}HOM effect, and its relevant features}\label{sec:SU2:eHOM}
In this section we briefly review the $SU(2)$ extended HOM (eHOM) effect \cite{eHOM_PRA:2022,eHOM_PRA:2025} and the salient features that we wish to generalize to $SU(N)$.
The former is governed by the symmetric $SU(2)$  50/50 BS matrix given by
\be{SU2:BS}
S_{N=2}=
\frac{1}{\sqrt{2}}\,
\left(\begin{array}{cc}
1 & 1 \\
1 & \om
\end{array}\right) 
=
 \frac{1}{\sqrt{2}}\,
\left(\begin{array}{cc}
1 & 1 \\
1 & -1
\end{array}\right), \qquad  \om =  e^{i 2 \pi /(N=2)} = -1.
\ee
The primary result found in \cite{eHOM_PRA:2022} was that for any input Fock state (FS) state 
$\ket{n_1,n_2}$, the amplitude $A$ for the output coincidence state defined by 
$\ket{\frac{n_1+n_2}{2}, \frac{n_1+n_2}{2}}$ (i.e. equal number of output photons in both ports), was zero \tit{iff both $n_1$ and $n_2$ were both odd}, and non-zero if  $n_1$ and $n_2$ were both \tit{even}. (Of course, trivially, if the total photon number $n \defn n_1+n_2$ were odd, there could not be equal number of photons both  output ports).
\begin{figure*}[ht]
\begin{center}
\begin{tabular}{ccc}
\includegraphics[width=3.25in,height=1.25in]{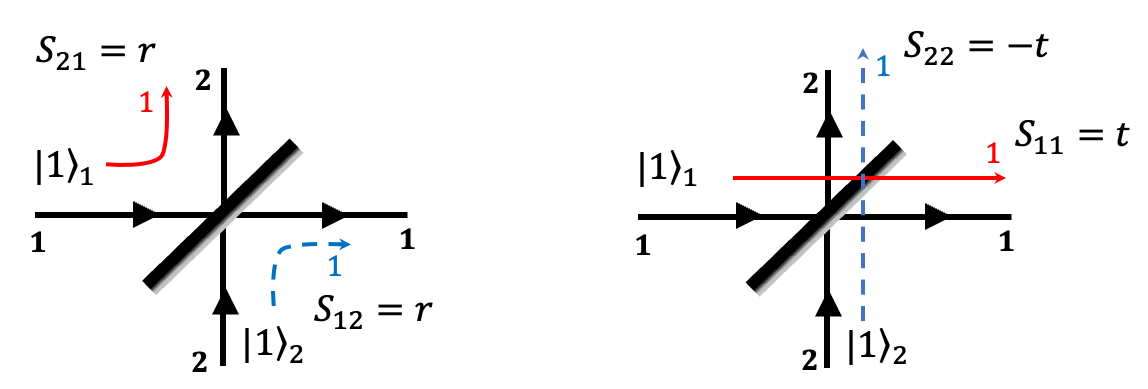} &
\hspace{0.5in} &
\includegraphics[width=3.25in,height=1.25in]{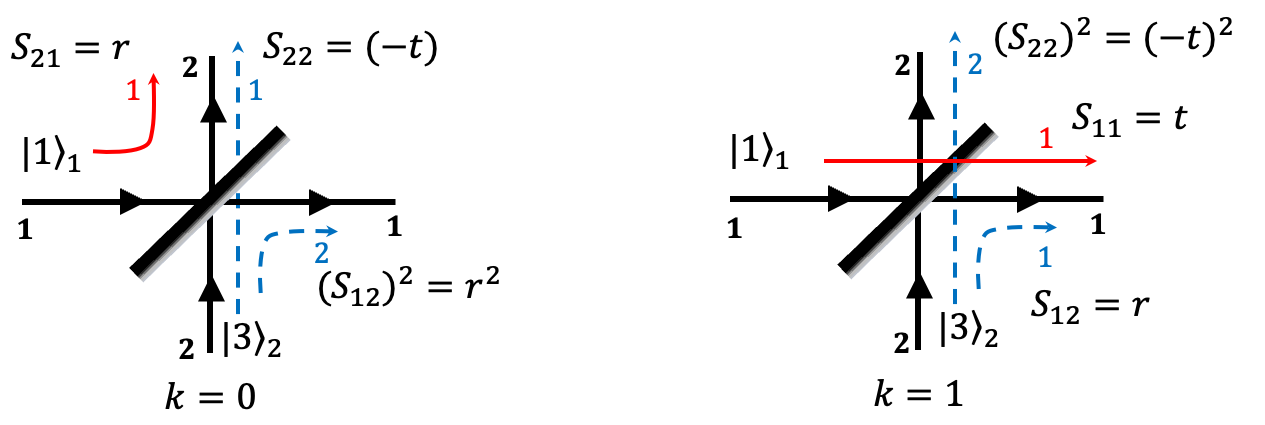}
\end{tabular}
\end{center} 
\caption{
(left) The zero amplitude $A=0$ two-photon HOM transition $\ket{1,1}\overset{S_2}{\to}\ket{1,1}$.
(right) The zero amplitude $A=0$ four-photon eHOM transition $\ket{1,3}\overset{S_2}{\to}\ket{2,2}$.
In both cases the total amplitude is given by $A = A_{k=0}+A_{k=1}$ where $k$ 
indicates the number $n_1$ of input photons in port-$1$ that are transmitted to output port-$2$.
Both sub-amplitudes  $A_{k=0}$ and $A_{k=1}$ have equal amplitudes, but opposite signs, and thus the pair cancels. to produce a total amplitude of $A=0$.
Here, $t=r=1/\sqrt{2}$ in the general $SU(2)$ beam splitter 
$S_2 = \scriptsize{\Big(\begin{array}{cc} t & r\\ r & -t \end{array}\Big)}$.
}
\label{fig:HOM:FS:FS:1:1:and:1:3}    
\end{figure*}
In \Fig{fig:HOM:FS:FS:1:1:and:1:3}(left) we illustrate the well known two-photon HOM effect  
\cite{HOM:1987} for the transition $\ket{1,1}\overset{S_2}{\to}\ket{1,1}$. Here, the total amplitude is given by $A=A_{k=0} + A_{k=1}=0$, where $k$ indicates the number $n_1$ of input photons in port-$1$ that are transmitted to output port-$1$. For a lossless symmetric (balanced) 50/50 BS, $A_{k=0} = -A_{k=1} = \tfrac{1}{\sqrt{2}}$, and so the pair of sub-amplitudes cancel. 
\smallskip

In \Fig{fig:HOM:FS:FS:1:1:and:1:3}(right) we illustrate the zero amplitude  four-photon eHOM transition 
$\ket{1,3}\overset{S_2}{\to}\ket{2,2}$.
In this case as well, there are only two sub-amplitudes $A_{k=0}$ and $A_{k=1}$ (where $k$ has the same meaning as before) again with equal magnitude, but opposite signs, so that $A = A_{k=0}+A_{k=1}=0$. The difference from the two-photon HOM effect is the value of the combinatorial coefficient 
$C_0=-C_1 \propto \binom{1}{1}\,\binom{3}{1}$ indicating the number of ways $n_1=1$ and $n_2=3$ can be scattered from their respective input ports to their respective output ports. (Here, we use $``\propto"$ since we have dropped unimportant $k$-independent constants that can be factored out of the zero amplitude $A=0$).
\smallskip

In both the HOM and eHOM case, we see the pair of canceling sub-amplitudes corresponds to a pair of complementary \tit{mirror-image}  diagrams having (i) the number of input photons in port-1 and port-2 that are respectively reflected/transmitted into output port-1 and port-2  in the left diagram, reversed with the number transmitted/reflected in the right diagram, and (ii) both diagrams having the same amplitude, but opposite signs. 
\smallskip

\begin{figure*}[ht!]
\begin{center}
\hspace{-0.55in}
\includegraphics[width=6.0in,height=1.5in]{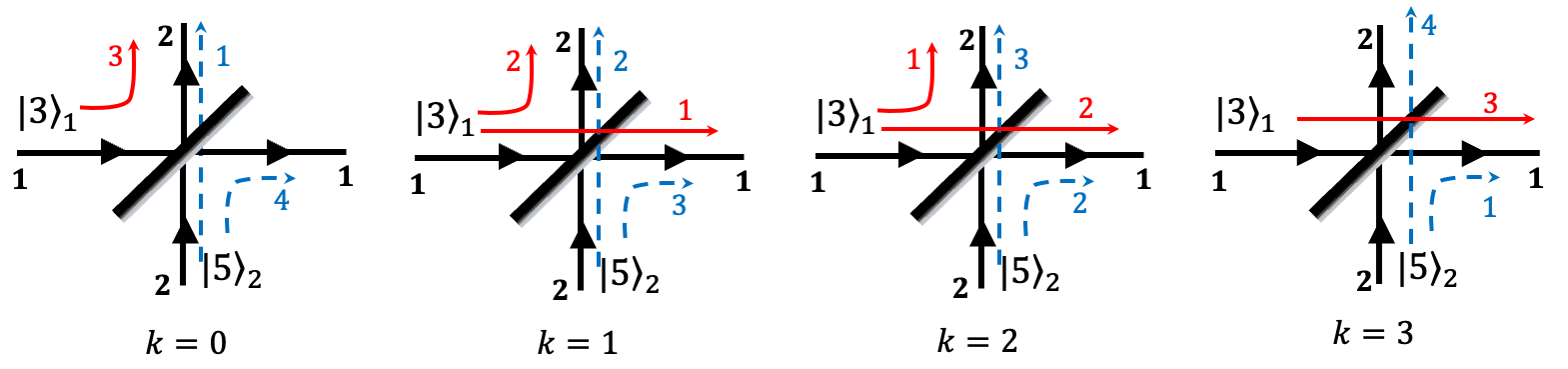}
\end{center} 
\caption{The zero amplitude $A=0$, 8-photon eHOM transition $\ket{3,5}\overset{S_2}{\to}\ket{4,4}$ illustrating the two pairs of scattering amplitudes, $(A_{k=0}, A_{k=3})$, and $(A_{k=1}, A_{k=2})$, 
each with with equal $k$-dependent magnitudes and opposite signs, that cancel in pairs, and contribute to the complete destructive interference on the eHOM coincident output state $\ket{4,4}$ via
$A =C_0\,(A_{k=0} +A_{k=3}) + C_1\,(A_{k=1}+A_{k=2})=0+0=0$. The coefficients $C_k$ are combinatorial factors with $C_0=C_3$ and $C_1=C_2$.
}
\label{fig:eHOM:FS:FS:3:5:in:4:4:out}    
\end{figure*}

The first encounter with more than two sub-amplitudes arises in the zero amplitude $A=0$, 8-photon eHOM transition $\ket{3,5}\overset{S_2}{\to}\ket{4,4}$ illustrated in \Fig{fig:eHOM:FS:FS:3:5:in:4:4:out}. 
Here the total amplitude is given by
$A =C_0\,(A_{k=0} +A_{k=3}) + C_1\,(A_{k=1}+A_{k=2})=0+0=0$ where again
$k\in\{0,1,2,3\}$ indicates the number $n_1$ of photons transmitting from input port-$1$ to output port-$1$.
Again, mirror-image diagrams cancel in pairs, e.g. the outer two diagrams $C_0(A_{k=0} +A_{k=3})=0$ and
the inner two diagrams  $C_1(A_{k=1} +A_{k=2})=0$.
In this case the combinatorial coefficients are different for the two pairs, with
$C_0=-C_3 \propto \binom{3}{3}\,\binom{5}{4}$ for the outer two diagrams and 
$C_1=-C_2 \propto \binom{3}{2}\,\binom{5}{3}$ for the inner two diagrams.
In the leftmost diagram $A_{k=0}$, the coefficient $C_0 \propto \binom{3}{3}\,\binom{5}{4}$ indicates the product of the number equivalent ways $\binom{3}{3}$ the $n_1=3$ \tit{indistinguishable} input photons in port-1 can reflect into output port-2, times the number equivalent ways $\binom{5}{4}$ four of the  $n_2=5$ 
indistinguishable  input photons in port-2 can reflect into output port-1.
\smallskip

In \cite{eHOM_PRA:2022, eHOM_PRA:2025} the authors showed that for $(n_1,n_2)$ both \tit{odd}, there will always be an even number $n_1+1$ of sub-amplitudes that will pair up in cancelling mirror-image diagrams
$C_k\,(A_k + A_{n_1-k}) = 0$ (where we have assumed, without loss of generality, that $n_1\le n_2$) with
$C_k = -C_{n_1-k}\propto \binom{n_1}{k}\,\binom{n_2}{(n_1+n_2)/2-k}$,
 generalizing the previous eHOM case of $\ket{3,5}\overset{S_2}{\to}\ket{4,4}$ to 
$\ket{n_1,n_2}\overset{S_2}{\to}\ket{\tfrac{n_1+n_2}{2},\tfrac{n_1+n_2}{2}}$. 
On the other hand, for the case of $(n_1,n_2)$ both \tit{even} their are an odd number $n_1+1$ sub-amplitudes with (i) the mirror-image diagrams now constructively interfering to a non-zero value, plus (ii) and additional lone ``center" sub-amplitude/diagram $A_{k=n_1/2}$ that cannot cancel with any other diagram, leading to an additional non-zero contribution to the total amplitude $A\ne 0$. This is the $SU(2)$ eHOM effect.

An implication of the eHOM effect is that for any odd-parity input state (consisting only of odd number of photons) entering port-1 of the 50/50 BS, then regardless of the state entering port-2, be it pure or mixed, there will always be a \tit{central nodal line} (CNL) of zeros in the probability distribution $P(m_1,m_2)$ of the output photons along the diagonal output states $\ket{m,m}$.
\begin{figure*}[th]
\begin{center}
\begin{tabular}{cccc}
\hspace{-0.65in}
\includegraphics[width=1.85in,height=1.35in]{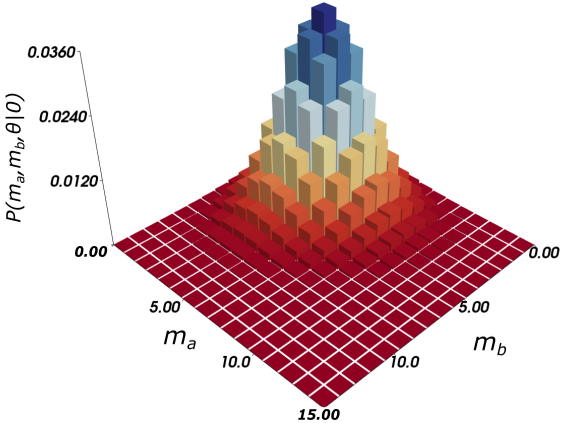}       &  
\includegraphics[width=1.85in,height=1.35in]{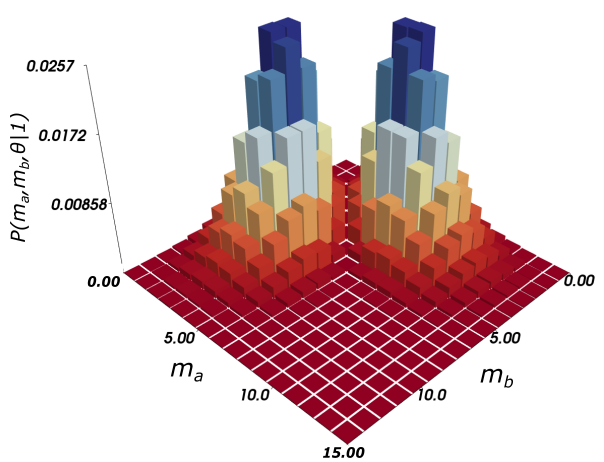}   & 
\includegraphics[width=1.85in,height=1.35in]{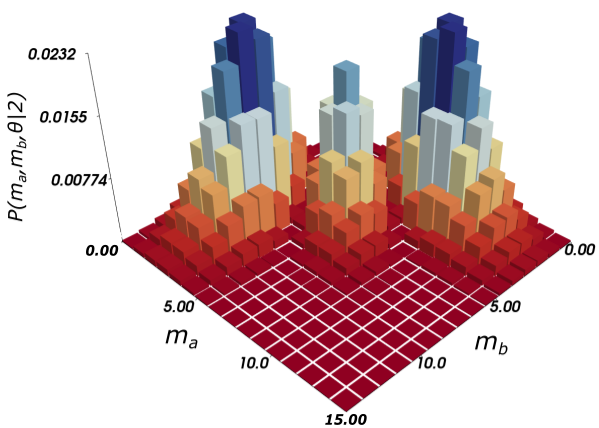} & 
\includegraphics[width=1.85in,height=1.35in]{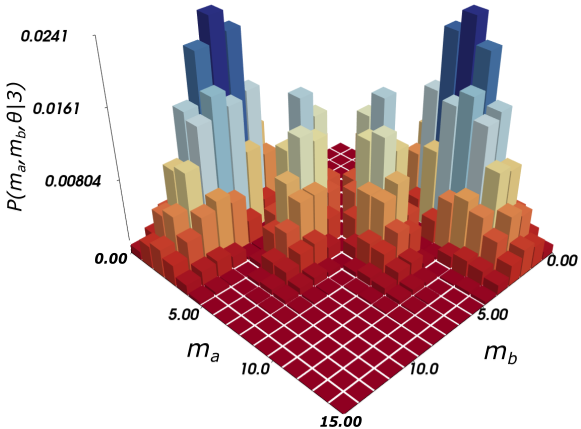} \\
\hspace{-0.65in}
\includegraphics[width=1.85in,height=1.35in]{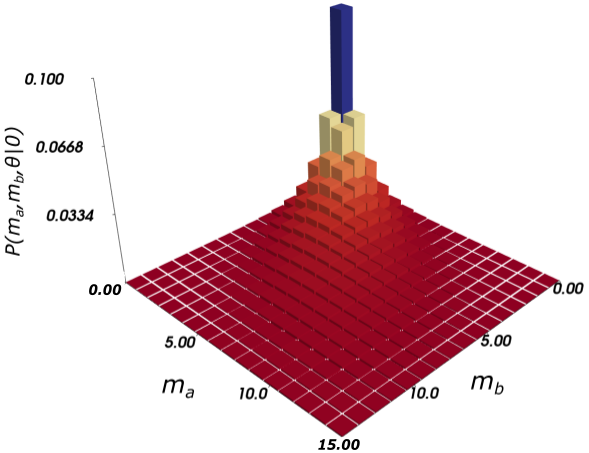}       &  
\includegraphics[width=1.85in,height=1.35in]{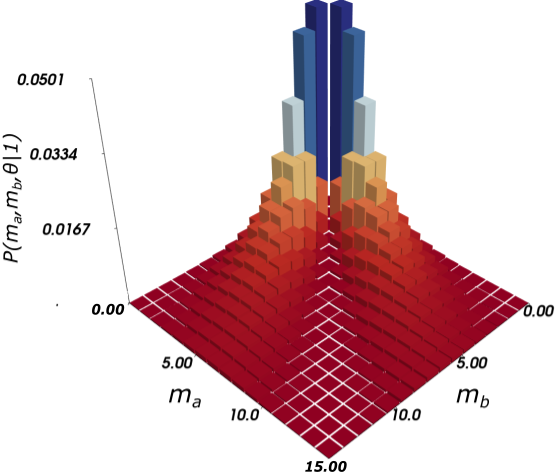}   & 
\includegraphics[width=1.85in,height=1.35in]{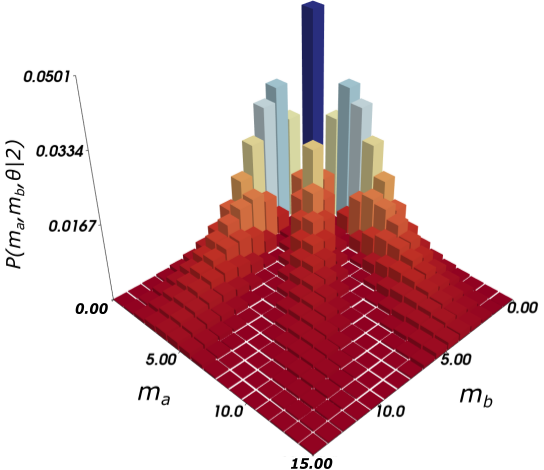} & 
\includegraphics[width=1.85in,height=1.35in]{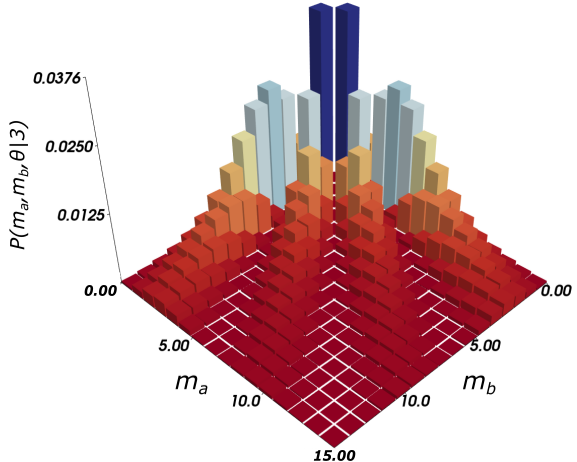}    
\end{tabular}
\end{center} 
\caption{
Joint output probability $P(m_1, m_2)$ 
to measure $m_1$ photons in mode-1 and $m_2$ photons in mode-2 from a 50:50 BS
for input  Fock number states (FS) $\ket{n}_1$ in mode-1, for $n_1=\{0,1,2,3\}$ (top row, left to right), 
and  an input coherent state (CS) $\ket{\beta}_2$ in mode-2, with mean number of photons 
with  $|\beta|^2=9$. 
An output central nodal line (CNL) of zeros for the input states $\ket{n,\beta}$ 
is observed for odd $n=\{1,3\}$ indicating destructive interference of coincidence 
detection on all output FS/FS $\ket{m,m}$.
No CNL is observed for the input states with even $n=\{0,2\}$, indicating non-zero coincidence detection.
(bottom row) Same as top row, but now with the CS mode-2 input state replaced by a mixed thermal state $\rho^{\trm{thermal}}_2$ of average photon number $\bar{n} =9$.
}
\label{fig:eHOM:fig2:fig3:left:column:1photon:CS}    
\end{figure*}
This is illustrate in \Fig{fig:eHOM:fig2:fig3:left:column:1photon:CS} for the top row with a Fock state/Coherent state (FS/CS) input $\ket{n,\beta}$ with $n\in\{0,1,2,3\}$, and a CS with mean number $|\beta|^2=9$.
For odd $n=\{1,3\}$ we observe a CNL which bifurcates the output probability distribution $P(m_1,m_2)$, with zeros along the FS/FS eHOM output coincident states $\ket{m,m}$. For even $n=\{0,2\}$ no such CNL is observed along the  central diagonal of the output probability distribution.
The bottom  row of \Fig{fig:eHOM:fig2:fig3:left:column:1photon:CS} is the same as top row, but now with the CS mode-2 input state is replaced by a mixed thermal state $\rho^{\trm{thermal}}_2$ of average photon number $\bar{n} =9$. The CNL is again observed for odd $n=\{1,3\}$, and not for even $n=\{0,2\}$.
\smallskip

The primary goals of the rest of this work are to explore the generalization of these two features of the $SU(2)$ eHOM effect to the symmetric $SU(N)$ BS; namely (i) the grouping of sub-amplitudes which separately sum to zero, leading to an overall zero amplitude $A=0$ on the generalized eHOM output state (with equal number of photons in each output port), and (ii) the possibilities of CNLs for larger values of $N$. 

\section{The $\mathbf{SU(N)}$ symmetric Beam Splitter}\label{sec:SUN:BS}
The $SU(N)$ symmetric beam splitter $S_N$ is given by \tit{real bordered-formed} matrix
\bsub 
\bea{SUN:BS}
\hspace{-0.65in}
(S_N)_{i,j} &=&\frac{1}{\sqrt{N}}\, \om^{(i-1)(j-1)} =  \frac{1}{\sqrt{N}}\, \om^{\Mod[(i-1)(j-1),N]}, \quad \om = e^{i 2 \pi/N}, \label{SUN:BS:line1} \\
\hspace{-0.75in}
S_3 &=& \frac{1}{\sqrt{3}}
\left(
\begin{array}{ccc}
 1 & 1 & 1 \\
 1 & \omega  & \omega ^2 \\
 1 & \omega ^2 & \omega  \\
\end{array}
\right), \;
S_4 = \frac{1}{\sqrt{4}}
\left(
\begin{array}{cccc}
 1 & 1 & 1 & 1 \\
 1 & \omega  & \omega ^2 & \omega ^3
   \\
 1 & \omega ^2 & 1 & \omega ^2 \\
 1 & \omega ^3 & \omega ^2 & \omega 
   \\
\end{array}
\right),\;
S_5= \frac{1}{\sqrt{5}}
\left(
\begin{array}{ccccc}
 1 & 1 & 1 & 1 & 1 \\
 1 & \omega  & \omega ^2 & \omega ^3
   & \omega ^4 \\
 1 & \omega ^2 & \omega ^4 & \omega 
   & \omega ^3 \\
 1 & \omega ^3 & \omega  & \omega ^4
   & \omega ^2 \\
 1 & \omega ^4 & \omega ^3 & \omega
   ^2 & \omega  \\
\end{array}
\right),\;
S_6 = \frac{1}{\sqrt{6}}
\left(
\begin{array}{cccccc}
 1 & 1 & 1 & 1 & 1 & 1 \\
 1 & \omega  & \omega ^2 & \omega ^3
   & \omega ^4 & \omega ^5 \\
 1 & \omega ^2 & \omega ^4 & 1 &
   \omega ^2 & \omega ^4 \\
 1 & \omega ^3 & 1 & \omega ^3 & 1 &
   \omega ^3 \\
 1 & \omega ^4 & \omega ^2 & 1 &
   \omega ^4 & \omega ^2 \\
 1 & \omega ^5 & \omega ^4 & \omega
   ^3 & \omega ^2 & \omega  \\
\end{array}
\right), \qquad \label{SUN:BS:line2}
\eea 
\esub
where $\{\om^p = e^{i 2 \pi\,p/N}\}$ are the \tit{N roots of unity}, and 
in \Eq{SUN:BS:line1} we have used the cyclical fact that $\om^{p+N} = \om^{p}$, so that only the \tit{integer} powers  $p\in\{0,1,2,\ldots,N-1\}$ of $\om^p$ that appear in $S_N$ when we take  the exponents as 
of the matrix elements as $\Mod_N$.
Except for the first row and first column, each row and column of $S$ satisfies the 
\tit{fundamental summation relation} (FSR)
\be{FSR}
\sum_{i=1}^N \om^{(i-1)} = 1+\om+\om^2+\ldots+\om^{N-1} = \frac{1-\om^N}{1-\om} = 0, \qquad \trm{since}\;\; \om^N=\om^0 = 1.
\ee
The FSR will play a fundamental role in subsequent analysis, since it determines the minimum number of terms necessary for destructive interference to occur. For $SU(2)$ we saw that the  FSR was simply 
$1+\om =0$ with $\om = e^{i 2 \pi/(N=2)} = -1$, which governed the ability for terms in the amplitude $A$ to cancel in pairs (modulo the permutation factors that multiply the pair of cancelling sub-amplitudes). 
From the FSR in \Eq{FSR} we can discern two facts about a zero amplitude for a given
transition $\ket{n_1,n_2,\ldots,n_N}\overset{S_N}{\to} \ket{m_1,m_2,\ldots,m_N}$
 based on the even/odd parity of $N$.

\begin{enumerate}
\item[(1)] For $N\in odd$ the only way for a group of terms to sum to zero is with the full FSR in  \Eq{FSR}, namely \tit{all $N$} powers of $\om^p$  must be involved, and be multiplied by  \tit{identical} (combinatorial) coefficients, $C\,(1+\om+\om^2+\ldots+\om^{N-1})=0$.
\item[(2)] For $N\in even$, we have the additional symmetry
\be{FSR:N:even}
\sum_{i=1}^N \om^{(i-1)} = 1+\om+\om^2+\ldots+\om^{N-1} = \frac{1-\om^N}{1-\om} 
=  \frac{1-\om^{N/2}}{1-\om} \, (1+\om^{N/2}) = 0, \quad \trm{since}\;\;  1+\om^{N/2} = 1+e^{i \pi}=0,
\ee
where the last term $(1+\om^{N/2})$ \tit{effectively acts} as an $SU(2)$ BS with 
$\om' = \om^{N/2} \Rightarrow (1+\om')=0$. That is, we only require at least \tit{two} terms having the factors $\om^0=1$ and $\om^{N/2}$ to have identical coefficients $C'$ in order for a pair of sub-amplitudes to cancel,  $C'\,(1+\om')=0$. 
\item[(3)] For $N\in even$ we can also group the terms in the FSR $\mS_N=0$ in terms of the even and odd exponents of $\om$ as
\bea{FSR:N:even:2}
\hspace{-0.75in}
\mS_N = \sum_{i=1}^N \om^{(i-1)} = 
1+\om+\om^2+\ldots+\om^{N-1} &=& 
(1 + \om^2 + \om^4 + \ldots + \om^{N-2}) + 
(\om^1 + \om^3 +\om^5 + \ldots + \om^{N-1}), \no
\hspace{-0.75in}
&=&
(1+\om)\,(1 + \om^2 + (\om^2)^2 + \ldots + (\om^2)^{N/2-1}), \quad \trm{define} \; \om' = \om^2, \no
\hspace{-0.75in}
&\equiv&
(1+\om)\,(1 + \om' + (\om')^2 + \ldots + (\om')^{N/2-1}) 
=  (1+\om)\,\sum_{i=1}^{N/2} (\om')^{(i-1)}  \no
&=&
 (1+\om)\frac{(1-\om'^{N/2})}{1-\om'} = 0, \quad \trm{since} \quad (\om')^{N/2}=(\om^2)^{N/2}=\om^N\equiv 1.
\eea
That is, for $N\in even$ the FSR factorizes as 
\be{FSR:N:Ndiv2}
\mS_N = (1+\om)\,\mS_{N/2}, \quad \trm{with}\quad 
 \om=e^{i 2 \pi /N} \;\trm{in}\; \mS_N \to \om'=e^{i 2 \pi /(N/2)} \;\trm{in}\; \mS_{N/2}. 
\ee
This implies that rather than requiring $N$ factors of $\om^p$ to have identical coefficients in order to have an zero amplitude via $\mS_N=0$, one only needs $N/2$ factors of $(\om')^p$ with the same coefficient in order to have a zero sub-amplitude via $\mS_{N/2}=0$, 
(i.e. all the even, or all the odd powers of $\om^p$).
Further, if $N$ contains a divisor of $2^q$ (e.g. 
$N=6=2*3$, $N=8 = 2^3$, $N=10 = 2*5$, $N=12 = 2^2*3$, etc\ldots) then $S_N \propto S_{N/2^q}$, and 
only $N/2^q$ factors of $\om'^p$ (with $\om' \defn \om^{2^q}$) are needed to have identical coefficients in order to obtain a zero amplitude
via $S_{N/2^q}$.
\end{enumerate} 

So far we have discussed FSRs for general $N$ that involve \tit{only} ``+" signs as in \Eq{FSR:N:even:2}.
However, as we shall see in later examples, we can \tit{possibly} have Alternating FSRs (AFSR) where the signs in the geometric series alternate between $\pm 1$. These can arise in \tit{factorized}  zero amplitudes $A=0$ involving larger values of $N$. As discussed in the previous paragraph, consider $N$ begin divisible by $2^q$ in its prime factorization, so that $N = 2^q\, 3^{q_3}\,5^{q_5}\,\ldots$ Then, it is possible that the factorized amplitude $A$ can contain a factor proportional to the AFSR $\mS_N^{(A)}$ defined by
\bsub
\bea{AFSR}
\mS_N^{(A)} &\defn& \sum_{i=1}^{N/2^q} (-1)^{(i-1)}\, \left( \om^{2^{(q-1)}} \right)^{(i-1)}
= 1 - \left( \om^{2^{(q-1)}} \right) + \left( \om^{2^{(q-1)}} \right)^{2} - \cdots + \left( \om^{2^{(q-1)}} \right)^{(N/2^q)-1}, \label{AFSR:line1} \\
 \left( \om^{2^{(q-1)}} \right)\,  \mS_N^{(A)}  &=& 
 \left( \om^{2^{(q-1)}} \right) - \left( \om^{2^{(q-1)}} \right)^{2} + \cdots 
 - \left( \om^{2^{(q-1)}} \right)^{(N/2^q)-1} +  \left( \om^{2^{(q-1)}} \right)^{(N/2^q)}, \label{AFSR:line2} \\
 \Rightarrow  \mS_N^{(A)} &=& \frac{1+ \left( \om^{2^{(q-1)}} \right)^{(N/2^q)}}{1+\om^{2^{(q-1)}} } 
 =  \frac{1+ \om^{N/2}}{1+\om^{2^{(q-1)}} }  = 0, \quad \trm{since}\;\; \om^{N/2}\equiv (-1). \label{AFSR:line3}
\eea 
\esub
Note that series in \Eq{AFSR:line1} terminates with a $+$ sign since $N/2^q\in odd$ (since we have factored out all powers of $2$ from $N$ in its prime factorization). 
The factor of $\om^{2^{(q-1)}}$ appears so that it yields the term
$\om^{N/2}$ in the numerator, which is then raised to the power of $N/2^q$ when the geometric series is summed in  \Eq{AFSR:line3}. We shall see how this arises in the eHOM transitions (i.e. equal photons number in each output port) for the transitions $\ket{n_1,n_2,\ldots,n_N}\overset{S_N}{\to}\ket{\tfrac{n}{N}}^{\otimes N}$, with total input photon number $n\defn \sum_{i=1}^N n_i$, for the case $(N,n) = \{(12,12), (14,14)\}$ with $q=\{2,1\}$, respectively.
\smallskip

From the above considerations, we expect that there exists many more possibilities to obtain an overall zero output amplitude for $N$ even, over that of $N$ odd.
In the following, we will see specific examples of how the FSR for $N$, either even or odd, dictates the ability to have an overall zero destructive interference amplitude on a Fock output state, and how sub-amplitudes can sum zero in various subgroups characterized by the common multiplying coefficient. 

\section{Calculation of amplitude for the transition $\mathbf{\ket{n_1,n_2,\ldots,n_N}\overset{S_N}{\to} \ket{m_1,m_2,\ldots,m_N}}$}\label{sec:calc:of:A}
In this section we calculate the amplitude for the transition 
$\ket{n_1,n_2,\ldots,n_N}\overset{S_N}{\to} \ket{m_1,m_2,\ldots,m_N}$ by two different methods.

\subsection{Arbitrary unitary matrix S}\label{subsec:Arb:S}
Let us first consider a general unitary matrix $S=\{S_{ij}\}$, $i,j\in\{1,2,\ldots,N\}$.
We take the transformation of the boson creation operators for an $N\to N$ port device as
\be{S:adag}
a^\dag_i \to \sum_{j=1}^N\,S_{ij}\,a^\dag_j. 
\ee
Thus, $S_{ij}$ is the amplitude for a singe photon entering input port-$i$ to scatter to output port-$j$,
and $(S_{ij})^k$ is the amplitude for a $k$ photons entering input port-$i$ to all scatter to output port-$j$.
For an input state $\ket{n_1,n_2,\ldots,n_N}$ with total photon number $n \defn \sum_{i=1}^N n_i$ 
the action of $S$ yields, after employing the \tit{multinomial theorem} \newline 
$(x_1+x_2+\cdots+x_N)^n = \sum_{k_1+k_2+\ldots+k_N = n} \frac{n!}{k_1!k_2!\cdots k_N!}\,x_1^{k_1} x_2^{k_2}\cdots x_N^{k_N}$,
\bea{S:in:to:out}
\hspace{-0.5in}
\ket{n_1,n_2,\ldots,n_N}&=& \frac{(a^\dag_1)^{n_1}}{\sqrt{n_1!}}\,  \frac{(a^\dag_2)^{n_2}}{\sqrt{n_2!}}\,\ldots
 \frac{(a^\dag_N)^{n_N}}{\sqrt{n_N!}}\,\ket{0}, \no
 \hspace{-0.5in}
 &\overset{S}{\to}& 
 \frac{\Pi_{j=1}^N (n_i!)}{\sqrt{\Pi_{j=1}^N (n_i!)}}
 \sum_{k_{11} + k_{11}+\ldots+k_{1N}=n_1} 
            \frac{(S_{11} a^\dag_1)^{k_{11}}}{k_{11}!}\, 
            \frac{(S_{12} a^\dag_2)^{k_{12}}}{k_{12}!}\, \cdots
            \frac{(S_{1N} a^\dag_N)^{k_{1N}}}{k_{1N}!}\,\no
\hspace{-0.5in}
&\times&
\hspace{.75in}
 \sum_{k_{21} + k_{22}+\ldots+k_{2N}=n_2} \,    
             \frac{(S_{21} a^\dag_1)^{k_{21}}}{k_{21}!}\, 
            \frac{(S_{22} a^\dag_2)^{k_{22}}}{k_{22}!}\, \cdots
            \frac{(S_{2N} a^\dag_N)^{k_{2N}}}{k_{2N}!}\,\no  
\hspace{-0.5in}
&\times& 
\hspace{1.25in}\vdots
\hspace{1.5in} \vdots\no
\hspace{-0.5in}
&\times&
\hspace{.65in}
 \sum_{k_{N1} + k_{N2}+\ldots+k_{NN}=n_N} \,    
             \frac{(S_{N1} a^\dag_1)^{k_{N1}}}{k_{N1}!}\, 
            \frac{(S_{N2} a^\dag_2)^{k_{N2}}}{k_{N2}!}\, \cdots
            \frac{(S_{NN} a^\dag_N)^{k_{NN}}}{k_{NN}!} \, \ket{0}.                           
\eea
For notational purposes, we will write the conditions on the sums  $\sum_{j} k_{i j} = n_i$ in each row above (as is common practice), as $|K_i| = n_i$. Projecting the above onto the output state
$\ket{m_1,m_2,\ldots,m_N}$, with the same total number photons as the input, 
$\sum_{i=1}^N n_i = n = \sum_{j=1}^N m_j$, introduces  Kronecker delta functions of the column sums $\sum_{i=1}^N k_{ij} = m_j$. Thus, if we define the $N\times N$, matrix $K/n = \{k_{ij}/n\}$ we see that it is 
\tit{doubly stochastic} in the sense that
\be{K}
K \defn \{k_{ij}\}, \quad \trm{with} \quad \sum_{j} k_{i j} = n_i, \quad \trm{and} \quad \sum_{i} k_{i j} = m_j, 
\quad \Rightarrow\quad \sum_{i=1}^N \sum_{j=1}^N k_{ij} = n = \sum_{i=1}^N n_i =  \sum_{j=1}^N m_j.
\ee
In words, the sum of the $i$-th row of $K$ equals $n_i$, 
and the sum of the $j$-th column of $K$ equals $m_j$,
and the sum of all the matrix elements of $K$ must equal the total number of input/output photons to the $N$-port device.
These are the required conditions the $N\times N$ matrix $K$ must fulfill 
for the input state $\ket{n_1,n_2,\ldots,n_N}$ to project onto the output state $\ket{m_1,m_2,\ldots,m_N}$ under the action of the unitary $S$.
\smallskip

From inspection of \Eq{S:in:to:out} we can now conclude the well-known result \cite{Scheel:2004,Scheel:2008} for the amplitude for the transition $\ket{1}^{\otimes N} \overset{S_N}{\to} \ket{1}^{\otimes N}$. Since each $n_i$ and $m_j$ are simply $1$, all the factorial denominators are simply unity. The resulting amplitude
$A = \bra{m_1,m_2,\ldots,m_N} S_N \ket{n_1,n_2,\ldots,n_N} = \trm{Perm}(S_N)$ is simply the 
\tit{permanent of the matrix $S_N$}. Let us illustrate this for the case of $N=3$. Then
\bsub
\bea{S:in:out:N:3}
\hspace{-0.5in}
S \ket{1,1,1} &=& 
(S_{11} a^\dag_1 +S_{12} a^\dag_2+S_{13} a^\dag_3)\,
(S_{21} a^\dag_1 +S_{22} a^\dag_2+S_{23} a^\dag_3)\,
(S_{31} a^\dag_1 +S_{32} a^\dag_2+S_{33} a^\dag_3)\,\ket{0}, \\
\hspace{-0.5in}
&=&
 \cdots +
 \Big(
 S_{11}\,S_{22}\,S_{33} + S_{12}\,S_{23}\,S_{31} + S_{13}\,S_{21}\,S_{32} +
 S_{11}\,S_{23}\,S_{32} + S_{12}\,S_{21}\,S_{33} + S_{13}\,S_{22}\,S_{31}
 \Big)\,a^\dag_1\,a^\dag_2\,a^\dag_3\,\ket{0}
+\cdots, \qquad \\
&=&  \cdots + \trm{Perm}(S)\ket{1,1,1} + \cdots 
\eea
\esub
We see that the total amplitude for the output state $\ket{1,1,1}$ is created by taking the sum of all the  $3!$ permutations of the integers $(1,2,3)$ in the factors $S_{1i}\,S_{2j}\,S_{3k}$, i.e. one term from each of the three parenthesis in the first line \Eq{S:in:out:N:3}, capable of creating the output state  $\ket{1,1,1}$.
\smallskip

For input states containing some $n>1$, the relationship of the total transition amplitude to the permanent of $S$ is more complicated, but was worked out by Scheel in 2004/2008  \cite{Scheel:2004,Scheel:2008}, and will be discussed in the next section. What is also non-trivial and non-obvious is the result proved by Lim and Beige in 2005 \cite{Lim:Beige:2005} (using a clever symmetry argument, discussed later) that for a symmetric $SU(N)$ BS, the amplitude for the transition
$\ket{1}^{\otimes N}\overset{S_N}{\to}\ket{1}^{\otimes N}$ is zero for $N\in even$, and non-zero
for $N\in odd$. We will explore and extend these results in subsequent sections.

\subsection{Symmetric $\mathbf{SU(N)}$ BS}\label{subsec:SUN:BS}
We now specialize to the case of the symmetric  $SU(N)$ BS, 
$S_N = \{S_{ij} = (\om_N)^{(i-1)(j-1)}/\sqrt{N}\}$, with $i,j~\in~\{1,2,\ldots,N\}$.
Since we are only interested in this work in transition amplitudes that are zero, from now on we will drop
all multiplicative factors that are independent of the summation variables $k_{ij}$ of the matrix elements of the $N\times N$ matrix $K$, since these simply factor out of the amplitude, and do not effect the amplitude taking the value of zero (of course, they would effect the value of the amplitude if it was non-zero).
\smallskip

Inserting $(S_N)_{ij} = (\om_N)^{(i-1)(j-1)}$ into \Eq{S:in:to:out} (i.e. dropping the factors of $1/\sqrt{N}$) we see that 
\bsub
\bea{A:SUN}
A &=& \bra{m_1,m_2,\ldots,m_N} S_N \ket{n_1,n_2,\ldots,n_N}, \no
&\propto& 
\sum_{|K_1|=n_1}\,\sum_{|K_2|=n_2} \cdots \sum_{|K_N|=n_N} 
\frac{ \om^{\sum_{ij}^N (i-1)(j-1) k_{ij}},\label{A:SUN:line1} }{\Pi_{i j}^N k_{ij}!}
\label{A:SUN:line1} \\
&\propto& 
\sum_{|K_1|=n_1}\,\sum_{|K_2|=n_2} \cdots \sum_{|K_N|=n_N} 
\frac{ \om^{ \sum_{ij}^N i j k_{ij} } }{\Pi_{i j}^N k_{ij}!}.
\label{A:SUN:line2} 
\eea 
\esub
The last line \Eq{A:SUN:line2}  follows from the first line \Eq{A:SUN:line1} by noting that we can write the exponent of $\om$ as
\bsub
\bea{exp:om}
\sum_{ij}^N (i-1)(j-1) k_{ij} &=& 
\sum_{ij}^N i j k_{ij} 
- \sum_{i}^N i \sum_{j}^N k_{ij}
-\sum_{j}^N j \sum_{i}^N k_{ij}
+\sum_{ij}^N  k_{ij}, \no
&=& 
\sum_{ij}^N i j k_{ij} 
- \sum_{i}^N i\, n_i
-\sum_{j}^N j\, m_j
+ n. \label{exp:om:line1} \\
&\Rightarrow& \om^{\sum_{ij}^N (i-1)(j-1) k_{ij}} \propto \om^{\sum_{ij}^N i j\, k_{ij}}, \label{exp:om:line2} 
\eea
\esub
Here, the last three terms in \Eq{exp:om:line1} are now independent of the summation variables $k_{ij}$ (since they have been summed over to give $n_i, m_j$ and $n$, respectively), and thus can be factored out of the over amplitude - again, without effecting a sought for amplitude $A=0$.
Further, we can always interpret any exponent of $\om$ as $\Mod_N$.
Note that we can also write the exponent of $\om$ as the \tit{point multiplication} $\odot$ 
(element-by-element), of the matrix $(IJ) \defn \{i j\}$ with the matrix 
$K= \{k_{ij}\}$, i.e. 
\be{IJK}
(IJ)\odot K\equiv \Mod[\,\sum_{ij}^N i j\, k_{ij}, N\,]
\ee
From \Eq{exp:om:line2}, \Eq{IJK} determines the exponent ($\Mod_N$) of $\om$, and hence will play an important role, along with the product of  factorial denominators 
$\Pi_{ij}^N \,k_{ij}$ in \Eq{A:SUN:line2},  
in determining which sub-amplitudes of the total amplitude $A$ will sum to zero separately in groups.

\subsection{An exhaustive search method to evaluate a zero amplitude $\mathbf{A=0}$ \\ for the transition $\mathbf{ \bra{m_1,m_2,\ldots,m_N} S \ket{n_1,n_2,\ldots,n_N}}$, and the JKN estimate  for the \\ number of valid $\mathbf{K}$ matrices, satisfying the row-sums and column-sum conditions}\label{subsec:brute:force}
In order to calculate \Eq{A:SUN:line2} for the amplitude $A$ we need to 
(i) form the  $N\times N$ matrices $K$, and 
(ii) ensure that the sum of each row $i$ sums to input photon number $n_i$, and 
the sum of each column $j$ sums to the output photon number $m_j$.
\smallskip

We can form the matrices $K$ by an exhaustive enumeration of the potential candidates that will  subsequently each be tested for the validity condition in (ii) as follows.
For each  photon number $n_i$ in the input state $\ket{n_1,n_2,\ldots,n_N}$ we
we form the partition $P_N(n_i)$ of $n_i$, i.e. we solve the Diophantine equation 
$k_{i1} +  k_{i2} +\cdots+k_{iN} = n_i$ for the $N$ integers $\{k_{i,1}, k_{i,2},\ldots, k_{iN}\}$.
This is also known as the \tit{weak partition of $n$}, i.e. the number of distinct ordered sets of $N$ non-negative integers that sum to $n$.
An element of this partition will form the $i$-th row of $N\times N$ matrix $K$ candidate.
We then do this for each of the $N$ rows $i\in\{1,2,\ldots,N\}$, to form the full matrix $K$.
\begin{table}[!ht]
\begin{tabular}{|c|c|c|c|c|c|c|c|c|c|c|c|} \hline
   {}     & \bf{n}=1 & \bf{n}=2 &  \bf{n}=3 & \bf{n}=4 & \bf{n}=5 & \bf{n}=6 & \bf{n}=7 & \bf{n}=8 & \bf{n}=9 & \bf{n}=10 \\ \hline
 \bf{N=2} & 2  & 3 & 4 & 5 & 6 & 7 & 8 & 9 & 10 & 11 \\  \hline
 \bf{N=3} & 3  & 6 & 10 & 15 & 21 & 28 & 36 & 45 & 55 & 66 \\ \hline
 \bf{N=4} & 4  & 10 & 20 & 35 & 56 & 84 & 120 & 165 & 220 & 286 \\ \hline
 \bf{N=5} & 5  & 15 & 35 & 70 & 126 & 210 & 330 & 495 & 715 & 1001 \\ \hline
 \bf{N=6} & 6  & 21 & 56 & 126 & 252 & 462 & 792 & 1287 & 2002 & 3003 \\ \hline
 \bf{N=7} & 7  & 28 & 84 & 210 & 462 & 924 & 1716 & 3003 & 5005 & 8008 \\ \hline
 \bf{N=8} & 8  & 36 & 120 & 330 & 792 & 1716 & 3432 & 6435 & 11440 & 19448 \\ \hline
 \bf{N=9} & 9  & 45 & 165 & 495 & 1287 & 3003 & 6435 & 12870 & 24310 & 43758 \\ \hline
 \bf{N=10} & 10  & 55 & 220 & 715 & 2002 & 5005 & 11440 & 24310 & 48620 & 92378 \\ \hline
 \bf{N=11} & 11  & 66 & 286 & 1001 & 3003 & 8008 & 19448 & 43758 & 92378 & 184756 \\ \hline
 \bf{N=12} & 12  & 78 & 364 & 1365 & 4368 & 12376 & 31824 & 75582 & 167960 & 352716 \\ \hline
\end{tabular}
\caption{The number of partitions $|P_N(n)| = \scriptsize{\left(\begin{array}{c} n + N-1 \\n\end{array}\right)}$  of the total number of input/output photons $n$ for a given number of input/output ports $N$.}
\label{tbl:PNn}
\end{table}
\smallskip

In \Tbl{tbl:PNn} we list the number  $|P_N(n)|$ of $N$-vectors in a partition of an input photon number $n$ for a given $N$ (input modes). This is given by the following formula
\be{PNn:formula}
|P_N(n)| = \left(\begin{array}{c} n + N-1 \\n\end{array}\right) = 
 \left(\begin{array}{c} n + N-1 \\N-1\end{array}\right).
\ee 
We can interpret this formula as placing $n$ ``stars" in $N$ bins. To divide the stars into $N$ bins, we need $N-1$ separators (``bars"). The total number items to arrange, i.e. the number of ``stars + bars," is $n+N-1$. 
Taking $n$ ``stars" at a time (or equivalently $N-1$ ``bars" at a time) yields the number of combinations given by \Eq{PNn:formula}.
\smallskip

The total number of possible candidates to be searched and check for the row-sum and column-sum validity in (ii) above is then the product of the number of all these partitions, i.e 
$|P_N(\n\defn\{n_1,n_2,\ldots,n_N\})| \defn \Pi_{i=1}^N\, P_N(n_i)$. \tit{Mathematica} can analyze a few 10s of millions of total candidates analytically (i.e. as a function of $\om$) in a reasonable amount of time (10s of mins to roughly an hour or two). The actual number of \tit{valid} $K$ matrices satisfying the row/column sum conditions is drastically smaller, but can be in the range of 10s to a few 1000s. 
\smallskip

The number $\Om(\n,\m)$ of non-negative integer matrices with given 
row sums $\n\defn\{n_1,n_2,\ldots,n_N\}$,
and column sums $\m\defn\{m_1,m_2,\ldots,m_N\}$ (and uniformly sampling from them),
appears in a variety of problems in mathematics and statistics, but no closed-form expression for it is known, so one must rely on
approximations of various kinds.
Here we use an approximate formula for $\Om(\n,\m)$ by Jerdee, Kirley and Newman (JKN)\cite{Jerdee:2022} , adapted to square matrices. 
\bsub
\bea{Om:JKN}
\Om(\n,\m)&\simeq& \left(\begin{array}{c}n + N\,\alpha_c -1 \\N\end{array}\right)^{-1}\,
\prod_{i=1}^N \left(\begin{array}{c}n_i + N\,\alpha_c -1 \\n_i\end{array}\right)\,
\prod_{j=1}^N \left(\begin{array}{c}m_j + N -1 \\m_j\end{array}\right), \label{Om:JKN:line1} \\
n \defn \sum_{i=1}^N n_i &\equiv& \sum_{j=1}^N m_j, \quad
\alpha_c = \frac{n^2 - n + (n^2-c^2)/N}{c^2-N}, \quad c^2\defn \sum_{j=1}^N m_j^2,  \label{Om:JKN:line2} \\
\Om^{(\trm{sym})}_{\n,\m} &\defn& \half\, \big( \Om(\n,\m) + \Om(\m,\n) \big),  \label{Om:JKN:line3}
\eea
\esub
where $\simeq$ indicates that we should round the result on the rhs of \Eq{Om:JKN} to the nearest integer.
We see from the terms in the products in \Eq{Om:JKN} that the last term (involving the output photon numbers $m_j$ in the $j$-th port)  is simply 
$|P_N(\m)|$, the total number of possible partitions to search for that satisfies the column-sum condition, independent of the row sum condition. The middle term (involving the input number of photons  $n_i$ in the $i$-th port) is essentially  $|P_N(\n)|$  (the total number of possible partitions to search for that satisfies the row-sum condition, independent of the column sum condition), except that the number of ports $N$ has been modified to a non-integer number of ports $N\to N\,\alpha_c$. Similarly, the denominator (first term in \Eq{Om:JKN}) is essentially  $|P_N(n)|$, again with  $N\to N\,\alpha_c$, and $n\defn\sum_{i=1}^N n_i$ the total number of input/output photons to the $N$-port beam splitter. 
\smallskip

As described in JLK \cite{Jerdee:2022} the purpose of $\alpha_c$ is to approximately match the expectation values of the row sum $\m_j$ and the covariances between the row-sum $\trm{cov}(m_j, m_{j'})$, which can be computed using an ansatz for the conditional probability $P(\n|\m)\simeq P(\n|\alpha_c)$ of finding a matrix with given row-sums $\n$, given the column-sum $\m$, in terms of a variable, non-integer column-sum $\alpha_c$.
The computation of the expectations and the variances can be computed analytically, leading to 
a condition that is satisfied by the value of $\alpha_c$ in \Eq{Om:JKN:line2},
 leading  formally to a non-integer number of input ports $N\to N\,\alpha_c$.
  $\Om_{\n,\m}^{(\trm{sym})}$ defined in \Eq{Om:JKN:line3}, is the JKN-recommended formula for the average of 
  $\Om(\n,\m)$ using an ansatz for $P(\n|\m)\simeq P(\n|\alpha_c)$, and 
  $\Om(\m,\n)$ using an ansatz for $P(\m|\n)\simeq P(\m|\alpha_c)$, where the roles of the row-sums $\n$ and column-sums $\m$ have been swapped.
 \smallskip
 
 The JKN formula  $\Om_{\n,\m}^{(\trm{sym})}$ is fast, and fairly accurate, even for large values of $n$ and $N$ (see Jerdee \tit{et al.} \cite{Jerdee:2022} for comparison tables with other known approximation formulas from the literature). 
 For the zero amplitude $A=0$,  $N=4$ transition $\ket{\n}\overset{S_4}{\to}\ket{\m}$ given by 
 $\n=\{7,7,7,7\}=\m$ our exhaustive search method yields $207,360,000$ possible candidate $K$ matrices to search through (taking $4,722$ secs in \tit{Mathematica}), with only $381,424$ actually valid $K$ matrices ($\sim 0.184\%$) satisfying the requisite row-sums and column-sums conditions.  Using $\Om_{\n,\m}^{(\trm{sym})}$ in \Eq{Om:JKN:line3} yields an estimate of $376,888$ valid $K$ matrices, which is only shy by roughly $1.25\%$ of the exact value.
 \smallskip
 
 Even with the same total number $n$ of input photons leading to the same output state, the distribution of the input photon number $\n$ drastically alters the possible number of $K$ candidates to search through, as well as the actual number of valid $K$ matrices. For the  $A=0$,  $N=4$ transition
 $\ket{0,0,14,14}\overset{S_4}{\to}\ket{7,7,7,7}$, the candidate number of searchable $K$ matrices is $462,400$ (taking $10.3$ secs in \tit{Mathematica}), with only $344$ actual valid $K$ matrices. Using the JKN formulas above we find and estimate of $\Om(\n,\m)=213$, $\Om(\m,\n)=345$, and the average $\Om_{\n,\m}^{(\trm{sym})} = 279$.
\smallskip

Recall, that our goal in this work is to compute the zero amplitude $A$ \tit{analytically} as a function of $\om$, factor it to examine its structure and relationship to the FSR discussed above, and then classify them into groups which sum separately to zero. Only afterwards do we substitute in the numerical value of $\om = e^{i 2 \pi/N}$ to double check that $A=0$ numerically.
But this is typically an afterthought, since we can see from the polynomial structure in $\om$ whether or not the amplitude $A$ will be zero.

\subsection{Scheel's method \cite{Scheel:2004,Scheel:2008} to compute the transition amplitude $\mathbf{A = \bra{m_1,m_2,\ldots,m_N} S \ket{n_1,n_2,\ldots,n_N}}$ \\ by means of a permanent of matrix $\mathbf{\L}$ whose matrix elements are taken from $\mathbf{S}$}\label{subsec:Schee}
Scheel  \cite{Scheel:2004,Scheel:2008} details a method to compute the transition amplitude 
$A = \bra{m_1,m_2,\ldots,m_N} S \ket{n_1,n_2,\ldots,n_N}$ by means of a permanent of an associated matrix $\L$ with matrix elements taken from $S$.
First, some definitions. The permanent Perm$(\L)$ of an $n\times n$ matrix $\L$,
of total input/output photon number $n=\sum_{i=1}^N n_i = \sum_{j=1}^N m_j$, 
is given by
\be{}
\trm{Perm}(\L) = \sum_{\sigma\in \mS_n}\, \prod_{i=1}^n \L_{i \sigma_i}, 
\ee
where $\mS_n$ is the group of $n!$ permutations of the integers $\{1,2,\ldots,n\}$, and $\sigma_i$ is the $i$-th element in the permutation $\sigma$. For example, for $\mS_3$ with $\sigma = (2,3,1)$, we have 
$\sigma_2=3$.  Perm$(\L)$ has the same decomposition as the Det$(\L)$ except all minus signs are replaced by plus signs. Thus, for example, permanent of a 
$3\times 3$ matrix $\L$ is given by 
\be{perm:n3}
\trm{Perm}
\left(
\begin{array}{ccc}
\L_{11} & \L_{12} &\L_{13} \\
\L_{21} & \L_{22} &\L_{23} \\
\L_{31} & \L_{32} &\L_{33} 
\end{array}
\right)
= \L_{11}\,\L_{22}\,\L_{33}+\L_{12}\,\L_{23}\,\L_{31}+\L_{13}\,\L_{21}\,\L_{32}
+  \L_{11}\,\L_{23}\,\L_{32}+\L_{12}\,\L_{21}\,\L_{33}+\L_{13}\,\L_{22}\,\L_{31}. 
\ee
Each term in the sum, e.g. $\L_{12}\,\L_{23}\,\L_{31}$, is called a \tit{diagonal}, and contains exactly $n$ terms in the product. Perm$(\L)$ is then given by the sum of all the possible diagonals.
\smallskip

As we saw previously, the essential part of the transition amplitude 
$A = \bra{m_1,m_2,\ldots,m_N} S \ket{n_1,n_2,\ldots,n_N}$
is given by 
$\prod_{i,j}^N \L^{k_{ij}}_{ij}$, which is a product of exactly $n$ factors, with $K=\{k_{ij}\}$ satisfying the row-sum and column-sum conditions discussed earlier. The key insight is that 
$\prod_{i,j}^N \L^{k_{ij}}_{ij}$ is a diagonal of the following matrix constructed from the matrix elements of $N\times N$ symmetric BS matrix $S_N$
\bea{Lambda:defn}
\L[1^{m_1}, 2^{m_2},\ldots,N^{m_N} | 1^{n_1}, 2^{n_2},\ldots,N^{n_N}],
\eea
and that the amplitude $A$ is given by the permanent of this matrix \cite{Scheel:2004,Scheel:2008} via
\be{A:eq:Perm:Lambda}
A\defn \bra{m_1,m_2,\ldots,m_N} \L \ket{n_1,n_2,\ldots,n_N} 
=
\frac{\trm{Perm}(\L[1^{m_1}, 2^{m_2},\ldots,N^{m_N} | 1^{n_1}, 2^{n_2},\ldots,N^{n_N}])}
{\sqrt{\prod_i^N n_i! \prod_i^N m_i!}}.
\ee
The key symmetry idea  is that if we take the permanent of the matrix in \Eq{Lambda:defn}, then out of all the possible permutations of the column indices, we observe that $\prod_j n_j!$ of those permutations are identical. Similarly there are $\prod_i m_i!$ ways of distributing the row indices. Hence, not all diagonals are distinct from each other, and only 
$\frac{(\prod_i m_i) (\prod_j n_j) }{\prod_{ij} k_{ij}}$ terms actually lead to the \tit{same} diagonal.
This accounts for the denominator in \Eq{A:eq:Perm:Lambda} (where the factor $\prod_{ij} k_{ij}$ has cancelled with other factors already present in the multinomial decomposition 
of the amplitude $A$). From now on when we write ``$A=\PermL$" we will drop the denominator factors in 
\Eq{A:eq:Perm:Lambda}, since we primarily interested in whether or not the amplitude $A$ is zero, vs its actual value (if non-zero). Thus, in reality we actually have $A\propto\PermL$.
\smallskip

The matrix in \Eq{Lambda:defn} is constructed from the matrix elements of $\L$ by the following procedure due to Scheel \cite{Scheel:2004,Scheel:2008}.
The matrix element $\L_{1\bullet}$ appears $m_1$ times in each column, 
$\L_{2\bullet}$ appears $m_2$ times in each column, $\cdots$, 
$\L_{N\bullet}$ appears $m_N$ times in each column. 
Then, in the first $m_1$ rows, $\L_{1 1}$ appears $n_1$ times in each of those  rows, 
followed by $\L_{1 2}$ appearing $n_2$ times, $\cdots$, 
followed by $\L_{1 N}$ appearing $n_N$ times.
We then repeat this procedure for the 
next $m_2$ rows containing $\L_{2 1}$ $n_1$ times, etc\ldots,
until the final $m_N$ rows containing $\L_{N N}$ $n_N$ times. 
Thus, each row index $i$ occurs $m_i$ times, and each column index $j$ appears $n_j$ times.
A couple of examples for $N=3$ with different number of total photons $n$ will help illustrate the construction.
\bsub
\bea{A:eq:Perm:L:examples}
A=\bra{0,2,1} \L \ket{1,1,1}
&\Leftrightarrow& \trm{Perm}(\L[1^0, 2^2 ,3^1|1^1, 2^1, 3^1])= 
\trm{Perm}\left(
\begin{array}{ccc}
\L_{2 1} & \L_{2 2} & \L_{2 3} \\
\L_{2 1}  & \L_{2 2} & \L_{2 3} \\
\L_{3 1}  &\L_{3 2} & \L_{3 3}
\end{array}
\right), \label{A:eq:Perm:L:examples:line1} \\
A=\bra{2,2,2} \L \ket{1,2,3}
&\Leftrightarrow& \trm{Perm}(\L[1^2, 2^2 ,3^2|1^1, 2^2, 3^3]) = 
\trm{Perm}\left(
\begin{array}{cccccc}
\L_{1 1} & \L_{1 2} & \L_{1 2} &\L_{1 3} &\L_{1 3} &\L_{1 3} \\
\L_{1 1} & \L_{1 2} & \L_{1 2} &\L_{1 3} &\L_{1 3} &\L_{1 3} \\
\L_{2 1} & \L_{2 2} & \L_{2 2} &\L_{2 3} &\L_{2 3} &\L_{2 3} \\
\L_{2 1} & \L_{2 2} & \L_{2 2} &\L_{2 3} &\L_{2 3} &\L_{2 3} \\
\L_{3 1} & \L_{3 2} & \L_{3 2} &\L_{3 3} &\L_{3 3} &\L_{3 3} \\
\L_{3 1} & \L_{3 2} & \L_{3 2} &\L_{3 3} &\L_{3 3} &\L_{3 3} 
\end{array}
\right).  \label{A:eq:Perm:L:examples:line2} 
\eea
\esub
In \Eq{A:eq:Perm:L:examples:line1} the total number of photons is $n=3$, and thus we need to take the permanent of of an $3\times 3$ matrix. Similarly, the total number of photons in 
\Eq{A:eq:Perm:L:examples:line2} is $n=6$, and hence we need to compute the permanent of a $6\times 6$ matrix. In general, the permanent of an $n\times n$ matrix contains $n!$ terms in its expansion, so for large input number of photons, even for small $N$, this computation grows prohibitally costly.
\smallskip

There is an alternative, easy to describe/code algorithm to construct $\L$ due to Aaronson and Arkhipov \cite{Aaronson:2010} and Chabaud \tit{et al.} \cite{Chabaud:2022}. Since \tit{Mathematica} lists are row-based, we will use the Chabaud method, although the Aaronson method, simply performs the construction using columns first. Both are equivalent to the method above due to Scheel \cite{Scheel:2004,Scheel:2008}.
\smallskip

The Chabuad construction proceeds in two steps, and can be visualized as 
$S_N\overset{\n}{\to}\L_{\n}\overset{\m}{\to}\L_{\m\n}\equiv \L(S_N)$.
Here as usual, we are considering the transition $\ket{\n}\overset{S_N}{\to}\ket{\m}$ with
$\ket{\n} = \ket{n_1,n_2,\ldots,n_N}$ and $n\defn\sum_{i=1}^N n_i$.
\begin{quote}
\begin{enumerate}
\item[Step 1:] $S_N\overset{\n}{\to}\L_{\n}$: create an $N\times n$ matrix $\L_{\n}$ by repeating the $i$-th row of $S_N$, $n_i$ times (if $n_i=0$, skip the $i$-th row of $S_N$).
 \item[Step 2:] $\L_{\n}\overset{\m}{\to}\L_{\m\n}\equiv \L(S_N)$: 
 now create the $n\times n$ matrix $\L_{\m\n}$ by repeating the $j$-th column of $\L_{\n}$, $m_j$ times (if $m_j=0$, skip the $j$-th column of $\L_{\n}$).
\end{enumerate}
\end{quote}

In \App{app:LChabaud}
\Fig{fig:Lambda:Chabaud} shows the \tit{Mathematica} code to implement the Chabaud construction of 
$\L(S_N)$ (the output $\L$), consisting essentially of two simple \ttt{Do} (or \ttt{For}) loops. This code is easily translatable into other programable languages, such as \tit{Python}.
\smallskip

In the next section on results, we will investigate both Scheel's method to compute the transition matrix element $A = \bra{m_1,m_2,\ldots,m_N} S \ket{n_1,n_2,\ldots,n_N}$, as well as the exhaustive search method. Again, the point is not just to compute the zero amplitude $A=0$, 
but to also understand the detailed destructive interference structure analytically, i.e. as a function of $\om$ for the symmetric $SU(N)$ beam splitter $S_N$.

\section{The cancellation of groups of sub-amplitudes summing separately zero within a total zero amplitude 
$\mathbf{A=0}$ transition}\label{sec:Results:cancellation:in:groups}
In this section we present results for the zero amplitudes for various illustrative cases within $N=3,4$, focusing on how and when groups of sub-amplitudes separately sum to zero within a total zero amplitude $A=0$. At the center of these results is how groups of sub-amplitudes, with equal coefficients, collect to yield an FSR whose value is zero when evaluated on $\om = e^{i 2\pi /N}$.

\subsection{The gHOM effect  $\mathbf{\ket{1}^{\otimes N}\overset{S_N}{\to}\ket{1}^{\otimes N}}$  
for $\mathbf{N=\{2,3,4,\ldots,14\}}$:\,}\label{subsec:1sto1s}
The amplitude $A$  for the transition 
$\ket{1}^{\otimes N}\overset{S_N}{\to}\ket{1}^{\otimes N}$ was studied by 
Lim and Beige in 2005 \cite{Lim:Beige:2005} who proved by a clever symmetry argument (without having to compute Perm$(\L)$ explicitly, and which we will discuss in the next section) that $A=0$ 
iff $N\in even$, and non-zero if $N\in odd$. 
While their symmetry argument tells us when $A=\Perm(S_N)=0$, it does not inform us how the total destructive interference comes about through the cancellation of sub-amplitudes/diagrams.
\begin{table}[!ht]
\hspace{-0.65in}
\begin{tabular}{|c|c|c|} \hline
\bf{N} & $\bf{A\boldsymbol{\propto}\bf{Perm}(\L)} $ & $\mathbf{A(\boldsymbol{\om=}e^{i 2 \pi/N})}$\\ \hline
2 & $(1+\om)$ & 0\\ \hline
3 & $\om\,(1+\om)$ & -3\\ \hline
4 & $(1+2\,\om)\,(1+\om^2)$ & 0\\ \hline
5 & $4+5\,(\om+\om^2+\om^3+\om^4)$ & -5\\ \hline
6 & $ (1+\om^3)\,(4+3\,(\om+\om^2)$ & 0\\ \hline
7 & $6+7\,(\om+\om^2+\om^3+\om^4+\om^5+\om^6)$ & -105\\ \hline
8 & $(1+\om^4)\,(89+72\,\om+82\,\om^2+72\,\om^3)$ & 0\\ \hline
9 & $(486+504\,\om+504\,\om^2+485\,\om^3+504\,\om^4+504\,\om^5+485\,\om^6+504\,\om^7+504\,\om^8)$ & 81\\ \hline
10 & $(1+\om^5)\,\big(916+905(\om+\om^2+\om^3+\om^4)$ & 0\\ \hline
11 & $(22030+21989\,(\om+ \om^2+ \om^3+ \om^4+ \om^5+ \om^6+ \om^7+ \om^8+ \om^9+ \om^{10})$ & 6765\\ \hline
12 & $(1+\om^6)\,(1884+1966\,\om+1883\,\om^2+1968\,\om^3+1883\,\om^4+1966\,\om^5)$ & 0\\ \hline
13 & $3350796 +3349567\,(\om+ \om^2+ \om^3+ \om^4+ \om^5+ \om^6+ \om^7+ \om^8+ \om^9+ \om^{10} + \om^{11}+ \om^{12})$ & 175747\\ \hline
14 & $(1+\om^7)\,(1985502+1985683\,(\om+\om^2+\om^3+\om^4+\om^5 + \om^6)$ & 0\\ \hline
\end{tabular}
\caption{$\om$ dependence for the amplitude $A$ (dropping all numerical prefactors) for the transition $\ket{1}^{\otimes N}\overset{S_N}{\to}\ket{1}^{\otimes N}$. }
\label{tbl:A:1sto1s}
\end{table}
In \Tbl{tbl:A:1sto1s} we symbolically compute the $\om$ dependence of the amplitude $A$ (dropping all numerical prefactors) for  Perm$(\L)$ discussed in the previous section, for $N\in\{2,3,,4\ldots,14\}$.
\smallskip

We observe several interesting features. 
\begin{quote}
\begin{enumerate}
\item[(1)] Since all the factorial denominators are simply $1$ for a single photon in each input/output port,
all terms in the full amplitude have the \tit{same} coefficient.  However, sub-amplitudes can still form subgroups of terms that can sum to zero by the FSR discussed in \Eq{FSR:N:even}, \Eq{FSR:N:even:2} and \Eq{FSR:N:Ndiv2}.
\item[(2)] Lim and Beige's results is seen to explicitly hold, since as discussed in \Eq{FSR:N:even} for the FSR, with  $N\in even$, the amplitude 
$A\propto \trm{Perm}(S_N)\propto (1+\om^{N/2}) \overset{\om=e^{i 2 \pi/N}}{\longrightarrow}=0$
since $1+(e^{i 2 \pi/N})^{N/2} = 1+e^{i \pi} = 0$.
\item[(3)] For $N$ \tit{odd}, it is curious how $A\propto \trm{Perm}(S_N)$ ``just fails" to be proportional to a full FSR. For example, for $N\in\{3,5,7,11,13\}$ (i.e. skipping $N=9$) the coefficient multiplying all non-zero powers of $\om$ are the same, and nearly identical, but different than the coefficient of $\om^0=1$.
Thus adding and subtracting this coefficient, gives a non-zero result proportional to $\om^0$.
For example, for $N=5$, 
$A\propto 4+5\,(\om+\om^2+\om^3+\om^4) =  \big((4+1) -1\big)+5\,(\om+\om^2+\om^3+\om^4)=
-1+ 5\, (1+\om+\om^2+\om^3+\om^4) \overset{FSR_{N=5}}{=} -1$.
Note that $N=9=3^2$ does not fit this pattern, which we conjecture might be related to purely odd prime decomposition of $N$ (i.e. containing \tit{no} powers of $2$). In general, for $N$ \tit{odd} the only way 
Perm$(S_N)$ could be zero, is if it involves the \tit{full} FSR expression $\sum_{i=1}^{N (odd)} \om^{i-1}=0$, which we observe from \Tbl{tbl:A:1sto1s} that  it (``barely") does not.
\end{enumerate}
\end{quote}

\subsection{A deeper inspection of the cancellations in $\mathbf{A=0}$ for the $\mathbf{N=4}$ transition 
$\mathbf{\ket{1111}\overset{S_4}{\to}\ket{1111}}$}\label{subsec:N4:1111:1111}
While the calculation of Scheel's permanent in the previous section tells us why the amplitude $A$ is zero for the $N\in even$ transitions, it does \tit{not} present any insight as to how the total amplitude may become zero, through of groups of sub-amplitudes summing separately to zero. Thus, in this section we use the exhaustive search method discussed previously to inspect the valid $K$ matrices for the  transition
$\ket{1111}\overset{S_4}{\to}\ket{1111}$, and observe how they are associated with the powers $p$ of 
$\om^p$. In \Eq{N4:om0:1111:1111:om:0} - \Eq{N4:om0:1111:1111:om:3} we show the matrices $K$ formed from the $4!=24$~permutations 
\bsub
\bea{N4:om0:1111:1111}
\hspace{-0.75in}
&{}& \om^0:\;
\scriptsize{
\left\{
\begin{array}{cccc}
 \left(
\begin{array}{cccc}
 0 & 0 & 0 & 1 \\
 0 & 0 & 1 & 0 \\
 0 & 1 & 0 & 0 \\
 1 & 0 & 0 & 0 \\
\end{array}
\right), & 
\left(
\begin{array}{cccc}
 0 & 0 & 0 & 1 \\
 1 & 0 & 0 & 0 \\
 0 & 1 & 0 & 0 \\
 0 & 0 & 1 & 0 \\
\end{array}
\right), 
& \left(
\begin{array}{cccc}
 0 & 1 & 0 & 0 \\
 0 & 0 & 1 & 0 \\
 0 & 0 & 0 & 1 \\
 1 & 0 & 0 & 0 \\
\end{array}
\right), & 
\left(
\begin{array}{cccc}
 0 & 1 & 0 & 0 \\
 1 & 0 & 0 & 0 \\
 0 & 0 & 0 & 1 \\
 0 & 0 & 1 & 0 \\
\end{array}
\right) 
\end{array}
\right\}, 
} \label{N4:om0:1111:1111:om:0} \\
\hspace{-0.75in}
&{}& \om^2:\;
\scriptsize{
\left\{
\begin{array}{cccc}
 \left(
\begin{array}{cccc}
 0 & 0 & 1 & 0 \\
 0 & 0 & 0 & 1 \\
 1 & 0 & 0 & 0 \\
 0 & 1 & 0 & 0 \\
\end{array}
\right),
 & \left(
\begin{array}{cccc}
 0 & 0 & 1 & 0 \\
 0 & 1 & 0 & 0 \\
 1 & 0 & 0 & 0 \\
 0 & 0 & 0 & 1 \\
\end{array}
\right),
 & \left(
\begin{array}{cccc}
 1 & 0 & 0 & 0 \\
 0 & 0 & 0 & 1 \\
 0 & 0 & 1 & 0 \\
 0 & 1 & 0 & 0 \\
\end{array}
\right),
 & \left(
\begin{array}{cccc}
 1 & 0 & 0 & 0 \\
 0 & 1 & 0 & 0 \\
 0 & 0 & 1 & 0 \\
 0 & 0 & 0 & 1 \\
\end{array}
\right)
\end{array}
\right\}, 
} \label{N4:om0:1111:1111:om:2}  \\
\hspace{-0.5in}
&{}& \no
\hspace{-0.5in}
&{}& \no
\hspace{-0.5in}
&{}& 
\om^1:\;
\scriptsize{
\left\{
\begin{array}{cccccccc}
\left(
\begin{array}{cccc}
 0 & 0 & 0 & 1 \\
 0 & 0 & 1 & 0 \\
 1 & 0 & 0 & 0 \\
 0 & 1 & 0 & 0 \\
\end{array}
\right),
 & \left(
\begin{array}{cccc}
 0 & 0 & 0 & 1 \\
 0 & 1 & 0 & 0 \\
 0 & 0 & 1 & 0 \\
 1 & 0 & 0 & 0 \\
\end{array}
\right),
 & \left(
\begin{array}{cccc}
 0 & 0 & 1 & 0 \\
 0 & 0 & 0 & 1 \\
 0 & 1 & 0 & 0 \\
 1 & 0 & 0 & 0 \\
\end{array}
\right),
 & \left(
\begin{array}{cccc}
 0 & 0 & 1 & 0 \\
 1 & 0 & 0 & 0 \\
 0 & 0 & 0 & 1 \\
 0 & 1 & 0 & 0 \\
\end{array}
\right),
 & \left(
\begin{array}{cccc}
 0 & 1 & 0 & 0 \\
 0 & 0 & 0 & 1 \\
 1 & 0 & 0 & 0 \\
 0 & 0 & 1 & 0 \\
\end{array}
\right),
 & \left(
\begin{array}{cccc}
 0 & 1 & 0 & 0 \\
 1 & 0 & 0 & 0 \\
 0 & 0 & 1 & 0 \\
 0 & 0 & 0 & 1 \\
\end{array}
\right),
 & \left(
\begin{array}{cccc}
 1 & 0 & 0 & 0 \\
 0 & 0 & 1 & 0 \\
 0 & 1 & 0 & 0 \\
 0 & 0 & 0 & 1 \\
\end{array}
\right), &
 \left(
\begin{array}{cccc}
 1 & 0 & 0 & 0 \\
 0 & 1 & 0 & 0 \\
 0 & 0 & 0 & 1 \\
 0 & 0 & 1 & 0 \\
\end{array}
\right) 
 \end{array}
\right\}, 
}\qquad\quad  \label{N4:om0:1111:1111:om:1}  \\
\hspace{-0.5in}
&{}& 
\om^3:\;
\scriptsize{
\left\{
\begin{array}{cccccccc}
\left(
\begin{array}{cccc}
 0 & 0 & 0 & 1 \\
 0 & 1 & 0 & 0 \\
 1 & 0 & 0 & 0 \\
 0 & 0 & 1 & 0 \\
\end{array}
\right),
 & \left(
\begin{array}{cccc}
 0 & 0 & 0 & 1 \\
 1 & 0 & 0 & 0 \\
 0 & 0 & 1 & 0 \\
 0 & 1 & 0 & 0 \\
\end{array}
\right),
 & \left(
\begin{array}{cccc}
 0 & 0 & 1 & 0 \\
 0 & 1 & 0 & 0 \\
 0 & 0 & 0 & 1 \\
 1 & 0 & 0 & 0 \\
\end{array}
\right),
 & \left(
\begin{array}{cccc}
 0 & 0 & 1 & 0 \\
 1 & 0 & 0 & 0 \\
 0 & 1 & 0 & 0 \\
 0 & 0 & 0 & 1 \\
\end{array}
\right),
 & \left(
\begin{array}{cccc}
 0 & 1 & 0 & 0 \\
 0 & 0 & 0 & 1 \\
 0 & 0 & 1 & 0 \\
 1 & 0 & 0 & 0 \\
\end{array}
\right),
 & \left(
\begin{array}{cccc}
 0 & 1 & 0 & 0 \\
 0 & 0 & 1 & 0 \\
 1 & 0 & 0 & 0 \\
 0 & 0 & 0 & 1 \\
\end{array}
\right),
 & \left(
\begin{array}{cccc}
 1 & 0 & 0 & 0 \\
 0 & 0 & 0 & 1 \\
 0 & 1 & 0 & 0 \\
 0 & 0 & 1 & 0 \\
\end{array}
\right),
 & \left(
\begin{array}{cccc}
 1 & 0 & 0 & 0 \\
 0 & 0 & 1 & 0 \\
 0 & 0 & 0 & 1 \\
 0 & 1 & 0 & 0 \\
\end{array}
\right)
 \end{array}
\right\}. 
}\qquad\quad  \label{N4:om0:1111:1111:om:3} 
\eea
\esub
of the row $(1,0,0,0)$, whose 
point-product  $|IJ\odot K| \defn \Mod[ \prod_{i,j}^N i j k_{ij}, 4]$,
with 
$IJ[4] = \{ \Mod[i j, 4] \}=
\tiny{
\left(
\begin{array}{cccc}
 1 & 2 & 3 & 0 \\
 2 & 0 & 2 & 0 \\
 3 & 2 & 1 & 0 \\
 0 & 0 & 0 & 0 \\
\end{array}
\right),
}$ 
yields the exponent $p$ of $\om^p$ associated with the matrix $K$.
For example, using the first matrix from  \Eq{N4:om0:1111:1111:om:0} and \Eq{N4:om0:1111:1111:om:2},
and similarly the first matrix from  \Eq{N4:om0:1111:1111:om:1} and \Eq{N4:om0:1111:1111:om:3},
we have
\bsub
\bea{N:4:1111:1111:IJK}
\hspace{-0.5in}
\om^0:\;\; IJ\odot K &=& 
\left(
\begin{array}{cccc}
 0 & 0 & 0 & 0 \\
 0 & 0 & 2 & 0 \\
 0 & 2 & 0 & 0 \\
 0 & 0 & 0 & 0 \\
\end{array}
\right) \Rightarrow |IJ\odot K| = \Mod[4,4]=0, 
\;\;
\om^2:\;\; IJ\odot K = 
\left(
\begin{array}{cccc}
 0 & 0 & 3 & 0 \\
 0 & 0 & 0 & 0 \\
 3 & 0 & 0 & 0 \\
 0 & 0 & 0 & 0 \\
\end{array}
\right)  \Rightarrow |IJ\odot K| = \Mod[6,4]=2, \\
\hspace{-0.5in}
\om^1:\;\; IJ\odot K &=& 
\left(
\begin{array}{cccc}
 0 & 0 & 3 & 0 \\
 0 & 0 & 0 & 0 \\
 0 & 2 & 0 & 0 \\
 0 & 0 & 0 & 0 \\
\end{array}
\right) \Rightarrow |IJ\odot K| = \Mod[5,4]=1, 
\;\;
\om^3:\;\; IJ\odot K = 
\left(
\begin{array}{cccc}
 0 & 0 & 3 & 0 \\
 0 & 0 & 0 & 0 \\
 0 & 0 & 0 & 0 \\
 0 & 0 & 0 & 0 \\
\end{array}
\right)  \Rightarrow |IJ\odot K| = \Mod[3,4]=3. \qquad\;\; 
\eea
\esub
The relevant point is that while all the $K$ matrices have the same coefficient (here $1$) multiplying them, there are  cancellations between the 4 matrices in \Eq{N4:om0:1111:1111:om:0} associated with $\om^0$, and the 4 matrices in \Eq{N4:om0:1111:1111:om:2} associated with $\om^2$, \tit{both with equal coefficients}, 
which sum to $1+\om^2=0$ for any pair between the two sets. Thus, the two sets cancel as a group,
 which we call \tit{4-element bipartite} cancellations.
 Similarly, we have the \tit{8-element bipartite} cancellations between the two sets  of $8$ matrices in
 \Eq{N4:om0:1111:1111:om:1} associated with $\om^1$, and 
  in \Eq{N4:om0:1111:1111:om:3} associated with $\om^3$, such that 
  $\om+\om^3 = \om\,(1+\om^2)=0$. So these \tit{two separate groups} cancel \tit{separately}.
  In other words, the cancellation of sub-amplitudes for this transitions \tit{cancel in two groups associated with the sum of the even and odd powers of $\om^p$}.
 \smallskip
 
 The above is illustrated graphically in  \Fig{fig:N4:1111:1111:om0:om2}, where each 
 no-zero matrix element  $k_{ij}\in K$ in the
 (top row) \Eq{N4:om0:1111:1111:om:0} associated with factor $\om^0$, and 
 (bottom row) \Eq{N4:om0:1111:1111:om:2} associated with factor $\om^2$,
 indicates a photon transmitting from 
 input port-$i$ to output port-$j$.
\begin{figure}[h]
\includegraphics[width=5.0in,height=2.0in]{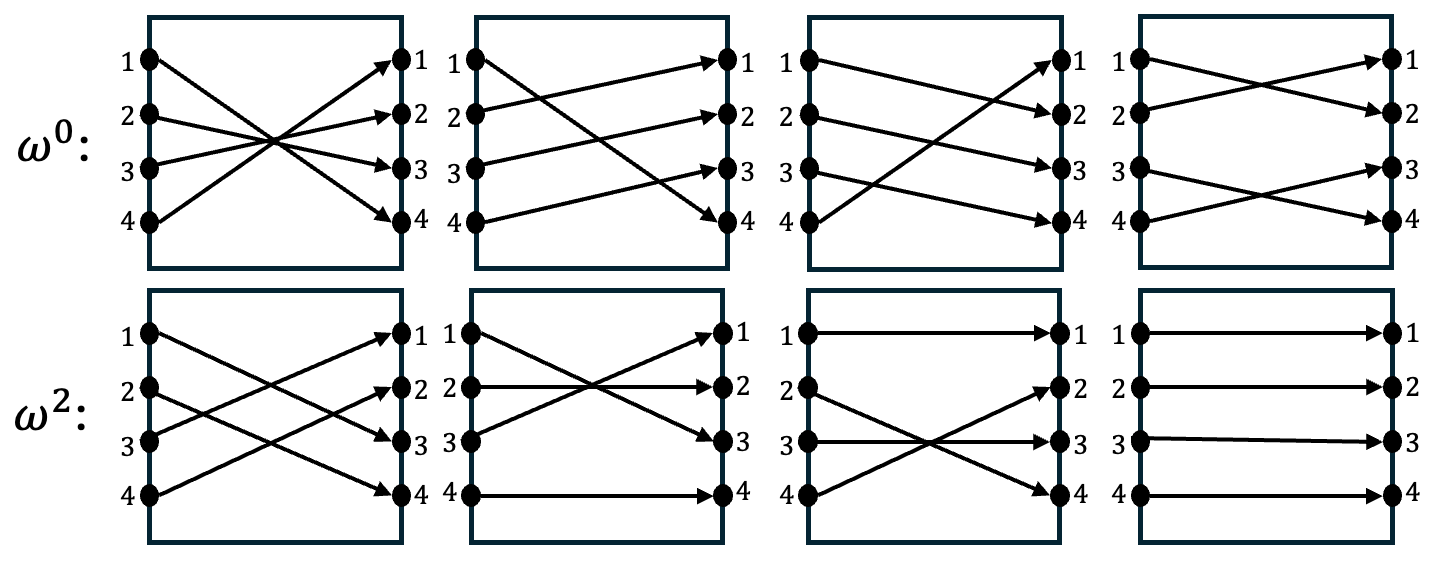} 
\caption{$N=4$ Scattering diagrams 
for the transition $\ket{1111}\overset{S_4}{\to}\ket{1111}$
for the $K$ matrices in 
(top row) \Eq{N4:om0:1111:1111:om:0} associated with factor $\om^0$, and 
(bottom row) \Eq{N4:om0:1111:1111:om:2} associated with factor $\om^2$. 
\tit{Any} pair of diagrams, one from each row, contributes a pair of sub-amplitude (with equal coefficients) which sums to $1+\om^2 = 0$, since $\om^2 = (e^{i 2 \pi/4})^2 = -1$ for $N=4$.. The two groups (top and bottom row) can be said to cancel 
as a \tit{4-bipartite group}.
}\label{fig:N4:1111:1111:om0:om2}
\end{figure}
\tit{Any} pair of diagrams, one from each row, contributes a pair of sub-amplitudes (with equal coefficients) which sums to $1+\om^2 = 0$, since $\om^2 = (e^{i 2 \pi/4})^2 = -1$ for $N=4$. The two groups (top and bottom row) can be said to cancel 
as a \tit{4-bipartite group}. The same could be drawn graphically for the two sets of 8-matrices in
\Eq{N4:om0:1111:1111:om:1} associated with factor $\om^1$, and 
 \Eq{N4:om0:1111:1111:om:3} associated with factor $\om^3$, with any pair cancelling as
 $\om(\,1+\om^2) = 0$. 

\subsection{An inspection of the cancellations in $\mathbf{A=0}$ for the $\mathbf{N=4}$ transition 
$\mathbf{\ket{3333}\overset{S_4}{\to}\ket{3333}}$}\label{subsec:N4:3333:3333}
It is instructive to look at case of higher multiphoton inputs to the symmetric BS, again with the goal of discerning what group of sub-amplitudes (diagrams) cancel in subgroups. An illustrative case is the
$N=4$ transition $\ket{3333}~\overset{S_4}{\to}~\ket{3333}$ which we find has
$A\propto (113+118\,\om)\,(1+\om^2)=0$. Since the number of partitions of $3$ is $|P_4(3)|=20$,
the total number of potential candidate $K$ matrices in our exhaustive search is $20^4=160,000$. However, we find that there are only a total of $2008$ valid $K$ matrices satisfying the required row-sum and column-sum conditions.
An example of three such valid $K$ matrices are
$\tiny{\left(
\begin{array}{cccc}
 0 & 0 & 0 & 3 \\
 0 & 0 & 3 & 0 \\
 3 & 0 & 0 & 0 \\
 0 & 3 & 0 & 0 \\
\end{array}
\right)}$, 
$\tiny{\left(
\begin{array}{cccc}
 0 & 0 & 0 & 3 \\
 0 & 1 & 2 & 0 \\
 1 & 2 & 0 & 0 \\
 2 & 0 & 1 & 0 \\
\end{array}
\right)}$, 
$\tiny{\left(
\begin{array}{cccc}
 0 & 0 & 0 & 3 \\
 0 & 1 & 2 & 0 \\
 2 & 1 & 0 & 0 \\
 1 & 1 & 1 & 0 \\
\end{array}
\right)}$.
The difference  now between the current transition $\ket{3333}\overset{S_4}{\to}\ket{3333}$ and the previous $\ket{1111}\overset{S_4}{\to}\ket{1111}$ is that for the former, the coefficients are no longer the same for all matrices, and this breaks the sub-amplitudes (diagrams) into groups governed by \tit{both} the power $p$ of $\om^p$, as well as by the value of their coefficients (since for terms to cancel, the coefficients - or combinatorial  factors  - must be \tit{identical}).

\bsub
\bea{N4:3333:3333}
\begin{array}{c}
\om^0/d \\
 \trm{\# terms}\;\; n_c
\end{array}
&:& \;\;
\left(
\begin{array}{cccccccccc}
 1 & \frac{1}{1296} & \frac{1}{6} & \frac{1}{144} & \frac{1}{48} & \frac{1}{16} &
   \frac{1}{2} & \frac{1}{4} & \frac{1}{24} & \frac{1}{8} \\
 4 & 4 & 8 & 24 & 32 & 36 & 40 & 48 & 80 & 208 \\
\end{array}
\right), \label{N4:3333:3333:om0} \\
%
\begin{array}{c}
\om^2/d \\
 \trm{\# terms}\;\; n_c
\end{array}
&:& \;\;
\left(
\begin{array}{cccccccccc}
 \omega ^2 & \frac{\omega ^2}{1296} & \frac{\omega ^2}{6} & \frac{\omega ^2}{144} &
   \frac{\omega ^2}{48} & \frac{\omega ^2}{16} & \frac{\omega ^2}{2} & \frac{\omega ^2}{4}
   & \frac{\omega ^2}{24} & \frac{\omega ^2}{8} \\
 4 & 4 & 8 & 24 & 32 & 36 & 40 & 48 & 80 & 208 \\
\end{array}
\right),  \label{N4:3333:3333:om2} \\
&{}& \no
&{}& \no
\begin{array}{c}
\om^1/d \\
 \trm{\# terms}\;\; n_c
\end{array}
&:& \;\;
\left(
\begin{array}{ccccccccc}
 \omega  & \frac{\omega }{1296} & \frac{\omega }{2} & \frac{\omega }{144} & \frac{\omega
   }{48} & \frac{\omega }{24} & \frac{\omega }{16} & \frac{\omega }{4} & \frac{\omega }{8}
   \\
 8 & 8 & 32 & 48 & 64 & 64 & 72 & 96 & 128 \\
\end{array}
\right),  \label{N4:3333:3333:om1} \\
\begin{array}{c}
\om^3/d \\
 \trm{\# terms}\;\; n_c
\end{array}
&:& \;\;
\left(
\begin{array}{ccccccccc}
 \omega ^3 & \frac{\omega ^3}{1296} & \frac{\omega ^3}{2} & \frac{\omega ^3}{144} &
   \frac{\omega ^3}{24} & \frac{\omega ^3}{48} & \frac{\omega ^3}{16} & \frac{\omega
   ^3}{4} & \frac{\omega ^3}{8} \\
 8 & 8 & 32 & 48 & 64 & 64 & 72 & 96 & 128 \\
\end{array}
\right).  \label{N4:3333:3333:om3}
\eea
\esub

In \Eq{N4:3333:3333:om0} and \Eq{N4:3333:3333:om2}  we list the $10$ distinct coefficients
of $\om^0/d$ and $\om^2/d$ (top row), and the number of times $n_c$ they occur (bottom row), respectively.
Similarly, in \Eq{N4:3333:3333:om1} and \Eq{N4:3333:3333:om3}  we list the $9$ distinct coefficients
of $\om^1/d$ and $\om^3/d$ (top row), and the number of times they occur (bottom row), respectively.

In this way we see that a pair of matching terms in each of the 10 columns of 
\Eq{N4:3333:3333:om0} and \Eq{N4:3333:3333:om2}, and similarly from the $9$ columns of
\Eq{N4:3333:3333:om1} and \Eq{N4:3333:3333:om3} can \tit{cancel in $n_c$-bipartite groups},
as $1+\om^2$, or  $\om\,(1+\om^2)$ respectively, 
where $n_c$ is the number of coefficients for the given term $\om^p/d$.
For example, for the coefficient $\frac{1}{1296}$ (second column) there are $n_c=4$ terms with factors 
$\om^0$ and $\om^2$, so this forms a 4-bipartite group of cancellations as  $1+\om^2=0$.
For the \tit{same} coefficient $\frac{1}{1296}$ there are $n_c=8$ terms with factors 
$\om^1$ and $\om^3$, so this forms a separate 8-bipartite group of cancellations as  $\om\,(1+\om^2)=0$.
The specific $K$ matrices associated with each group $\frac{\om^p}{1296}$ are show in
\Eq{N4:3333:3333:omp:d} below.
\be{N4:3333:3333:omp:d}
\om^0:\; 
\begin{array}{cc}
 \left(
\begin{array}{cccc}
 0 & 0 & 0 & 3 \\
 0 & 0 & 3 & 0 \\
 0 & 3 & 0 & 0 \\
 3 & 0 & 0 & 0 \\
\end{array}
\right) & \frac{1}{1296} \\
 \left(
\begin{array}{cccc}
 0 & 0 & 0 & 3 \\
 3 & 0 & 0 & 0 \\
 0 & 3 & 0 & 0 \\
 0 & 0 & 3 & 0 \\
\end{array}
\right) & \frac{1}{1296} \\
 \left(
\begin{array}{cccc}
 0 & 3 & 0 & 0 \\
 0 & 0 & 3 & 0 \\
 0 & 0 & 0 & 3 \\
 3 & 0 & 0 & 0 \\
\end{array}
\right) & \frac{1}{1296} \\
 \left(
\begin{array}{cccc}
 0 & 3 & 0 & 0 \\
 3 & 0 & 0 & 0 \\
 0 & 0 & 0 & 3 \\
 0 & 0 & 3 & 0 \\
\end{array}
\right) & \frac{1}{1296} \\
\end{array}, \quad
\om^2:\; 
\begin{array}{cc}
 \left(
\begin{array}{cccc}
 0 & 0 & 3 & 0 \\
 0 & 0 & 0 & 3 \\
 3 & 0 & 0 & 0 \\
 0 & 3 & 0 & 0 \\
\end{array}
\right) & \frac{\omega ^2}{1296} \\
 \left(
\begin{array}{cccc}
 0 & 0 & 3 & 0 \\
 0 & 3 & 0 & 0 \\
 3 & 0 & 0 & 0 \\
 0 & 0 & 0 & 3 \\
\end{array}
\right) & \frac{\omega ^2}{1296} \\
 \left(
\begin{array}{cccc}
 3 & 0 & 0 & 0 \\
 0 & 0 & 0 & 3 \\
 0 & 0 & 3 & 0 \\
 0 & 3 & 0 & 0 \\
\end{array}
\right) & \frac{\omega ^2}{1296} \\
 \left(
\begin{array}{cccc}
 3 & 0 & 0 & 0 \\
 0 & 3 & 0 & 0 \\
 0 & 0 & 3 & 0 \\
 0 & 0 & 0 & 3 \\
\end{array}
\right) & \frac{\omega ^2}{1296} \\
\end{array}, \quad
\om^1:\; 
{\tiny{
\begin{array}{cc}
 \left(
\begin{array}{cccc}
 0 & 0 & 0 & 3 \\
 0 & 3 & 0 & 0 \\
 3 & 0 & 0 & 0 \\
 0 & 0 & 3 & 0 \\
\end{array}
\right) & \frac{\omega }{1296} \\
 \left(
\begin{array}{cccc}
 0 & 0 & 0 & 3 \\
 3 & 0 & 0 & 0 \\
 0 & 0 & 3 & 0 \\
 0 & 3 & 0 & 0 \\
\end{array}
\right) & \frac{\omega }{1296} \\
 \left(
\begin{array}{cccc}
 0 & 0 & 3 & 0 \\
 0 & 3 & 0 & 0 \\
 0 & 0 & 0 & 3 \\
 3 & 0 & 0 & 0 \\
\end{array}
\right) & \frac{\omega }{1296} \\
 \left(
\begin{array}{cccc}
 0 & 0 & 3 & 0 \\
 3 & 0 & 0 & 0 \\
 0 & 3 & 0 & 0 \\
 0 & 0 & 0 & 3 \\
\end{array}
\right) & \frac{\omega }{1296} \\
 \left(
\begin{array}{cccc}
 0 & 3 & 0 & 0 \\
 0 & 0 & 0 & 3 \\
 0 & 0 & 3 & 0 \\
 3 & 0 & 0 & 0 \\
\end{array}
\right) & \frac{\omega }{1296} \\
 \left(
\begin{array}{cccc}
 0 & 3 & 0 & 0 \\
 0 & 0 & 3 & 0 \\
 3 & 0 & 0 & 0 \\
 0 & 0 & 0 & 3 \\
\end{array}
\right) & \frac{\omega }{1296} \\
 \left(
\begin{array}{cccc}
 3 & 0 & 0 & 0 \\
 0 & 0 & 0 & 3 \\
 0 & 3 & 0 & 0 \\
 0 & 0 & 3 & 0 \\
\end{array}
\right) & \frac{\omega }{1296} \\
 \left(
\begin{array}{cccc}
 3 & 0 & 0 & 0 \\
 0 & 0 & 3 & 0 \\
 0 & 0 & 0 & 3 \\
 0 & 3 & 0 & 0 \\
\end{array}
\right) & \frac{\omega }{1296} \\
\end{array}, \quad
}} 
\om^3:\; 
{\tiny{
\begin{array}{cc}
 \left(
\begin{array}{cccc}
 0 & 0 & 0 & 3 \\
 0 & 0 & 3 & 0 \\
 3 & 0 & 0 & 0 \\
 0 & 3 & 0 & 0 \\
\end{array}
\right) & \frac{\omega ^3}{1296} \\
 \left(
\begin{array}{cccc}
 0 & 0 & 0 & 3 \\
 0 & 3 & 0 & 0 \\
 0 & 0 & 3 & 0 \\
 3 & 0 & 0 & 0 \\
\end{array}
\right) & \frac{\omega ^3}{1296} \\
 \left(
\begin{array}{cccc}
 0 & 0 & 3 & 0 \\
 0 & 0 & 0 & 3 \\
 0 & 3 & 0 & 0 \\
 3 & 0 & 0 & 0 \\
\end{array}
\right) & \frac{\omega ^3}{1296} \\
 \left(
\begin{array}{cccc}
 0 & 0 & 3 & 0 \\
 3 & 0 & 0 & 0 \\
 0 & 0 & 0 & 3 \\
 0 & 3 & 0 & 0 \\
\end{array}
\right) & \frac{\omega ^3}{1296} \\
 \left(
\begin{array}{cccc}
 0 & 3 & 0 & 0 \\
 0 & 0 & 0 & 3 \\
 3 & 0 & 0 & 0 \\
 0 & 0 & 3 & 0 \\
\end{array}
\right) & \frac{\omega ^3}{1296} \\
 \left(
\begin{array}{cccc}
 0 & 3 & 0 & 0 \\
 3 & 0 & 0 & 0 \\
 0 & 0 & 3 & 0 \\
 0 & 0 & 0 & 3 \\
\end{array}
\right) & \frac{\omega ^3}{1296} \\
 \left(
\begin{array}{cccc}
 3 & 0 & 0 & 0 \\
 0 & 0 & 3 & 0 \\
 0 & 3 & 0 & 0 \\
 0 & 0 & 0 & 3 \\
\end{array}
\right) & \frac{\omega ^3}{1296} \\
 \left(
\begin{array}{cccc}
 3 & 0 & 0 & 0 \\
 0 & 3 & 0 & 0 \\
 0 & 0 & 0 & 3 \\
 0 & 0 & 3 & 0 \\
\end{array}
\right) & \frac{\omega ^3}{1296} \\
\end{array}.\quad
}} 
\ee
Again, $p\defn |IJ\odot K| = \Mod[ \prod_{ij}^N i j k_{i j}, 4]$ of every matrix $K$ in the group $\om^p$ (columns in \Eq{N4:3333:3333:omp:d}) determines its associated exponent $p\in\{0,1,2,3\}$. 
Thus, while all the  $K$ matrices in 
\Eq{N4:3333:3333:omp:d} contains four $3$s, it is their specific permutation that gives rise to the particular exponent $p$, which along with an identical (combinatorial) coefficient 
(here, $\frac{1}{1296}$), determines the particular $n_c$-partite group.

\subsection{An inspection of the cancellations in $\mathbf{A=0}$ for the $\mathbf{N=3}$ transition 
$\mathbf{\ket{012}\overset{S_3}{\to}\ket{111}}$  and similar transitions}\label{subsec:N3:012:111}
We saw earlier that the transitions 
$\ket{11\ldots1}\overset{S_N}{\to}\ket{11\ldots1}$ for $N$ odd had non-zero amplitudes. However, this does not imply that different inputs  cannot lead to $A=0$ on the same output state
$\ket{n_1,n_2,\ldots,n_N}\overset{S_N}{\to}\ket{11\ldots1}$. The simplest case to consider is the $N=3$ transition $\ket{012}\overset{S_3}{\to}\ket{111}$, with $18$ total candidate $K$ matrices, of which only $3$ are valid, and given by
\be{N3:012:111}
(\om^0, \om^1, \om^2)\leftrightarrow
\left(
\begin{array}{ccc}
 0 & 0 & 0 \\
 0 & 0 & 1 \\
 1 & 1 & 0 \\
\end{array}
\right),\left(
\begin{array}{ccc}
 0 & 0 & 0 \\
 0 & 1 & 0 \\
 1 & 0 & 1 \\
\end{array}
\right),\left(
\begin{array}{ccc}
 0 & 0 & 0 \\
 1 & 0 & 0 \\
 0 & 1 & 1 \\
\end{array}
\right)\quad \Rightarrow\quad p=\{0,1,2\}.
\ee
These three $K$ matrices, all with equal coefficients, sum to give $A\propto 1+\om+\om^2 = 0$, and therefore cancel as a 3-group. 
\begin{figure}[h]
\includegraphics[width=6.0in,height=1.25in]{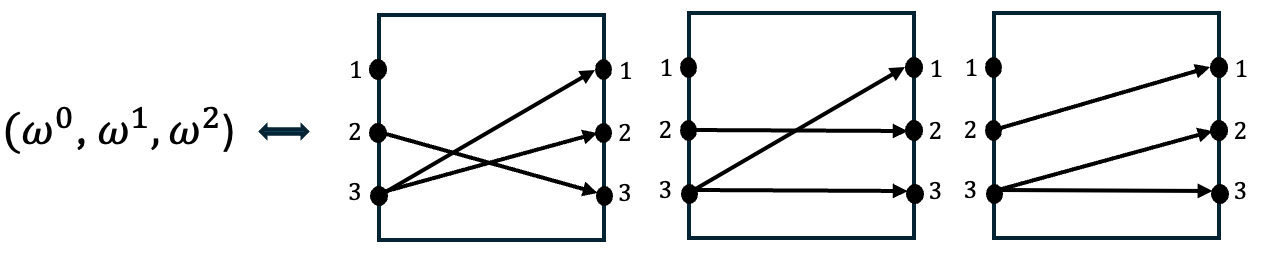} 
\caption{The three  scattering diagrams 
for the $N=3$ transition $\ket{012}\overset{S_3}{\to}\ket{111}$
for the $K$ matrices in 
 \Eq{N3:012:111} associated with factors $(\om^0, \om^1, \om^2)$. 
This group   can be said to cancel as a \tit{3-element group}.
}\label{fig:N4:1111:1111:om0:om2}
\end{figure}

Let us now consider increasing the total input/output photon number for $N=3$.
Of all possible $9$-photon inputs, i.e. for the transition 
$\ket{n_1,n_2,n_3}\overset{S_3}{\to}\ket{333}$, we obtained a zero amplitude $A=0$ for the inputs
listed in \Eq{N3:n1n2n3:333:A:eq:0}
\bsub
\be{N3:n1n2n3:333:A:eq:0}
\ket{n_1, n_2, n_3} = 
\left(
\begin{array}{ccc}
 0 & 1 & 8 \\
 0 & 2 & 7 \\
 0 & 4 & 5 \\
 1 & 2 & 6 \\
 1 & 3 & 5 \\
 2 & 3 & 4 \\
\end{array}
\right)
\overset{S_3}{\to}\ket{333} \Rightarrow A=0.
\ee
Of course, any of the $3!$ permutation of the order of the input photons $(n_1,n_2,n_3)$ leads to the same $A=0$ output on $\ket{333}$, since the BS is symmetric by construction.
\smallskip

As an example, for the input $\ket{2,3,4}$ (last line row in  \Eq{N3:n1n2n3:333:A:eq:0})
the $45$ valid $K$ matrices
break up into $6$ sub-groups with $6$ different coefficients $c$ such that 
$c\,(1+\om+\om^2+\om^3) = 0$, as shown in the top row of \Eq{N3:n1n2n3:333}, along with the number of times $n_c$ these groups appear (second row of \Eq{N3:n1n2n3:333}).
\be{N3:n1n2n3:333}
\begin{array}{c}
 c \\
\trm{\# terms}\, n_c\, \trm{with coefficient}\, c
\end{array}
: \;\;
\left(
\begin{array}{cccccc}
\frac{1}{72} & \frac{1}{24} & \frac{1}{12}& \frac{1}{8} & \frac{1}{4}&  \frac{1}{2} \\
2 & 3 & 2 & 5 & 2 & 1
\end{array}
\right) \quad \Rightarrow\quad c\, (\om^0+\om^1+\om^2)=0.
\ee
\esub
Note that the total sum of the number of coefficient $n_c$, (i.e. the sum of the second row in 
\Eq{N3:n1n2n3:333}, which is $15$), times the number of terms in the $N=3$ FSR required to allow for a cancellation, which is $3$ for $1+\om+\om^2=0$, is equal to the total number valid $K$ matrices, $45 = 15*3$. Since $A=0$ can only occur for an \tit{odd} $N$ if the full FSR is utilized, this statement is true in general for any $N$ odd.
\smallskip

Similarly, for the $N=3$, $12$-photons inputs for the transition 
$\ket{n_1,n_2,n_3}\overset{S_3}{\to}\ket{4,4,4}$ we obtain $A=0$ on the $10$ inputs in 
\Eq{N3:n1n2n3:444:A:eq:0}
\bsub
\be{N3:n1n2n3:444:A:eq:0}
\ket{n_1, n_2, n_3} = 
\left(
\begin{array}{ccc}
 0 & 1 & 11 \\
 0 & 2 & 10 \\
 0 & 4 & 8 \\
 0 & 5 & 7 \\
 1 & 2 & 9 \\
 1 & 3 & 8 \\
 1 & 5 & 6 \\
 2 & 3 & 7 \\
 2 & 4 & 6 \\
 3 & 4 & 5 \\
\end{array}
\right)
\overset{S_3}{\to}\ket{444} \Rightarrow A=0.
\ee
As an example, for the input $\ket{3,4,5}$ (last line row in  \Eq{N3:n1n2n3:444:A:eq:0})
the $105$ valid $K$ matrices
break up into $14$ sub-groups with $14$ different coefficients $c$ such that 
$c\,(1+\om+\om^2) = 0$, as shown in the top row of \Eq{N3:n1n2n3:444}, along with the number of times $n_c$ these groups appear (second row of \Eq{N3:n1n2n3:444}).
\be{N3:n1n2n3:444}
\hspace{-0.5in}
\begin{array}{c}
 c \\
\trm{\# terms}\, n_c\, \trm{with coefficient}\, c
\end{array}
: \;\;
\left(
\begin{array}{cccccccccccccc}
 \frac{1}{3456} & \frac{1}{864} & \frac{1}{576} & \frac{1}{288} & \frac{1}{216} &
   \frac{1}{192} & \frac{1}{144} & \frac{1}{96} & \frac{1}{72} & \frac{1}{48} & \frac{1}{32} &
   \frac{1}{24} & \frac{1}{16} & \frac{1}{8} \\
 2 & 2 & 2 & 2 & 2 & 2 & 4 & 3 & 4 & 4 & 2 & 3 & 2 & 1 \\
\end{array}
\right) \quad \Rightarrow\quad c\, (\om^0+\om^1+\om^2)=0.
\ee
\esub
Note that the total sum of the number of coefficient $n_c$, (i.e. the sum of the second row in 
\Eq{N3:n1n2n3:444}, which is $35$), times the number of terms in the $N=3$ FSR required to allow for a cancellation, which is $3$ for $1+\om+\om^2=0$, is again equal to the total number valid $K$ matrices, 
$105 = 35*3$.
\smallskip

\begin{figure*}[ht]
\begin{center}
\begin{tabular}{ccc}
\hspace{-0.5in}
\includegraphics[width=3.5in,height=2.5in]{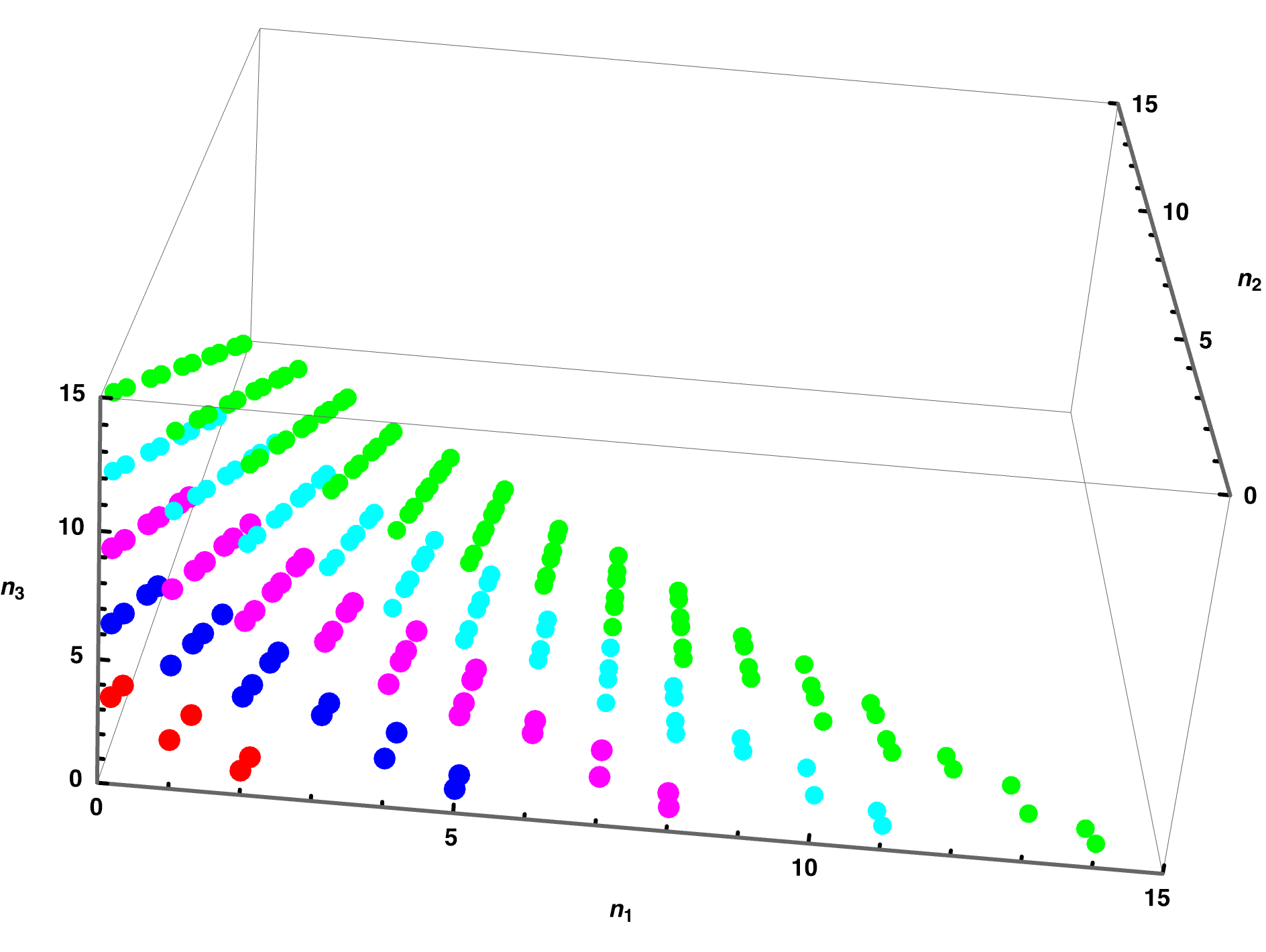} &
{\hspace{0.25in}} &
\includegraphics[width=3.5in,height=2.5in]{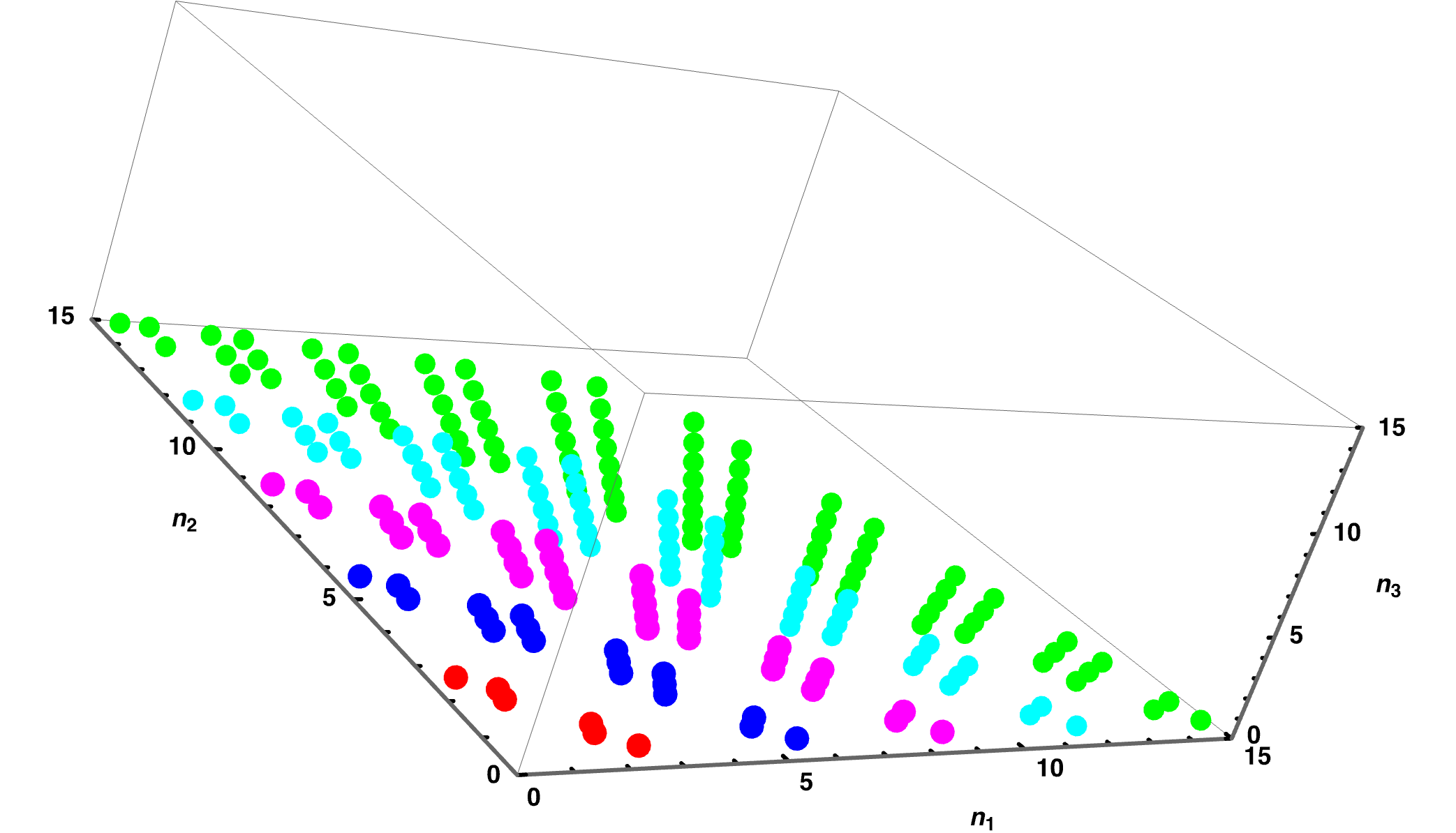}
\end{tabular}
\end{center} 
\caption{For $N=3$ we show (from two different perspectives) the inputs $\ket{n_1, n_2, n_3}$ yielding zero amplitude $A=0$ when projected onto the output state $\ket{n/3}^{\otimes 3}$ with equal number of photons in each output port, where $n=n_1+n_2+n_3$ is the total number of photons. We label the points as
$n=\{3,6,9,12,15\}$ with colors \{red, blue, magenta, cyan, green\}, with output states 
$\{\ket{111}, \ket{222}, \ket{333}, \ket{444}, \ket{5,5,5}\}$, respectively.
}
\label{fig:N3:111to444}    
\end{figure*}

In \Fig{fig:N3:111to444}  we show 
for $N=3$, the inputs $\ket{n_1, n_2, n_3}$ with zero amplitude $A=0$ when projected onto the output state $\ket{n/3}^{\otimes 3}$ with equal number of photons in each output port, where $n=n_1+n_2+n_3$ is the total number of input/output photons. We label the points as
$n=\{3,6,9,12, 15\}$ with colors \{red, blue, magenta, cyan, green\}, with output states 
$\{\ket{111}, \ket{222}, \ket{333}, \ket{444}, \ket{555}\}$, respectively. As discussed above, the input state
$\ket{111}$ is \tit{not} included. Even for this low value of $N$, and modestly low values of $n$, patterns for the zero amplitudes $A=0$ begin to emerge. 

\section{A symmetry for zero amplitudes $\mathbf{A=0}$  for eHOM transitions 
$\mathbf{\ket{\n}\overset{S_N}{\to}\ket{\tfrac{n}{N}}^{\otimes N}}$}\label{sec:psym}
In this section we consider zero amplitude $A=0$, eHOM  $S_N$ transitions 
$\ket{\n}\defn\ket{n_1,n_2,\ldots,n_N}\overset{S_N}{\to}\ket{\tfrac{n}{N}}^{\otimes N}$, with the \tit{coincident} output state $\ket{\m}\defn\ket{m_1,m_2,\ldots,m_N} = \ket{\tfrac{n}{N}}^{\otimes N}$, with 
$m_j\equiv \tfrac{n}{N}$ for all $j\in\{1,2,\ldots,N\}$, with $n\defn\sum_{i=1}^N n_i = \sum_{j=1}^N m_j$.
We develop a generalization of a symmetry argument employed by Lim and Beige \cite{Lim:Beige:2005} that those authors employed for the case of $\ket{\n}=\ket{\m} = \ket{1}^{\otimes N}$ to show that $A=0$ when 
$N\in even$, and $A\ne 0$ if~$N~\in~odd$.
\smallskip

First, we recall a property of permanents. If $D$ is a square $n\times n$ diagonal matrix with entries $d_i$, and $\L$ is a general  $n\times n$ matrix, then 
\be{Perm:properties}
\trm{Perm}(D\,\L) =\trm{Perm}(\L\,D) = \left(\prod_{i=1}^N d_i\right) \,\trm{Perm}(\L).
\ee
Note, that for \tit{determinants} this property is true for \tit{any} $n\times n$ matrix $D$, not just those that are diagonal. However, for permanents, this latter property holds \tit{only} for diagonal $D$.
\begin{figure}[h]
\includegraphics[width=4.65in,height=2.75in]{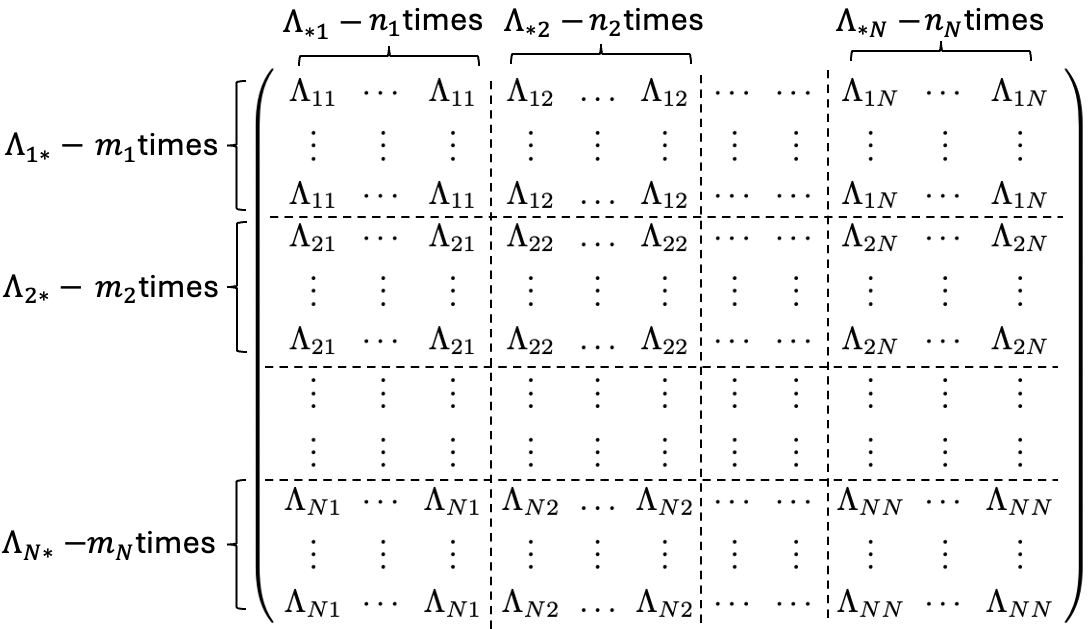} 
\caption{Form of $\L$ matrix for the general transition 
$\ket{\n}\overset{S_N}{\to}\ket{\m}$
with total photon number $n~\defn~\sum_{i=1}^N n_i = \sum_{j=1}^N m_j$.
}\label{fig:Lambda:Matrix}
\end{figure}

In \Fig{fig:Lambda:Matrix} we show the form of Scheel's matrix $\L$  such that $A\propto\trm{Perm}(\L)$ for the general transition 
$\ket{\n}\overset{S_N}{\to}\ket{\m}$
with total photon number $n~\defn~\sum_{i=1}^N n_i = \sum_{j=1}^N m_j$.
Let us consider two diagonal matrices $D_L$ and $D_R$, multiplying $\L$ from the left and from the right respectively, defined as 
\bsub
\bea{DL:DR}
(D_L)_{ik} &=& \om^{(i-1)}\,\delta_{ik},\quad  
(D_R)_{\l j} = \delta_{\l j}\,\om^{(j-1)}, \label{DL:DR:line1} \\
\trm{Perm}(D_L\,\L\,D_R) &=& 
\left(\prod_{i=1}^N \om^{(i-1)\, m_i}\right)\, 
\left(\prod_{j=1}^N \om^{(j-1)\,n_j }\right)\, 
\trm{Perm}(\L) = 
 \left(\prod_{i=1}^N \om^{i\,(n_i + m_i) - 2\,n}\right)\, \trm{Perm}(\L), \no
 &=&  \left( \om^{\big[\sum_{i=1}^N i\,(n_i + m_i)\big] - 2\,n} \right)\, \trm{Perm}(\L) 
 \equiv \om^{p_{sym}}\,\trm{Perm}(\L) , 
 \label{DL:DR:line2}
\eea
where we have used $n\defn\sum_{j=1}^N n_j = \sum_{i=1}^N m_i$, and have also defined
\be{psym:defn}
p_{sym} \defn \Mod\Big[ \big[\sum_{i=1}^N i\,(n_i + m_i)\big] - 2\,n, N\Big].
\ee
\esub

\subsection{Application to Lim and Beige's \tit{generalized HOM case}: 
$\mathbf{\ket{1}^{\otimes N}\overset{S_N}{\to}\ket{1}^{\otimes N}}$}\label{subsec:psym:case}
Lim and Beige \cite{Lim:Beige:2005} considered multiplying only by $D_L$, and considered the case that $n_i = m_j=1$ for $i,j~\in~\{1,2,\ldots,N\}$, i.e. the transition of all-ones in and all-ones out, namely
$\ket{1}^{\otimes N}\overset{S_N}{\to}\ket{1}^{\otimes N}$. Thus, their pre-multiplication factor in 
\Eq{DL:DR:line2} was $\prod_{i=1}^N \om^{(i-1)} = \om^{\sum_{i=1}^N i} = \om^{(N+1)\,N/2} = (-1)^{N+1}$. Thus, $\trm{Perm}(D_L\,\L) = (-1)^{N+1}\,\trm{Perm}(\L)$. 
Now the crucial observation is that for the all-ones transition 
$\ket{1}^{\otimes N}\overset{S_N}{\to}\ket{1}^{\otimes N}$, we have that $\L\equiv S_N$, so that 
$(\L)_{ij}\equiv (S_N)_{ij} =\tfrac{1}{\sqrt{N}}\om^{(i-1)(j-1)}$. 
Therefore, $(D_L\,\L)_{ij} = \sum_{k=1}^N (D_L)_{i k}\,(S_N)_{k j} 
= \tfrac{1}{\sqrt{N}} \om^{(i-1) j} \equiv \tfrac{1}{\sqrt{N}} \om^{(i-1) [(j+1)-1]}$ 
$\defn (S_N)_{i,j'} = (\L)_{i j'}$ where $j'\defn j+1$. That is, multiplication of $\L=S_N$ by $D_L$ from the left is just a permutation of the columns of $\L=S_N$, so that we \tit{also have}
$\trm{Perm}(D_L\,\L) \equiv \trm{Perm}(\L)$. Combining these two results we have
\bsub
\bea{Lim:Beige:result}
\trm{For}\; \ket{1}^{\otimes N}\overset{S_N}{\to}\ket{1}^{\otimes N}: &\quad&
(-1)^{N+1}\,\trm{Perm}(\L)=\trm{Perm}(D_L\,\L) \equiv \trm{Perm}(\L), \label{Lim:Beige:result:line1} \\
\trm{For}\, N\in odd: &\quad& \hspace{0.28in}\trm{Perm}(\L) = \trm{Perm}(\L), \quad \hspace{.25in}\trm{an identity}, \label{Lim:Beige:result:line2}  \\
\trm{For}\, N\in even: &\quad& (-1)\trm{Perm}(\L) = \trm{Perm}(\L), \quad\Rightarrow\quad  
\trm{Perm}(\L)=0. \label{Lim:Beige:result:line3} 
\eea
\esub
Thus, Lim and Beige \cite{Lim:Beige:2005} showed from \Eq{Lim:Beige:result:line3}, 
that if $N$ is \tit{even}, $A=\trm{Perm}(\L)=0$ for the
\tit{generalized HOM} transition $\ket{1}^{\otimes N}\overset{S_N}{\to}\ket{1}^{\otimes N}$.
For $N$ \tit{odd}, we obtain from \Eq{Lim:Beige:result:line2} only a trivial identity $\trm{Perm}(\L)=\trm{Perm}(\L)$, but in  actuality, we observe in symbolic calculations that  $A=\trm{Perm}(\L)\ne 0$.
\smallskip

\subsection{A numerical investigation of the  \tit{generalized eHOM} transitions: 
$\mathbf{\ket{n_1,n_2,\ldots,n_N}\overset{S_N}{\to}\ketnNN}$}\label{subsec:Tables}
We now perform a numerical investigation of the generalized eHOM transitions 
for $N\in\{3,4,\ldots,15\}$, 
whose features we explain analytically in the next section.
\smallskip

In  \Tbl{Table:N:odd} and \Tbl{Table:N:even} we show the results of the eHOM transitions
$\ket{\n}\overset{S_N}{\to}\ketnNN\defn\ket{\m}$, for $N\in odd$ and $N\in even$, respectively, for the $|P_N(n)|$ input states $\ket{\n} = \ket{n_1,n_2,\ldots,n_N}$ 
with \tit{eHOM coincident} output state $\ket{\m} = \ket{\tfrac{n}{N}}^{\otimes N}$
(without loss of generality taking 
$0\le n_1 \le n_2\le \ldots n_N\le N$ due to the invariance of $A$ with respect to permutations of the input and/or output states). The total photon number is given by $n=\sum_{i=1}^N n_i$.
\smallskip
\begin{table}[!ht]
\begin{tabular}{|c|c|c|c|c|c|}\hline
\multicolumn{6}{|c|}{$\mathbf{N\in odd}$}\\ \hline
\multicolumn{1}{|c|}{ N} & 
\multicolumn{1}{|c|}{ n } & 
\multicolumn{1}{|c|}{\bf output state $\mathbf{\ket{\m}}$} & 
\multicolumn{1}{|c|}{$\mathbf{|P^{(\tbf{sorted})}_N(n)|}$} &
\multicolumn{1}{|c|}{\bf \# $\mathbf{A=0}$} &
\multicolumn{1}{|c|}{\bf  \# $\mathbf{A\ne 0}$}  \\ \hline \hline
3 & 3 &  $\ket{1,1,1}$ & $3$ & 1 & 2 \\
3 & 6 &  $\ket{2,2,2}$ & $7$ & 3 & 4 \\
3 & 9 &  $\ket{3,3,3}$ & $12$ & 6 & 6 \\
3 & 12 &  $\ket{4,4,4}$ & $19$ & 10 & 9 \\
3 & 15 &  $\ket{5,5,5}$ & $27$ & 15& 12 \\ \hline
 5 & 5 &  $\ket{1,1,1,1,1}$ & $7$ & 5 & 2 \\
 5 & 10 &  $\ket{2,2,2,2,2}$ & $30$ & 24 & 6 \\
 5 & 15 &  $\ket{3,3,3,3,3}$ & $84$ & 67 & 17 \\ 
 5 & $20^*$ &  $\ket{4,4,4,4,4}$ & $192$ & $5^*$ & $2^*$ \\ \hline
 7 & 7 &  $\ket{1}^{\otimes 7}$ & $15$ & 12 & 3 \\
 7 & 14 &  $\ket{2}^{\otimes 7}$ & $105$ & 89 & 16 \\ \hline
 9 & 9 &  $\ket{1}^{\otimes 9}$ & $30$ & 25 & 5 \\
 9 & $18^*$ &  $\ket{2}^{\otimes 9}$ & $318$ & $11^*$ & $2^*$ \\ \hline
11 & 11 &  $\ket{1}^{\otimes 11}$ & $56$ & 51 & 5 \\ \hline
13 & 13 &  $\ket{1}^{\otimes 13}$ & $101$ & 93 & 8 \\ \hline
15 & $15^*$ &  $\ket{1}^{\otimes 15}$ & $176$ & $21^*$ & $1^*$ \\ \hline
\hline
\end{tabular}
\caption{Number of $A=0$ and $A\ne 0$ for $SU(N)$ eHOM transitions
$\ket{\n}\overset{S_N}{\to}\ket{\tfrac{n}{N}}^{\otimes N}\defn\ket{\m}$, 
with $N\in odd.$ Note: ${}*$ indicates, that runs were too time intensive, and only partial results of the full number of inputs $|P^{(\tbf{sorted})}_N(n)|$ are reported. $P^{(\tbf{sorted})}_N(n)$ indicates that (without loss of generality) we only consider the inputs $\ket{\n}$ with $0\le n_1 \le n_2\le \ldots n_N\le N$. 
For \tit{all} inputs examined in the above Table for $N\in odd$ we observed that 
$\trm{Perm}(D_L \L D_R) \equiv \trm{Perm}(\L)$ \tit{analytically} as a function of $\om$, and that 
$p_{sym} \ne 0 \Rightarrow A=0$, and $p_{sym} = 0 \Rightarrow A\ne 0$.}
\label{Table:N:odd}
\end{table}
For $N\in odd$ in \Tbl{Table:N:odd} we indicate the number of transitions out of $|P_N(n)|$ with $A=0$, and with $A\ne 0$. For \tit{all} the inputs examined in  \Tbl{Table:N:odd}  we observed that 
$\trm{Perm}(D_L \L D_R) \equiv \trm{Perm}(\L)$ \tit{analytically} as a function of $\om$, and that 
$p_{sym} \ne 0 \Rightarrow A=0$, and $p_{sym} = 0 \Rightarrow A\ne 0$, where 
$p_{sym}$ is defined in \Eq{psym:defn}.
\smallskip

\begin{table}[!ht]
\hspace{-0.5in}
\begin{tabular}{|c|c|c|c|c|c|c|c|c|c|}\hline
\multicolumn{10}{|c|}{$\mathbf{N\in even}$}\\ \hline
\multicolumn{1}{|c|}{ N} & 
\multicolumn{1}{|c|}{ n} & 
\multicolumn{1}{|c|}{\bf output state $\mathbf{\ket{\m}}$} & 
\multicolumn{1}{|c|}{$\mathbf{|P^{(\tbf{sorted})}_N(n)|}$} &
\multicolumn{1}{|c|}{\bf \# $\mathbf{A=0}$} &
\multicolumn{1}{|c|}{\bf  $\mathbf{p^{A=0}_{sym}}$} &
\multicolumn{1}{|c|}{\bf \# $\mathbf{A\ne 0}$} &
\multicolumn{1}{|c|}{\bf  $\mathbf{p^{A\ne 0}_{sym}}$} &
\multicolumn{1}{|c|}{\bf \#  $\mathbf{\Delta\trm{Perm}\L\ne0}$} &
\multicolumn{1}{|c|}{\bf $\mathbf{(-1)^{(N-1)\,\tfrac{n}{M}}}$}  \\ \hline \hline
4 & 4 &  $\ket{1,1,1,1}$   & $5$ & 4 & 0,1,3 & 1 & 2 & 1   & -1 \\
4 & 8 &  $\ket{2,2,2,2}$   & $15$ & 10 & 1,2,3 & 5 & 0 & 0 & 1\\
4 & 12 &  $\ket{3,3,3,3}$ & $34$ & 26 & 0,1,3 & 8 & 2 & 8 & -1 \\
4 & 16 &  $\ket{4,4,4,4}$ & $64$ & 46 & 1,2,3 & 18 & 0 & 0& 1 \\ \hline
6 & 6          &   $\ket{1}^{\otimes 6}$   & $11$ & 8 & 0,1,2,5 & 3 & 3 & 6 & -1\\
6 & 12        &   $\ket{2}^{\otimes 6}$   & $58$ & 45 & 1,2,3,4,5 & 13 & 0 & 0 & 1 \\
6 & $18^*$ &   $\ket{3}^{\otimes 6}$   & $199$ & $24^*$ & 0,1,2,4,5 & $7^*$ & 3 & $16^*$ & -1 \\ \hline
8 & 8          &   $\ket{1}^{\otimes 8}$   & $22$   & 19 & 0,1,2,3,5,6,7 & 3 & 4 & 3& -1 \\
8 & 16        &   $\ket{2}^{\otimes 8}$   & $186$ & 161 & 1,2,3,4,5,6,7 & 25 & 0 & 0 & 1\\ \hline
10 & 10      &   $\ket{1}^{\otimes 10}$   & $42$   & 38 & 0,1,2,3,4,6,7,8,9 & 4 & 5 & 3 & -1 \\
12 & 12      &   $\ket{1}^{\otimes 12}$   & $77$   & 71 & 0,1,2,3,5,6,7,8,9,10,11 & 6 & 6 & 16 & -1 \\
14 & 14     &   $\ket{1}^{\otimes 14}$   & $135$   & 125 & 0,1,2,3,5,6,7,8,9,10,11,12,13 & 10 & 7 & 73 & -1 \\
\hline
\end{tabular}
\caption{Number of $A=0$ and $A\ne 0$ for $SU(N)$ eHOM transitions
$\ket{\n}\overset{S_N}{\to}\ket{\tfrac{n}{N}}^{\otimes N}\defn\ket{\m}$, 
with $N\in odd.$ Note: ${}*$ indicates, that runs were too time intensive, and only partial results of the full number of inputs $|P^{(\tbf{sorted})}_N(n)|$ are reported. $P^{(\tbf{sorted})}_N(n)$ indicates that (without loss of generality) we only consider the inputs $\ket{\n}$ with $0\le n_1 \le n_2\le \ldots n_N\le N$. 
$p^{A=0}_{sym}$ and $p^{A\ne 0}_{sym}$ indicate the values of $\om^{p^{sym}}$ that occur when $A=0$ and $A\ne 0$, respectively. 
The second to the last column indicates the number of times that 
$\Delta\trm{Perm}\L\defn\trm{Perm}(D_L \L D_R) - \trm{Perm}(\L)\ne0$, regardless if $A=0$ or $A\ne 0$.
The last column indicates the parity of $(-1)^{(N-1)\,\tfrac{n}{M}}$, for which we see that 
$\Delta\trm{Perm}\ne0$ whenever $N\in even$ 
and the photon number  $m=\tfrac{n}{N}\in odd$ in each  mode of the eHOM output state, i.e. 
$(-1)^{(N-1)\,\tfrac{n}{M}}=(-1)$.
(Note: $(N,n)=(14,14)$ took 9525 secs to complete, 2.65 hrs).
} 
\label{Table:N:even}
\end{table}

For $N\in even$ in  \Tbl{Table:N:even}, we indicate the same quantities as in  \Tbl{Table:N:odd} for $N\in odd$, but now  additionally indicate the values of $p_{sym}$ that appear for $A=0$, and separately for $A\ne0$, which we designate as $p^{A=0}_{sym}$ and $p^{A\neq 0}_{sym}$, respectively. 
The penultimate column of  \Tbl{Table:N:even} indicates the number of times that 
$\trm{Perm}(D_L \L D_R) \ne \trm{Perm}(\L)$, regardless if $A=0$ or $A\ne 0$. 
The last column indicates the sign factor $(-1)^{(N-1)\tfrac{n}{N}}$ which we will discuss in more detail in the next section.
\smallskip

Note that for $N\in even$ there are cases where $\trm{Perm}(D_L \L D_R) \equiv \trm{Perm}(\L)$ for all input states, regardless if $A=0$ or $A\ne 0$, indicated by a $0$ in the penultimate column; $(N,n) = \{(4,8), (4,16), (6,12), (8,16) \}$. Otherwise, in the majority of case there are many instances where both
$\trm{Perm}(D_L \L D_R)$ equals, and not equals $\trm{Perm}(\L)$ within the $|P_N(n)|$ input states for a given $(N,n)$. Also note that for a given $(N,n)$, the values of $p^{A\neq0}_{sym}$ are single integers, that \tit{most often} do not appear also in $p^{A=0}_{sym}$. However, there are isolated instances where they appear in both, e.g.  $(N,n)=\{(12,12), (14,14)\}$ with $p^{sym} = \{6, 7\}$, respectively. However, in both these latter  cases we observed $\trm{Perm}(D_L \L D_R) \ne \trm{Perm}(\L)$. 
\smallskip

In the next section we develop a symmetry constraint on the value of $A=\PermL$ from which we can analytically explain all the features observed in  \Tbl{Table:N:odd} and  \Tbl{Table:N:even} above.

\section{A symmetry constraint on $\mathbf{A=\trm{PERM}(\L)}$  \\ for the \tit{generalized \lowercase{e}HOM} transitions: 
$\mathbf{\ket{n_1,n_2,\ldots,n_N}\overset{S_N}{\to}\ketnNN}$}\label{sec:sym:constraint}
In this section we develop a symmetry constraint on the value of $\PermL$, from  additional auxiliary matrices 
$\L''$ and $\L'''$ formed from operations on $\L' \defn (D_L\,\L\,D_R)$. We first describe a procedure which takes
$\L' \to \L'' \to \L''' \equiv \L$.

\subsection{The procedure to take $\mathbf{\L'=(D_L\,\L\,D_R) \to \L'' \to \L''' \equiv \L}$}\label{subsec:procedure}

The following procedure, also verified symbolically in \tit{Mathematica}, converts $\L'\defn D_L \L D_R$ into $\L$, 
for  arbitrary $N$.
\begin{quote}
\begin{enumerate}
\item[Step 1:] After forming the matrices $\L$ and  $\L'$, convert all exponents $p$ of $\om^p$ in each of the matrix elements to modulo $N$, i.e. $\om^p \to \om^{\Mod[p,N]}$.
\item[Step 2:] Let the total photon number $n\defn \sum_{i=1}^N n_i$, be an integer multiple of $N$, i.e. $m\defn \tfrac{n}{N}$, appropriate for the eHOM coincident output state 
$\ket{m}^{\otimes N} \defn \ket{\tfrac{n}{N}}^{\otimes N}$.  
\item[Steps 3.i:] Multiply every row in  the $m_i=m$ block of rows of $\L'$ by $\om^{N+1-i}$, i.e. 
\item[Step 3.1:] Multiply each of the \tit{first} set of  $m_1=m$ rows of $\L'$ by $\om^N=1$.
\item[Step 3.2:] Multiply each of the \tit{second} set of  $m_2=m$ rows of $\L'$ by $\om^{N-1}$.
\item[Step 3.3:] Multiply each of the \tit{third} set of $m_3=m$ rows of $\L'$ by $\om^{N-2}$.
\item[Step 3.i:] Repeat this procedure until you...
\item[Step 3.N:] Multiply each of the \tit{last} set of $m_N=m$ rows of $\L'$ by $\om^{1}$. 
\newline Call this  matrix $\L''$. 
\newline Once again, set $\om^p \to \om^{\Mod[p,N]}$ in matrix elements of $\L''$.
\item[Step 4:] Now, define the final matrix $\L'''$ by permuting the rows of  $\L''$ \tit{downwards} $m$-times so that the bottom $m$ rows cycle to the top $m$ rows (in \tit{Mathematica} this operation is $\L''' = \ttt{RotateRight}[\L'', m]$). 
\item[Step 5:]  The end result of this procedure is that 
one has $\L'''\equiv \L$ which implies $\trm{Perm}(\L''') = \trm{Perm}(\L)$.
%
%
\item[Step 6:] From the multiplication of rows of $\L''$ by of powers of $\om$ in the procedure above to obtain 
$\L'\to\L''\to\L'''$,  we additionally have that
 $\trm{Perm}(\L''') = \om^{m\,\sum_{i=0}^{N-1} i} \,\trm{Perm}(\L') =(e^{i 2 \pi/N})^{m(N-1)N/2}\, \,\trm{Perm}(\L')
 = (-1)^{(N-1) m}\,\trm{Perm}(\L')$, or equivalently
 $\underline{\trm{Perm}(\L')} =  (-1)^{(N-1) m}\,\trm{Perm}(\L''') \equiv \underline{(-1)^{(N-1) m}\,\trm{Perm}(\L)}$ (last equality using Step 5).
\end{enumerate}
\end{quote}

\subsection{Constraint on zero amplitude $\mathbf{A=\trm{Perm}(\L)=0}$ eHOM transitions 
$\mathbf{\ket{\n}\overset{S_N}{\to}\ket{\tfrac{n}{N}}^{\otimes N}}$,
\\and analytic proof of the results presented in \Tbl{Table:N:odd} and  \Tbl{Table:N:even} }\label{subsec:Constraint:proof}
An illustration of the procedure in Steps 1 - Step 5 above
is shown below for the $N=4$, $n=8$ zero amplitude $A=\trm{Perm}(\L)=0$ transition 
$\ket{1,2,2,3}\overset{S_4}{\to}\ket{2,2,2,2}$, with the appropriate $\L$ matrix, and transformation of the matrices $\L'\to\L''\to\L'''\equiv\L$. 
\bsub
\bea{procedure:example}
\hspace{-0.5in}
\L &=&
\left(
\begin{array}{cccccccc}
1 & 1 & 1 & 1 & 1 & 1 & 1 & 1 \\
1 & 1 & 1 & 1 & 1 & 1 & 1 & 1 \\
1 & \om & \om & \om^2 & \om^2 & \om^3 & \om^3 & \om^3 \\
1 & \om & \om & \om^2 & \om^2 & \om^3 & \om^3 & \om^3 \\
1 & \om^2 & \om^2 & 1 & 1 & \om^2 & \om^2 & \om^2 \\
1 & \om^2 & \om^2 & 1 & 1 & \om^2 & \om^2 & \om^2 \\
1 & \om^3 & \om^3 &  \om^2 & \om^2 & \om & \om & \om \\
1 & \om^3 & \om^3 &  \om^2 & \om^2 & \om & \om & \om 
\end{array}
\right), \quad
\L'\defn D_L\,\L\,D_R = 
\left(
\begin{array}{cccccccc}
1 & \om & \om & \om^2 & \om^2 & \om^3 & \om^3 & \om^3 \\
1 & \om & \om & \om^2 & \om^2 & \om^3 & \om^3 & \om^3 \\
\om & \om^3 & \om^3 & \om & \om & \om^3 & \om^3 & \om^3 \\
\om & \om^3 & \om^3 & \om & \om & \om^3 & \om^3 & \om^3 \\
 \om^2 & \om & \om &  1 & 1 & \om^3 &  \om^3 &  \om^3 \\
 \om^2 & \om & \om &  1 & 1 & \om^3 &  \om^3 &  \om^3 \\
 \om^3 & \om^3 & \om^3 &  \om^3 & \om^3 & \om^3 &  \om^3 &  \om^3 \\
 \om^3 & \om^3 & \om^3 &  \om^3 & \om^3 & \om^3 &  \om^3 &  \om^3 
\end{array}
\right),\qquad\; \label{procedure:example:line1} \\
&{}&\no
&{}&\no
\hspace{-0.5in}
 \overunderset{\begin{array}{c}\to r_1\times\,\om^4 \\ \to r_2\times\,\om^4 \\ \to r_3\times\,\om^3 \\ \to r_4\times\,\om^3 \end{array}}
{\begin{array}{c}\to r_5\times\,\om^2 \\ \to r_6\times\,\om^2 \\ \to r_7\times\,\om^1 \\ \to r_8\times\,\om^1\end{array}}{\L'\longrightarrow}
 \L'' &=& 
\left(
\begin{array}{cccccccc}
1 & \om & \om & \om^2 & \om^2 & \om^3 & \om^3 & \om^3 \\
1 & \om & \om & \om^2 & \om^2 & \om^3 & \om^3 & \om^3 \\
1 & \om^2 & \om^2 & 1 & 1 & \om^2 & \om^2 & \om^2 \\
1 & \om^2 & \om^2 & 1 & 1 & \om^2 & \om^2 & \om^2 \\
1 & \om^3 & \om^3 &  \om^2 & \om^2 & \om & \om & \om \\
1 & \om^3 & \om^3 &  \om^2 & \om^2 & \om & \om & \om \\
1 & 1 & 1 & 1 & 1 & 1 & 1 & 1 \\
1 & 1 & 1 & 1 & 1 & 1 & 1 & 1 
\end{array}
\right),  \label{procedure:example:line2} 
\to \L''' = \ttt{RotateRight[$\L''$,2]} =
\left(
\begin{array}{cccccccc}
1 & 1 & 1 & 1 & 1 & 1 & 1 & 1 \\
1 & 1 & 1 & 1 & 1 & 1 & 1 & 1 \\
1 & \om & \om & \om^2 & \om^2 & \om^3 & \om^3 & \om^3 \\
1 & \om & \om & \om^2 & \om^2 & \om^3 & \om^3 & \om^3 \\
1 & \om^2 & \om^2 & 1 & 1 & \om^2 & \om^2 & \om^2 \\
1 & \om^2 & \om^2 & 1 & 1 & \om^2 & \om^2 & \om^2 \\
1 & \om^3 & \om^3 &  \om^2 & \om^2 & \om & \om & \om \\
1 & \om^3 & \om^3 &  \om^2 & \om^2 & \om & \om & \om
\end{array}
\right)\equiv \L, \qquad\;\;  
\eea
\esub
where in all the matrices above, we have modded each exponent by $N=4$, i.e.
 $\om^p\to \om^{\Mod[p,4]}$.
 \smallskip
 
 The strategy now for constructing a constraint on  $\trm{Perm}(\L)$ is to use the
 auxiliary matrices $\L'$, $\L''$ and $\L'''$, created from $\L$ by either multiplying rows by powers of $\om$, 
 and/or permuting rows and/or columns. In \Eq{procedure:example:line1} we show $\L$ for the 
 $(N,n) = (4,8)$ eHOM transtion
 $\ket{1,2,2,3}\overset{S_4}{\to}\ket{2,2,2,2}$, and the associated matrix $\L'=D_L\,\L\,D_R$, where
 we recall \Eq{DL:DR:line1} that $(D_L)_{ik} = \om^{(i-1)}\,\delta_{ik}$ and $(D_L)_{\l j} = \om^{(j-1)}\,\delta_{\l j}$ are diagonal matrices multiplying $\L$ from the left and right, respectively. 
 Since multiplication of a general matrix $\L$ by a diagonal matrix scales $\trm{Perm}(\L)$ by the determinant of the diagonal matrix, we have as shown in \Eq{DL:DR:line2} and \Eq{psym:defn}, that
 \be{Lprime:constraint}
 \trm{Perm}(\L') = \om^{p_{sym}}\,\trm{Perm}(\L), \qquad 
 p_{sym} \defn \Mod\Big[ \big[\sum_{i=1}^N i\,(n_i + m_i)\big], N\Big],
 \ee
  for eHOM transitions.
 \smallskip
 
 In \Eq{procedure:example:line2} we show the construction of a second auxiliary matrix $\L'''$, made from $\L'$ by
 first multiplying  each block of $m\defn\tfrac{n}{N}=\tfrac{8}{4}=2$ rows of $\L'$ by the factors 
 $\{\om^N=1, \om^{N-1},  \om^{N-2},\ldots, \om^{1}\} = \{\om^4=1,\om^3,\om^2,\om^1\}$.
 We subsequently cyclically permute the rows of $\L''$ \tit{downward} 
 so that the bottom $m$ rows cycle to the top $m$ rows, creating $\L''' \equiv \L$.  
 Now, permuting the rows of $\L'$ does not change the value of the permanent of $\L''$.
 From the multiplication $\L'$ by  the factors of $\om$ we have
 \bea{Lprime3:constraint}
 \trm{Perm}(\L)&\equiv& \trm{Perm}(\L''') = \om^{m\,\sum_{i=0}^{N-1} i} \,\trm{Perm}(\L') = 
 \om^{m(N-1)N/2} \,\trm{Perm}(\L') = (-1)^{(N-1) m} \, \trm{Perm}(\L'), \no
 \trm{or equivalently:}\quad \trm{Perm}(\L') &=&  (-1)^{(N-1) m}\,\trm{Perm}(\L).
 \eea
 Equating $\trm{Perm}(\L')$ from \Eq{Lprime:constraint} and \Eq{Lprime3:constraint} we arrive at a 
 \tit{constraint} on the value of  $\trm{Perm}(\L)$
 \be{L:constraint}
 (-1)^{(N-1) m}\,\trm{Perm}(\L) =  \om^{p_{sym}}\,\trm{Perm}(\L).
 \ee
\tit{This is one of the main analytical results of this work}, and below we show how it explains the results shown in 
\Tbl{Table:N:odd} and  \Tbl{Table:N:odd} for $N\in odd$ and $N\in even$, respectively.

\subsection{\bf Consequences of the constraint on $\mathbf{A=\trm{Perm}(\L)}$, \Eq{L:constraint}:}
\subsubsection{$N\in odd$}
For $N\in odd \Rightarrow (N-1)\in even$. Therefore, regardless of the value of $m\defn\tfrac{n}{N}$, we have
$(-1)^{(N-1) m}\equiv 1$ on the lhs of \Eq{L:constraint}. Thus, we have
\be{L:constraint:N:odd}
N\in odd\; \Rightarrow \trm{Perm}(\L) =  \om^{p_{sym}}\,\trm{Perm}(\L) 
\quad \Rightarrow\quad \trm{if}\;\; p_{sym}\ne 0  \Rightarrow \trm{Perm}(\L)=0.
\ee
The last implication in  \Eq{L:constraint:N:odd} is born out for \tit{all} the $A=0$ results  
shown in \Tbl{Table:N:odd}.
\smallskip

Note that if  $p_{sym}=0$, all we can conclude from \Eq{L:constraint:N:odd} is the identity
$\trm{Perm}(\L) = \trm{Perm}(\L)$. However, in this case, all the results in \Tbl{Table:N:odd} yielded 
$A=\trm{Perm}(\L)\ne 0$ (which need not necessarily be the case since we could have the possibility of $0=0$).
This \tit{general trend} of $\trm{Perm}(\L) = \trm{Perm}(\L)$ from \Eq{L:constraint} leading to $A=\trm{Perm}(\L)\ne 0$ arises in \tit{most}, but \tit{not all} cases, when we examine $N\in even$ below.

\subsubsection{$N\in even,\; m\in even$}
For $N\in even \Rightarrow (N-1)\in odd$. Therefore, $(-1)^{(N-1) m}\equiv (-1)^{m}$ on the lhs of 
 \Eq{L:constraint}, and thus depends on the \tit{parity} of $m = \tfrac{n}{M}$. 
 For $m\in even$  we have $(-1)^{m}\equiv 1$ so that, once again, we have
 \be{L:constraint:N:even:m:even}
(N,m)\in (even,even)\; \Rightarrow \trm{Perm}(\L) =  \om^{p_{sym}}\,\trm{Perm}(\L) 
\quad \Rightarrow\quad \trm{if}\;\; p_{sym}\ne 0  \Rightarrow \trm{Perm}(\L)=0.
\ee
This is borne out in \Tbl{Table:N:even} where in the last column, rows with $(-1)^{(N-1) m}=1$ are associated with $m\in even$ and $A=\trm{Perm}(\L)=0$ is associated with $p_{sym}\ne 0$.
Note that once again, in this case, we also have $p_{sym}= 0 \Rightarrow A\ne 0$, as in the $N\in odd$ cases.

\subsubsection{$N\in even,\; m\in odd,\; p_{sym}\ne N/2$}
For $N\in even \Rightarrow (N-1)\; \trm{and}\; m\in odd$, we have $(-1)^{(N-1) m}\equiv (-1)$.
\Eq{L:constraint} then yields the constraint
\bsub
\bea{L:constraint:N:even:m:odd}
(N,m)\in (even,odd)\; \Rightarrow (-1)\,\trm{Perm}(\L) &=&  \om^{p_{sym}}\,\trm{Perm}(\L), \label{L:constraint:N:even:m:odd:line1} \\ 
\equiv \om^{N/2}\,\trm{Perm}(\L) &=&  \om^{p_{sym}}\,\trm{Perm}(\L), \quad
\Rightarrow\quad \trm{if}\;\; p_{sym}\ne N/2  \Rightarrow \trm{Perm}(\L)=0,
\label{L:constraint:N:even:m:odd:line2}
\eea
\esub
where we have used $\om^{N/2}\equiv (-1)$ in \Eq{L:constraint:N:even:m:odd:line2}.
This latter conclusion in \Eq{L:constraint:N:even:m:odd:line2} is borne out in column headed by 
$p_{sym}^{A=0}$ in \Tbl{Table:N:even} where we observe that in rows with $m\in odd$ eHOM output states, 
$p_{sym} = N/2$ \tit{does not} occur.
\smallskip

Note that in particular, if we consider the transitions $\ket{m}^{\otimes N}\overset{S_N}{\to}\ket{m}^{\otimes N}$ with $m=\tfrac{n}{N}$, then $p_{sym} = \sum_{i=1}^{N-1} i (m+m)=(2m)\half N (N-1)$ so that 
$\om^{p_{sym}} = (-1)^{(2m) (N-1)}\equiv 1$. Thus, for $(N,m)\in (even,odd)$
the constraint  \Eq{L:constraint} yields $-\trm{Perm}(\L)=\trm{Perm}(\L)\Rightarrow A=\trm{Perm}(\L)=0$.
This result generalizes the Lim and Beige gHOM result  \cite{Lim:Beige:2005} 
that $\trm{Perm}(\L)=0$ for transitions 
$\ket{1}^{\otimes N}\overset{S_N}{\to}\ket{1}^{\otimes N}$ with $N\in even$ and $m=1$, to  arbitrary $m\in odd$ (e.g. for $N=4$, $A=0$ for $\ket{3}^{\otimes 4}\overset{S_4}{\to}\ket{3}^{\otimes 4}$, $\ket{5}^{\otimes 4}\overset{S_4}{\to}\ket{5}^{\otimes 4}$, \ldots).

\subsubsection{$N\in even,\; m\in odd,\; p_{sym}=N/2$}
For the case of $N\in even,\; m\in odd,\; p_{sym}=N/2$, \Eq{L:constraint} once again only yields an identity
\be{L:constraint:Identity}
(N,m)\in (even,odd)\;\trm{and}\; p_{sym} = N/2 \quad\Rightarrow\quad \trm{Perm}(\L)=\trm{Perm}(\L).
\ee
However, this time, \tit{for all but 2 instances (which breaks the ``general trend")} studied in \Tbl{Table:N:even}, we have $A\ne0$.
That is, for the cases $(N,n,p_{sym}) = \{(12,12,6), (14,14,7)\}$ we  have $p_{sym} = N/2$ \tit{also} leading to a zero amplitude  $A=0$. So, unfornately, we \tit{cannot} conclude, \tit{in general}, 
that for $(N,m)\in (even,odd), \trm{and}\; p_{sym} = N/2$ implies that $A\ne0$.
These two exceptional cases are worth exploring, since they involved $A=\trm{Perm}(\L)$ containing a factor which is an AFSR (alternating fundamental summation relation), discussed previously in 
\Eq{AFSR:line1}-\Eq{AFSR:line3}.
We examine these two special cases where $p_{sym}=N/2\; \trm{and}\;  A=0$ below. 
\smallskip

\begin{quote}
\begin{enumerate}
\item[Case 1:] $(N,n) = (12,12)$;\;  
$\ket{0}^{\otimes 5}\ket{1}^{\otimes 4}\ket{2}^{\otimes 2}\ket{4}\overset{S_{12}}{\to}\ket{1}^{\otimes 12}$ 
$\Rightarrow\; A=\trm{Perm}(\L)\propto 1-\om^2 + (\om^2)^2 $ 
$=\sum_{i=1}^{N/2^2} (-\om^2)^{i-1} = \frac{1+(\om^2)^{N/4}}{1+\om^2} =
\frac{1+\om^{N/2}}{1+\om^2} = 0$, since $\om^{N/2}=(-1)$. 
\newline
This is just \Eq{AFSR:line1} and \Eq{AFSR:line3} with $q=2$, i.e. $N=12 = 2^{q=2}\, 3^1$.
\vspace{0.1in}
\item[Case 2:] $(N,n) = (14,14)$;\;  
$\ket{0}^{\otimes 8}\ket{1}\ket{2}^{\otimes 2}\ket{3}^{\otimes 3}\overset{S_{14}}{\to}\ket{1}^{\otimes 14}$ 
$\Rightarrow\; A=\trm{Perm}(\L)\propto 1-\om + \om^2 - \om^3+ \om^4 - \om^5+ \om^6 $ 
$=\sum_{i=1}^{N/2^1} (-\om)^{i-1} = \frac{1+\om^{N/2}}{1+\om}  = 0$, since $\om^{N/2}=(-1)$. 
\newline
This is just \Eq{AFSR:line1} and \Eq{AFSR:line3} with $q=1$, i.e. $N=14 =2^{q=1}\,7^1$.
\end{enumerate}
\end{quote}

Thus, as discussed after \Eq{AFSR:line1}-\Eq{AFSR:line3}, when the prime factorization of $N$ is given by
$N\defn 2^q\,3^{q_3}\,5^{q_5}\ldots$, it is \tit{possible} that when $p_{sym}=N/2$, we \tit{might have} $A=\trm{Perm}(\L)$ proportional to an AFSR (times another polynomial in $\om$ that does \tit{not} evaluate to zero), such that this AFSR evaluates to zero. In fact, this situation occurred \tit{only once} in each of the two cases discussed above. From \Tbl{Table:N:even},
$(N,n)=(12,12)$ had $71$ cases of $A=0$ and $6$ cases of $A\ne0$, while
 $(N,n)=(14,14)$ had $125$ cases of $A=0$ and $10$ cases of $A\ne0$, with $p_{sym}=N/2$ 
 occurring \tit{only once} in the respective $A=0$ cases.

\subsubsection{Generalization of the Lim and Beige result to the transitions 
$\ket{m\defn\tfrac{n}{N}}^{\otimes N}\overset{S_N}{\to}\ket{m=\tfrac{n}{N}}^{\otimes N}$}
 Lim and Beige \cite{Lim:Beige:2005} showed that $A=0$ for transitions (what they called the 
 \tit{generalized HOM effect})
 $\ket{1}^{\otimes N}\overset{S_N}{\to}\ket{1}^{\otimes N}$ for $N\in even$, from a constraint equation that yielded $(-1)^{(N-1)}\,\trm{Perm}(\L) = \trm{Perm}(\L)$.
 For $N\in odd$ their constraint reduced to $\trm{Perm}(\L) = \trm{Perm}(\L)$, which they did not claim led to $A\ne 0$, but which in fact is borne out from all our symbolic and numerical investigations.
 \smallskip
 
 Let us now consider a generalized multiphoton input version of Lim and Beige, 
 namely the particular ``diagonal" eHOM transitions
 $\ket{m\defn\tfrac{n}{N}}^{\otimes N}\overset{S_N}{\to}\ket{m=\tfrac{n}{N}}^{\otimes N}$.
 Let us calculate $p_{sym}$ explicitly on the rhs of  \Eq{L:constraint}.
 We have $p_{sym} = \sum_{i=1}^N i (n_i + m_i) = 2 m \,\sum_{i=1}^N i  = 2 m (N+1) N/2$.
 Thus $\om^{p_{sym}} = \om^{ 2 m (N+1) N/2} = (-1)^{(2m) (N+1)}\equiv 1$, for \tit{all} $N$ and $m$.
 Therefore the rhs of  \Eq{L:constraint} is $\om^{p_{sym}}\,\trm{Perm}(\L) \to \trm{Perm}(\L)$.
 \smallskip
 
 Now the lhs of  \Eq{L:constraint} is $(-1)^{(N-1) m} =\trm{Perm}(\L) \to \pm \trm{Perm}(\L)$, with the 
 $-$ sign arising solely from $(N,m)\in(even, odd)$. In this latter case
 the constraint \Eq{L:constraint} becomes $-\trm{Perm}(\L) = \trm{Perm}(\L) \Rightarrow A=0$, which is borne out in \Tbl{Table:N:even}, e.g. $A=0$ for transitions
 $(N,n)=(4,4):    \ket{1}^{\otimes 4}\overset{S_4}{\to}\ket{1}^{\otimes 4}$, 
 $(N,n)=(4,12): \ket{3}^{\otimes 4}\overset{S_4}{\to}\ket{3}^{\otimes 4}$,
 $(N,n)=(6,6): \ket{1}^{\otimes 6}\overset{S_6}{\to}\ket{1}^{\otimes 6}$, 
 (N,n)=(6,18): $\ket{3}^{\otimes 6}\overset{S_6}{\to}\ket{3}^{\otimes 6}$, etc\ldots
 \smallskip
 
 For the case  $(N,m)\in(even, even)$ the constraint \Eq{L:constraint} only yields the identity
 $\trm{Perm}(\L) = \trm{Perm}(\L)$ which is not required, but is associated with $A\ne 0$, which is borne out in 
\Tbl{Table:N:even}, e.g. $A\ne0$ for transitions
$(N,n)=(4,8): \ket{2}^{\otimes 4}\overset{S_4}{\to}\ket{2}^{\otimes 4}$, 
 $(N,n)=(4,16): \ket{4}^{\otimes 4}\overset{S_4}{\to}\ket{4}^{\otimes 4}$,
 $(N,n)=(6,12): \ket{2}^{\otimes 6}\overset{S_6}{\to}\ket{2}^{\otimes 6}$, 
 $(N,n)=(8,16): \ket{2}^{\otimes 8}\overset{S_8}{\to}\ket{2}^{\otimes 8}$.
 \smallskip
 
 For $N\in odd$, $(N-1)\in even$, regardless of the value of $m$, we obtain $(-1)^{(N-1) m}=1$ and the constraint yields the identity $\trm{Perm}(\L) =\trm{Perm}(\L) $, which follows the general trend that $A\ne 0$. 
\smallskip
 
 Thus, we can conclude that for the eHOM ``diagonal" transitions 
 $\ket{m\defn\tfrac{n}{N}}^{\otimes N}\overset{S_N}{\to}\ket{m=\tfrac{n}{N}}^{\otimes N}$ with $N\in even$,
 that we obtain zero amplitude $A=0$ if $m\in odd$, and observe $A\ne 0$ for $m\in even$.
 On the other hand, for $N\in odd$, we always observe $A\ne 0$ for these transitions.
 These results are the \tit{generalization} of the results of Lim and Beige \cite{Lim:Beige:2005} to 
 $m=\tfrac{n}{N}\ge 1$ photons in each input/output port.
 
\subsubsection{Reduced constraint equation \Eq{L:constraint}}
As one last discussion, we can rewrite the $\trm{Perm}(\L)$ constraint equation \Eq{L:constraint} so that it is depends solely on the input photon numbers $\{n_i\}$, and is independent of $m\defn\tfrac{n}{N}$, the photon number in each output port of the eHOM coincident state.
Let us write $p_{sym}$ as
\be{p:sym:tilde}
p_{sym} = \sum_{i=1}^N i\,(n_i + m_i) \equiv \sum_{i=1}^N i\,n_i + m \sum_{i=1}^N i 
=  \sum_{i=1}^N i\,n_i + m (N+1)N/2 \equiv \tilde{p}_{sym} + m (N+1)N/2, \quad
\tilde{p}_{sym}\defn \sum_{i=1}^N i\,n_i .
\ee
Thus the constraint \Eq{L:constraint} reduces to
\bea{p:tilde:sym}
(-1)^{(N-1) m}\,\trm{Perm}(\L) &=& \om^{p_{sym}}\, \trm{Perm}(\L), \no
&=& (-1)^{(N+1) m}\,\om^{\tilde{p}_{sym}}\trm{Perm}(\L) \no
\Rightarrow \trm{Perm}(\L) &=& \om^{\tilde{p}_{sym}}\trm{Perm}(\L), \qquad  
\tilde{p}_{sym}=\sum_{i=1}^N i\,n_i \overset{\Mod_N}{\sim} \sum_{i=1}^{(N-1)} i\,n_i,
\eea
since $(-1)^{(N+1) m} = (-1)^{(N-1) m}\,(-1)^2 =  (-1)^{(N-1) m}$ cancels from both sides.
Note that \Eq{p:tilde:sym} is independent of the output photon number $m=\tfrac{n}{N}$ in the eHOM coincident state, and only depends on the input photon numbers $\{n_i\}$.
Thus, we also can state the constraint that $\tilde{p}_{sym}\ne 0 \Rightarrow A=\trm{Perm}(\L)=0$.
Also note that the sum in $\tilde{p}_{sym}$ only needs to go to $(N-1)$ vs $N$ since the last term in the sum is $\Mod[ N\,n_N,N]=0$.
\smallskip

We can check \Eq{p:tilde:sym} in the special cases above where $p_{sym}=N/2$, yet we get $A=0$ (vs the general trend of $A\ne0$ observed for the constraint yielding the identity $\trm{Perm}(\L) = \trm{Perm}(\L)$).
For example in Case 1: with $(N,n)=(12,12)$ and $A=0$,  the input state was  
$\ket{\n} = \ket{0}^{\otimes 5}\ket{1}^{\otimes 4}\ket{2}^{\otimes 2}\ket{4}$ so that
$\tilde{p}_{sym}=\sum_{i=1}^N i\, n_i = (6+7+8+9)(1) + (10+11)(2) + (12)(4) 
= 72 + (12)(4) = (12)(6+4)\overset{\Mod_{12}}{\sim} 0$.
Thus, $ \om^{\tilde{p}_{sym}=0}=1$, and the constraint reduces down once again to the identity
$\trm{Perm}(\L)=\trm{Perm}(\L)$, which we explained above yielded $A=0$ in this case (vs the trend of $A\ne0$ when we get an identity for the constraint) due to the presence of an AFSR \Eq{AFSR:line3}, 
with $N=12=2^{q=2}\,3^1$.
\smallskip

A similar calculation for Case 2 above with $(N,n)=(14,14)$ and $A=0$, with input state
$\ket{\n}=\ket{0}^{\otimes 8}\ket{1}\ket{2}^{\otimes 2}\ket{3}^{\otimes 3}$ yields
$\tilde{p}_{sym}=\sum_{i=1}^N i\, n_i = (9)(1) + (10+11)(2) + (12+13+14)(3) 
= 126 + (14)(3) = (14)(9 + 3)\overset{\Mod_{14}}{\sim} 0$. Thus, once again the constraint
reduces down  to the identity
$\trm{Perm}(\L)=\trm{Perm}(\L)$, which we explained above yielded $A=0$ (vs the trend of $A\ne0$ when we get an identity for the constraint) due to the presence of an AFSR  \Eq{AFSR:line3},  this time with $N=14=2^{q=1}\,7^1$.

 \subsubsection{All odd input photons for $N\in even$, with equal photon number in each output port \\ 
 and the generalization of the $SU(2)$ CNL effect}
 We now consider under what conditions does one obtain a zero amplitude $A=0$, for transitions where all the input photons are odd (and in general different) to the geHOM output state with an equal number of photons in each output port, for $(N,m) \in (even,odd)$. 
 Such a result would generalize the $SU(2)$ eHOM case of 
 $\ket{n_1,n_2}\overset{S_2}{\to}\ket{\tfrac{n_1+n_2}{2}, \tfrac{n_1+n_2}{2} }$ where $(n_1,n_2)\in(odd,odd)$, and $\tfrac{n_1+n_2}{2}\in odd$.
 \smallskip
 
 In general, we are considering the transitions $\ket{n_1,n_2,\ldots,n_N}\overset{S_N}{\to}\ket{m=\tfrac{n}{N}}^{\otimes N}$ where $n=\sum_{i=1}^N n_i$. We now consider the case when each $n_i = 2 n'_i+1$, so that
 $n=\sum_{i=1}^N (2 n'_i+1) = 2 n' + N$, where we have defined  $n'\defn\sum_{i=1}^N n'_i$.
 Therefore, $m=\tfrac{n}{N} \equiv m' +1\in \mathbb{Z}_{0+}$, defining $m'\defn\tfrac{n'}{N'}$, where since we are considering $N\in even$, we have written $N\equiv 2\,N'$, with $N'\in \mathbb{Z}_+$. Therefore, for $(N,m)\in (even,\mathbb{Z}_+)$ we are considering the transitions 
 $\bigotimes_{i=1}^N\ket{2 n'_i+1}\overset{S_N}{\to}\ket{m'+1}^{\otimes N}$.
 \smallskip
 
 Let us now consider the constraint \Eq{p:tilde:sym} $ \trm{Perm}(\L) = \om^{\tilde{p}_{sym}}\trm{Perm}(\L)$
 and compute 
 $\tilde{p}_{sym} = \sum_{i=1}^{N-1} i\,n_i = \sum_{i=1}^{N-1} i\,(2 n'_i +1) = 2\, \tilde{p}'_{sym} + \half N (N-1)$,
 where we have defined $\tilde{p}'_{sym}\defn \sum_{i=1}^{N-1} i\,n'_i$. 
 We now have that $\om^{\tilde{p}_{sym}} =  \om^{2\tilde{p}'_{sym}}\, \om^{N (N-1)/2} =  \om^{2\tilde{p}'_{sym}}\, (-1)^{(N-1)} \overset{N\in even}{\longrightarrow} (-1)\,\om^{2\tilde{p}'_{sym}}$. Thus, the constraint equation \Eq{p:tilde:sym} now reads as $ \trm{Perm}(\L) = -\om^{2\tilde{p}'_{sym}}\trm{Perm}(\L)$. We conclude then that as long as $\om^{2\tilde{p}'_{sym}}\ne (-1) \equiv \om^{\half N (2\l+1)}$ for some arbitrary integer $\l\in \mathbb{Z}_{0+}$, we have the lhs of the constraint not equal to the rhs of the constraint, which implies therefore that $A=\PermL = 0$. Lastly, since we have chosen $N=2 N' \in even$ the condition of the exponents reduces to
 $2\tilde{p}'_{sym}\ne \half N (2\l+1) =  N' (2\l+1)$. 
 Since $2 \tilde{p}'_{sym}\in even$ for \tit{any} value of $\tilde{p}'_{sym}$, we can ensure that the lhs is not equal to the rhs of the exponent constraint if we choose $N'\in odd$; namely that 
 $N\in even = 2\, (N'\in odd)$. 
 \smallskip

 As examples  of the above criteria, consider $N=6=2 * 3$. We already know from our 
 previous symmetry calculations for $(N,m)\in (even,odd)$ that we obtain a zero amplitude $A=0$ for the 
 equal all-odd input state transitions 
 $\ket{2\l+1}^{\otimes 6}\overset{S_6}{\to}\ket{2\l+1}^{\otimes 6}$ for $\l\in\mathbb{Z}_{0+}$.
 For $N=6$, we have also been able to explicitly compute (both symbolically and numerically) that the following transitions also have zero amplitude $A=0$, namely, 
 $\ket{1,1,1,3,3,3}\overset{S_6}{\to}\ket{2}^{\otimes 6}$ and 
 $\ket{1,1,3,3,5,5}\overset{S_6}{\to}\ket{3}^{\otimes 6}$.
 \smallskip
 
 For the case of $N=10=2 * 5$ we also have from  our previous symmetry calculation for $(N,m)\in (even,odd)$
 that we obtain a zero amplitude $A=0$ for the 
 equal all-odd input state transitions 
 $\ket{2\l+1}^{\otimes 10}\overset{S_{10}}{\to}\ket{2\l+1}^{\otimes 10}$ for $\l\in\mathbb{Z}_{0+}$.
 In addition, we have explicitly verified that one obtains a zero amplitude $A=0$ for the
 transitions $\ket{1,1,1,1,1,3,3,3,3,3}\overset{S_{10}}{\to}\ket{2}^{\otimes 10}$ and
  $\ket{1,1,1,1,1,1,3,3,3,5}\overset{S_{10}}{\to}\ket{2}^{\otimes 10}$. (Note, the latter two results involved  the symbolic computation of the permanent of a  $n\times n = 20\times 20$ matrix $\L$ as a function of $\om$, which took slightly over $11$ and  $10$ hours, respectively to compute symbolically. Both were proportional to the FSR
  $0=1+\om^5 = (1+\om)\,(1-\om+\om^2 - \om^3 + \om^4)$, where the second AFSR factor
  $(1-\om+\om^2 - \om^3 + \om^4)=0$ when evaluated on $\om = e^{i\,2\,\pi/10}$).
\smallskip
 
 The consequences of the above result is a follows. First, the $SU(2)$ eHOM effect 
  showed that one obtains a zero amplitude $A=0$ for the transitions 
  $\ket{n_1,n_2}\overset{S_2}{\to}\ket{\tfrac{n_1+n_2}{2},\tfrac{n_1+n_2}{2}}$ only when $(n_1,n_2)\in(odd,odd)$
 \cite{eHOM_PRA:2022,eHOM_PRA:2025}. Hence, if each input port contains only odd parity states (i.e. containing only odd numbers of photons), be they pure or mixed, with arbitrary quantum amplitudes, then the output probability distribution 
 $P(m_1,m_2)$ will exhibit a central nodal line (CNL) on the diagonal, i.e. $P(m,m) = 0$ for 
 $m\in\mathbb{Z}_{0+}$, bifurcating the output distribution as shown in \Fig{fig:eHOM:fig2:fig3:left:column:1photon:CS} . The $N=2$ case was special in the sense that if only the input port-1 contained an odd parity state, then \tit{regardless} of the state entering input port-2 (again, either pure or mixed), one obtains a CNL in the output probability distribution. This latter result stems from the additional trivial fact that  the even photon number Fock states in input port-2, and the odd photon Fock states in input port-1, i.e. $(n_1,n_2)\in(odd,even)$ do not produce an output coincident state since $\tfrac{n_1+n_2}{2}$ is then a half integer. 
 \smallskip
 
The result we proved is that for arbitrary $N\in even = 2* (N'\in odd)$, with all odd number of photons entering the input ports, there is a zero amplitude $A=0$ on the eHOM coincident output state (i.e. with equal number of photons in each of the output ports). As in the $SU(2)$ case (which is the particular case of the lowest even $N = 2 * 1$ with $N'=1\in odd$), we can now conclude that if only \tit{odd parity} states enter the input ports (again, be they pure or mixed, with arbitrary quantum amplitudes), then we obtain a CNL in the output probability distribution $P(m_1,m_2,\ldots,m_N)$ along the diagonal  $P(m,m,\ldots,m)$. Again, for those resulting input Fock states $\ket{n_1,n_2,\ldots,n_N}$ with total photon number
$n=\sum_{i=1}^N n_i$ not equal to an integer multiple of $N$ ($n\ne k\,N$ for $k\in\mathbb{Z}_+$), one simply does not have a projection onto the eHOM output coincident state, and so that amplitude will trivially be $A=0$.
Thus, we see that the  CNL feature first discussed for the $SU(2)$ symmetric BS does indeed generalize to  $N\in even = 2\, (N'\in odd)$, for arbitrary $N'\in \mathbb{Z}_{odd}$.

 \subsubsection{Other CNLs from the constraint \Eq{p:tilde:sym} for $\trm{Perm}(\L)$ involving $\tilde{p}_{sym}$}
 From the constraint \Eq{p:tilde:sym}  for $\trm{Perm}(\L)$ involving $p_{sym}$ we can construct CNLs  for non-odd parity input states, that yield zero amplitude $A=0$ on the eHOM output (``diagonal") states 
 $\ket{m=\tfrac{n}{N}}^{\otimes N}$. 
\begin{figure*}[ht]
\begin{center}
\begin{tabular}{ccc}
\hspace{-0.25in}
\includegraphics[width=3.25in,height=2.5in]{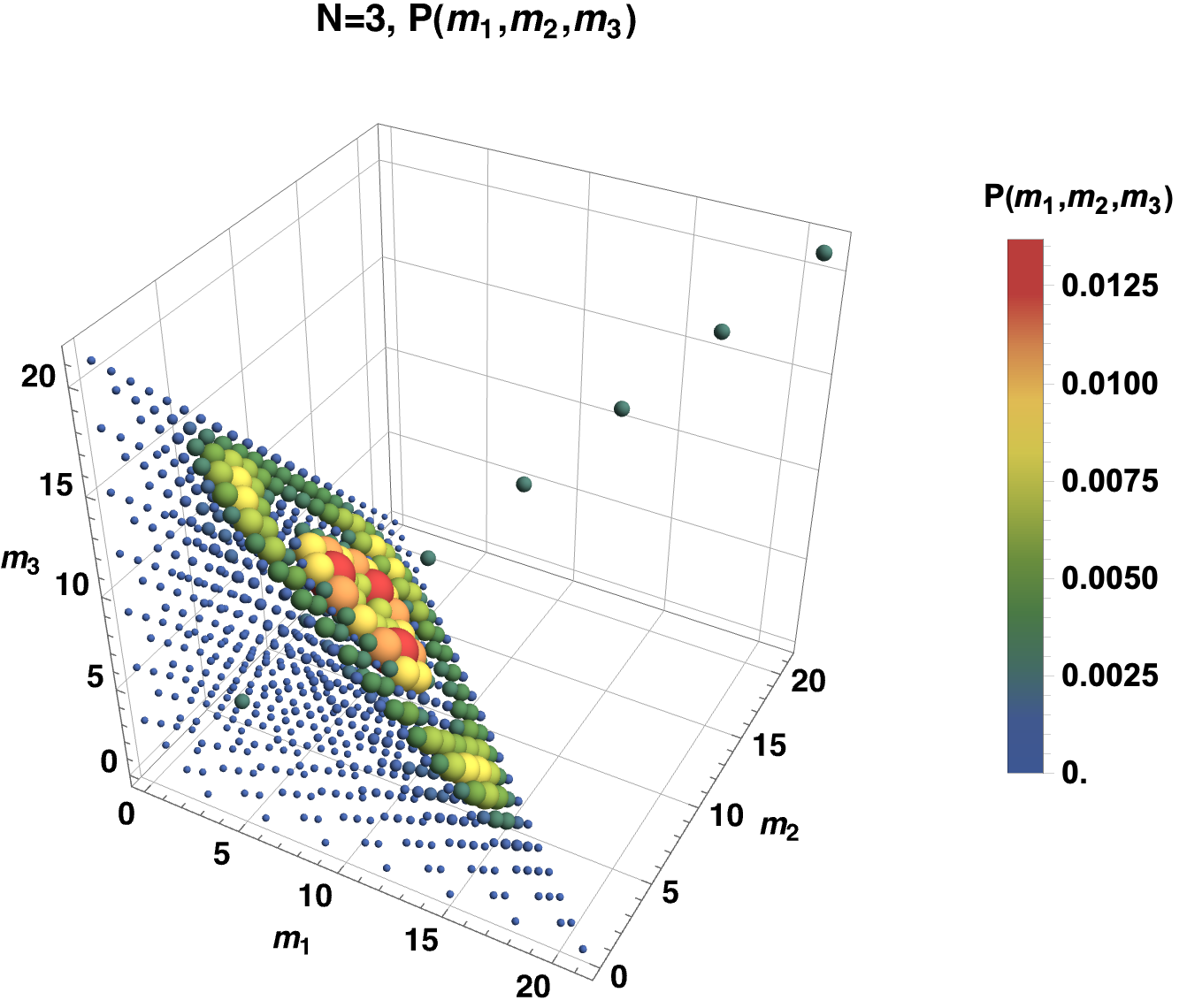} &
{\hspace{0.25in}} &
\includegraphics[width=3.25in,height=2.5in]{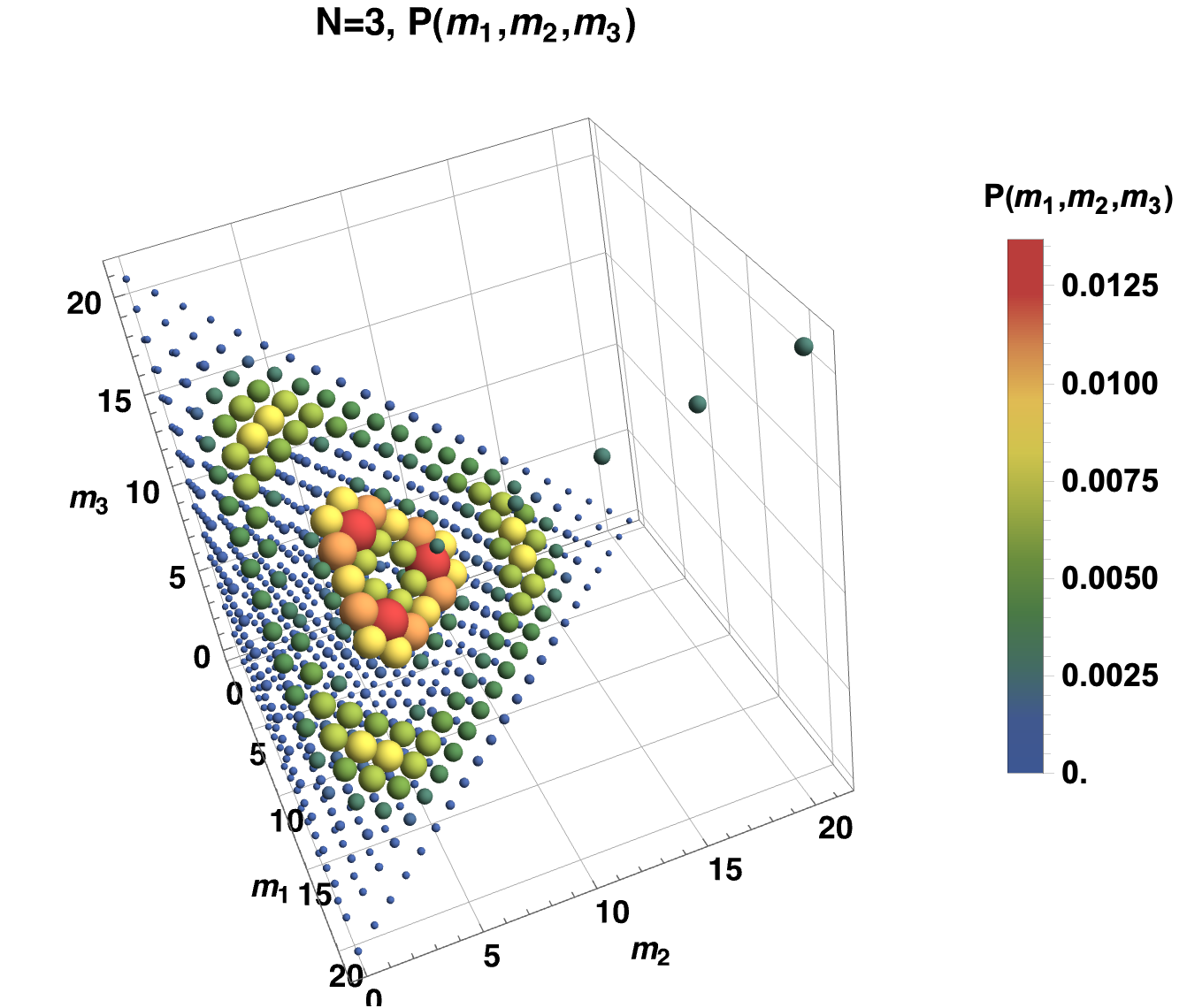}
\end{tabular}
\end{center} 
\caption{Output probability distribution $P(m_1,m_2,m_3) = |\bra{m_1,m_2,m_3}S_3\ket{\psi_3}|^2$
for the $N=3$ input state $\ket{\psi_3}~= ~\tfrac{1}{\sqrt{7}}\sum_{k=0}^{6}\,\ket{3k,1,2}$.
The blue dots along the diagonal $(m,m,m)$ shows the CNL where $P(m,m,m)=0$.
The color and diameter of the spheres are scaled to the value of $P(m_1,m_2,m_3)$.
(Note: output states $\ket{m_1,m_2,m_3}$ such that the total photon number is not a multiple of $3$,  
($\Mod[m_1+m_2+m_3,3]\ne 0$) are not plotted since they have a zero projection on any coincident output state $\ket{k+1}^{\otimes 3}$, and thus, trivially have  $P(m_1,m_2,m_3) =0$).
}
\label{fig:N3:CNL:instate:3k:1:2}    
\end{figure*}
 
 As a first example consider the $N=3$ input state $\ket{\psi_3} = \sum_{k=0}^\infty c_k\,\ket{3k,1,2}$.
 The CNL is then defined as the amplitude of $\ket{\psi_3}$ projected onto the eHOM output coincident states
 $\ket{k+1}^{\otimes 3}$ for $k\in \mathbb{Z}_{0+}$. From  \Eq{p:tilde:sym}, we will obtain $A=0$ if 
 $\Mod[\tilde{p}_sym,3]\ne 0$. Now, $\tilde{p}_{sym} = \sum_{i=1}^{N-1} i\,n_i = (1)(3k) + (2)(1)$ and 
 $\Mod[\tilde{p}_{sym},3] =$ $\Mod[(1)(3k) + (2)(1),3] =  \Mod[ (2)(1) ,3]=2\ne0$. Thus, for \tit{all $k$} we obtain 
$A=0$ for the $S_3$ transitions $\ket{3k,1,2}\overset{S_3}{\to}\ket{k+1}^{\otimes 3}$, and hence a zero amplitude (and hence probability) for each ${}^{\otimes 3}\IP{k+1}{\psi_3}$. This produces a CNL for the input state $\ket{\psi_3}$ \tit{regardless} of the coefficients $\{c_k\}$.
In \Fig{fig:N3:CNL:instate:3k:1:2}  we show the output probability distribution 
$P(m_1,m_2,m_3) = |\bra{m_1,m_2,m_3}S_3\ket{\psi_3}|^2$
for the $N=3$ input state $\ket{\psi_3}~= ~\tfrac{1}{\sqrt{7}}\sum_{k=0}^{6}\,\ket{3k,1,2}$.
The blue dots along the diagonal $(m,m,m)$ shows the CNL where $P(m,m,m)=0$.
(Note: output states $\ket{m_1,m_2,m_3}$ such that the total photon number is not a multiple of $3$,  
($\Mod[m_1+m_2+m_3,3]\ne 0$) are not plotted since they have a zero projection on any coincident output state $\ket{k+1}^{\otimes 3}$, and thus, trivially have  $P(m_1,m_2,m_3) =0$).
\smallskip

We can easily generalize this input state to 
$\ket{\psi_N} = \sum_{k=0}^\infty c_k\,\ket{N\,k}\ket{1}^{\otimes (N-2)}\ket{2}$ and consider the amplitudes $A$ for projection onto the ``diagonal" eHOM output states $\ket{k+1}^{\otimes N}$. Again we calculate 
$\Mod[\tilde{p}_{sym},N] = $ $\Mod[(1)(N\,k) + \left(\sum_{i=2}^{N-1} i \right)(1),N]=$ 
$\Mod[\half\,(N+1)(N-2),N] \ne 0$ for $N>2$ (i.e. one can show that in the last expression
$\half\,(N+1)(N-2)$ is \tit{never} an integer multiple of $N$, nor zero directly, once $N>2$).
Thus, for arbitrary $N$ we obtain $\tilde{p}_{sym}\ne 0$, which from \Eq{p:tilde:sym}  implies that 
$A=\trm{Perm}(\L) = 0$, and thus ${}^{\otimes N}\IP{k+1}{\psi_N} =0$, for \tit{all $k$}. Once again, this produces a CNL for the state $\ket{\psi_N}$  \tit{regardless} of the coefficients $\{c_k\}$.
\smallskip

The above are just two simple examples of how to construct superposition states with CNLs for arbitrary $N$.
Based on the distribution of the input photons, one can construct many more input states,
all with $\tilde{p}_{sym}\ne0$ for each term in the superposition, and hence $A=0$, 
that produce a CNL, regardless of the coefficients $\{c_k\}$, when projected onto the ``diagonal" eHOM output coincident states.

 \subsubsection{The net result of the analytic constraint \Eq{L:constraint} on $A=\trm{Perm}(\L)$}
 The conclusion of the above analysis is that the constraint equation \Eq{L:constraint} on $\trm{Perm}(\L)$ 
 \tit{does indeed} analytically explain all the results in 
 \Tbl{Table:N:odd} and \Tbl{Table:N:even} for $N\in odd$ and $N\in even$, respectively.
 \tit{This is one of the main results of this work}.

\section{Conclusions and Discussion}\label{sec:Conclusion}
In this investigation we have learned two essential points
for the ability to obtain zero amplitudes $A=0$ for the general transitions  $\ket{n_1, n_2,\ldots,n_N}\overset{S_N}{\to}\ket{m_1, m_2,\ldots,m_N}$ for a symmetric $SU(N)$ 
beam splitter (with matrix elements $(S_N)_{ij} = \frac{1}{\sqrt{N}}\,\om^{(i-1)(j-1)}$ 
composed of the roots of unity, with $\om=e^{i 2 \pi/N}$), 
preserving the total number of input/output photons.
\smallskip

First, the fundamental summation relationship (FSR), 
$\mS_N = \sum_{i=1}^N \om^{(i-1)} = 1 + \om + \om^2 + \cdots + \om^{N-1} \equiv 0$, governs the ability for sub-amplitudes of the total amplitude $A$ to group together and destructively interfere separately.
Such terms in the subgroups must all have the identical combinatorial factor coefficients multiplying them, in order for the sub-amplitudes in the group to add coherently to zero.
\smallskip

For $N$ \tit{odd}, the \tit{only} way subgroups of total amplitude can be zero, is if they are of the form
$c_i (1 + \om + \om^2 + \cdots + \om^{N-1}) = 0$ for some coefficient $c_i$; that is only if the full FSR is involved. In general, there is a set of distinct coefficients $\{c_i\}$.
\smallskip

However, for $N$ \tit{even}, there are many more possibilities for $A$ to be zero. First off,
$S_N = \frac{1-\om^N}{1-\om} = \frac{1-\om^{N/2}}{1-\om}\,(1+\om^{N/2})=0$ since
 $1+\om^{N/2} = 1+ e^{i \pi} = 0$. Thus, terms can cancel in groups of pairs as $c_i \, (1+\om^{N/2})$.
 Further, we can also group the even an odd powers of $\om$ in the FSR  as
 $S_N = \big(1+\om^2 + (\om^2)^2 + (\om^2)^{(N/2-1}\big) +
  \om\,\big(1+\om^2 + (\om^2)^2 + (\om^2)^{(N/2-1)}\big) =0$, for which
  $1+\om^2 + (\om^2)^2 + (\om^2)^{(N/2-1)}\equiv 1+\om^2 + \om'^2 + \om'^{(N/2-1)}=\mS_{N/2}=0$, since the latter is the FSR for $N/2$.  Depending on the power $q_2$ of the factor of $2$ in the prime factorization of $N$,  (i.e. for $N = 2^{q_2}\,3^{q_3}\,5^{q_5}\cdots$), this process can be repeated $q_2$ times, reducing the FSR $\mS_N$ to effectively the FSR for $\mS_{N/2^{q_2}}$, i.e.
  $0=\mS_N\propto \mS_{N/2^{q_2}}=0$. This drastically reduces the constraints for subgroups of 
  $N/2^{q_2}$ of amplitudes with the same coefficient, to ``group together" to form sub-amplitudes of the total amplitude $A$ that separately sum to zero.
\smallskip

Scheel \cite{Scheel:2004,Scheel:2008} has shown how the transition amplitude 
$A = \bra{n_1, n_2,\ldots,n_N} U_N  \ket{m_1, m_2,\ldots,m_N}$
for a general unitary matrix $U_N$, is equal to the permanent of an $n\times n$  
matrix (ignoring normalization factors), constructed from the matrix elements of  $U_N$, where 
$n$ is the total photon number of the inputs/outputs. While this is extremely useful for the computation of $A$, both analytically (as a function of $\om$) and numerically, a deeper insight in how the cancelation of scattering terms (diagrams) occurs can be obtained by the core expression for the amplitude for the symmetric $SU(N)$ BS, namely 
$A\propto \om^{\sum_{i j}^N i j k_{i j}}/\prod_{i j}^N (k_{ij} !)$.
The \tit{valid} $N\times N$ matrices $K = \{k_{ij}\}$ satisfy 
the row-sum $\sum_{j}^N k_{i j} = n_i$ and column-sum $\sum_{i}^N k_{i j} = m_i$ constraints such that
$\sum_{ij}^N k_{i j} = \sum_{i}^N n_i = \sum_{j}^N m_j\defn n$. In order for groups of sub-amplitudes to band together to destructively interfere they must have the same factorial denominator (coefficient), meaning that the $K$ matrices must all contain the same set of  integers greater than $2$ (since $0! = 1!=1$).
Within this subgroup with identical coefficient $c_i$, the placement of the integers $\{0,1,2,\ldots,N-1\}$ with $K$ determines the power $p$ of the term associated with the factor $c_i\,\om^p$. 
We obtain the exponent
via $p=|IJ \odot K| \defn \Mod[\sum_{i j}^N i j\, k_{i j}, N]$ where $IJ = \{ i j\}$ the matrix with matrix elements $i j$. Thus, matrices with the exact same set of integers in $K$ can give rise to different exponents $p$, and from the above discussion, and the parity (even or oddness) of $N$, dictate what subgroups can be formed, and potentially cancel separately within the total amplitude $A$.
\smallskip

In this work we concentrated on the generalized eHOM transitions 
$\ket{n_1,n_2,\ldots,n_N}\overSNto\ketnNN$. We symbolically and numerically investigate 
when the amplitude $A$ for these geHOM transitions were both zero and non-zero for 
$N\in\{3,4,\ldots,15\}$. We explained the features of these transitions 
by developing a symmetry constraint on the 
the value of $A=\PermL$, that generalized the one used by Lim and Beige \cite{Lim:Beige:2005} for the generalized HOM (gHOM) transitions $\ket{1}^{\otimes N}\overSNto\ket{1}^{\otimes N}$.
In particular, we could predict when a geHOM transition produced a zero amplitude $A=0$, noting the significant difference for the two cases when $N\in odd$ and when  $N\in even$. 
In the spirit of  the zero-transmission laws of Tichy \cite{Tichy:2010} (but by a different approach),
we analytically proved that the transitions
$\ketnNN\overSNto\ketnNN$ have zero amplitude $A=0$ for $N\in even$ and total photon number
$n\in odd$, which generalizes the gHOM effect of Lim and Beige \cite{Lim:Beige:2005}. Lastly, from our symmetry constraint on $A=\PermL$, we were able to construct multiphoton input states that produced a central nodal line (CNL) along the diagonal eHOM output states $\ketnNN$, generalizing the CNLs originally found in the $SU(2)$ 50/50 BS case \cite{eHOM_PRA:2022,eHOM_PRA:2025}. 
In particular, we showed that for even $N=2*N'$ where $N'\in odd$ there will be a central nodal line in the output probability distribution if only odd parity states enter the $SU(2\,N')$ BS, of which the eHOM CNL of the 
$N=2*(N'=1)$ BS is the lowest dimensional special case. 
The conclusion of these results is that many of the features of the $SU(2)$ symmetric BS eHOM transitions 
$\ket{n_1,n_2}\overset{S_2}{\to}\ket{\tfrac{n_1+n_2}{2}, \tfrac{n_1+n_2}{2}}$ have analogues in the generalized eHOM transitions $\ket{n_1,n_2,\ldots,n_N}\overSNto\ketnNN$, and can be unified in their understanding by an easily employed  symmetry constraint, which does involve the computation of $\PermL$, and only depends on the photon inputs $\{n_i\}$ and the number $m=\tfrac{n}{N}$ of output photons in each mode of the eHOM coincident state.
\smallskip

In this work we have considered an idealized, lossless symmetric $SU(N)$ beam splitter, which represents the maximum possible multiphoton interferences theoretically obtainable. Following the analysis in  \cite{eHOM_PRA:2025}, one could add the effects of losses, and imperfect detection to obtain experimental results that would be observed by a more realistic, lossless BS using imperfect detectors. Such considerations will be the focus of future research investigation.
In addition, in future work, we will explore the possibility of the construction of symmetry constraints for the boson sampling case (or variants thereof) in dimension $N$ \cite{Aaronson:2010} by examining the particular form of its permanent  $\PermL$.

\providecommand{\noopsort}[1]{}\providecommand{\singleletter}[1]{#1}%
%
\clearpage
\newpage
\appendix
 \section{\tit{Mathematica} code to construct $\L(S_N)$ via the method of Chabaud \tit{et al}. 
 \cite{Chabaud:2022}}\label{app:LChabaud}
\begin{figure}[!ht]
\includegraphics[width=6.0in,height=5.5in]{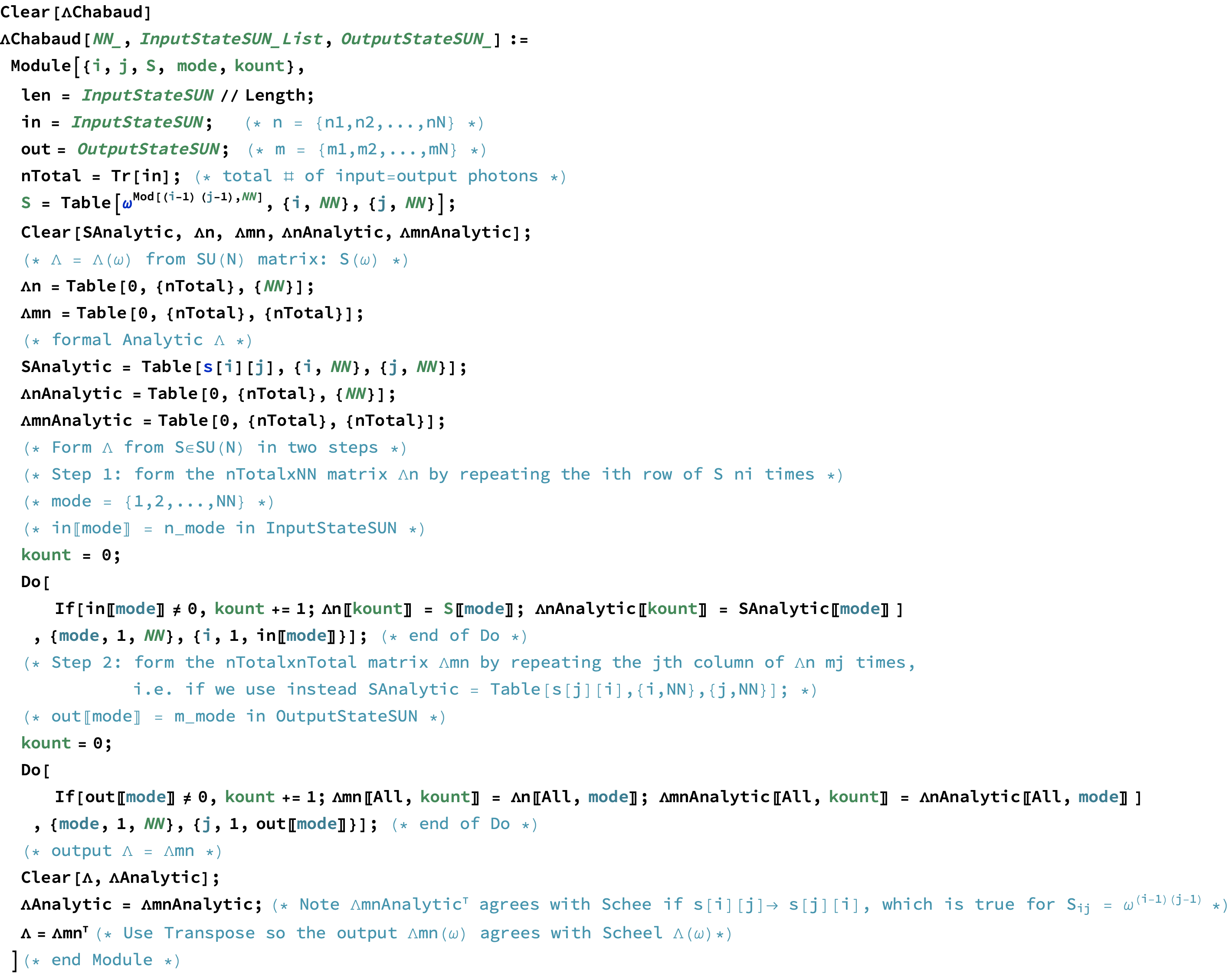} 
\caption{\tit{Mathematica} code $\L$\ttt{Chabaud} to compute $\L(S_N)$ via the method of Chabaud \tit{et al} \cite{Chabaud:2022}.
}\label{fig:Lambda:Chabaud}
\end{figure}

\Fig{fig:Lambda:Chabaud} shows the \tit{Mathematica} code to implement the Chabaud construction of 
$\L(S_N)$ (the output $\L$), consisting essentially of two \ttt{Do} (or \ttt{For})  loops. The code in \Fig{fig:Lambda:Chabaud} also constructs
$\L$\ttt{mnAnalytic}  with formal matrix elements \ttt{s[i][j]}, so that one can verify that the Chabaud constructing is explicitly following Steps 1 and 2.
\smallskip

\begin{figure}[!ht]
\begin{tabular}{cc}
\includegraphics[width=4.0in,height=2.35in]{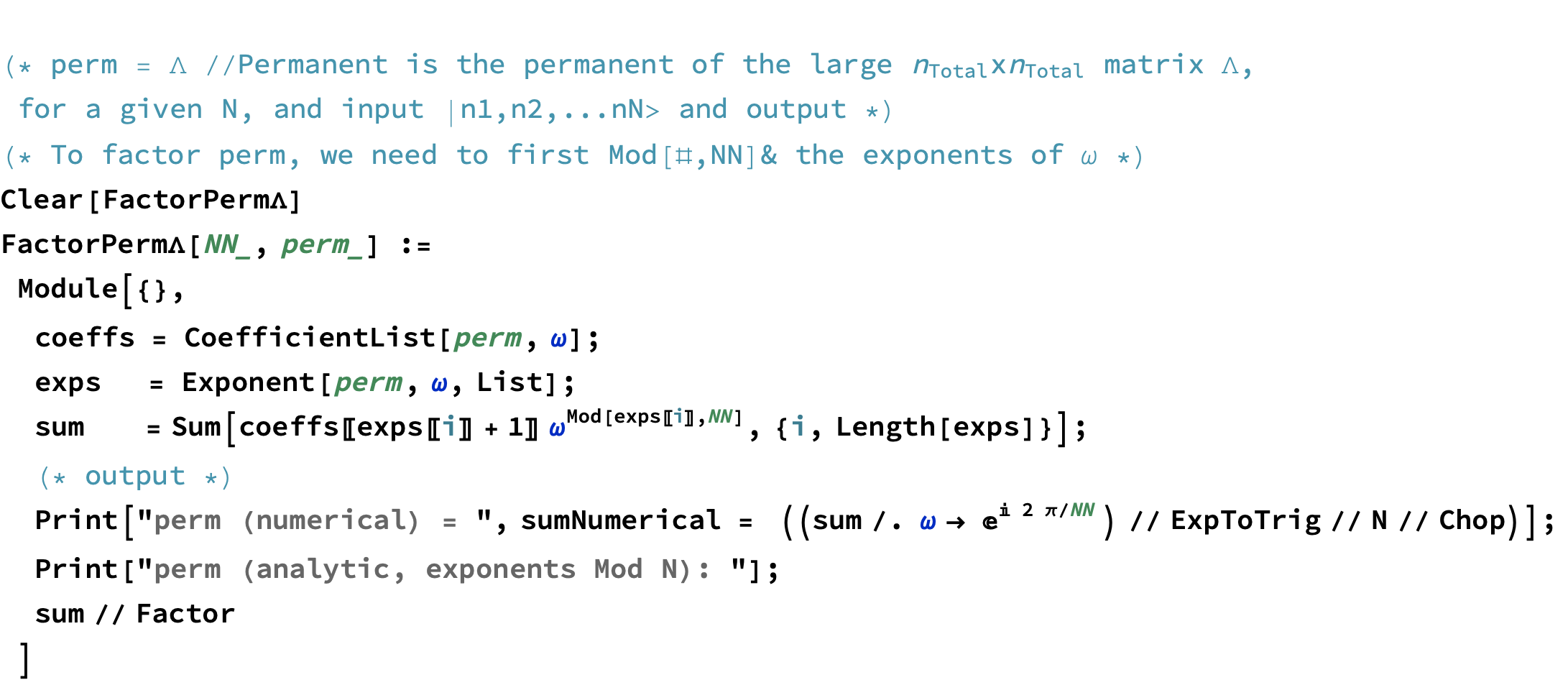} &
\includegraphics[width=3.0in,height=3.0in]{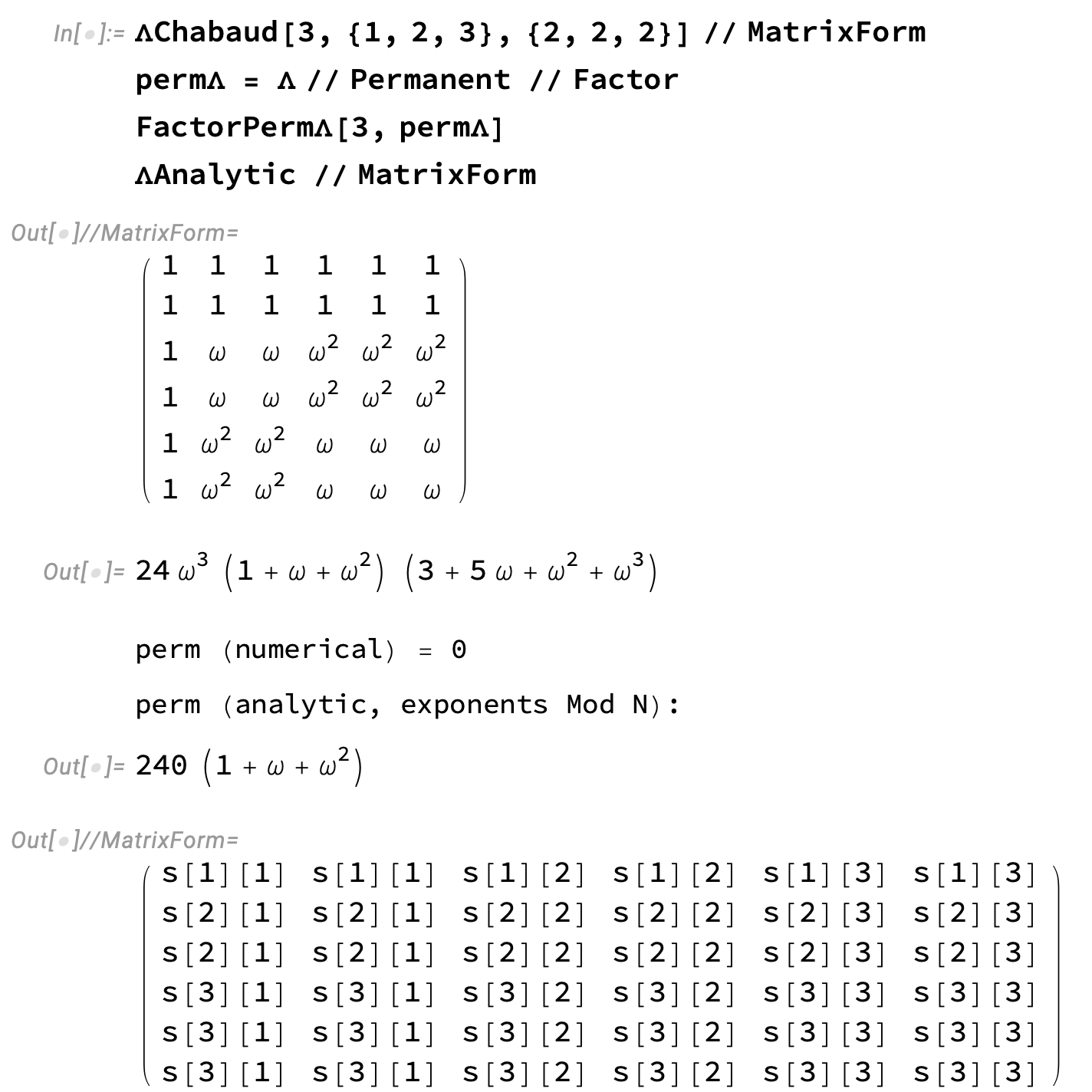}
\end{tabular}
\caption{(left)\tit{Mathematica} code  \texttt{FactorPerm}$\L$ to reduce exponents of $\om^p$ 
within Perm$(\L)$ by modulo $N$, and then factor the resulting expression.
(right) Testing the code for the amplitude $A=0$,  
$N=3$ transition $\ket{1,2,3}\overset{S_3}{\to}\ket{2,2,2}$.
}\label{fig:test:Lambda:Chabaud}
\end{figure}

After calling the code $\L$\ttt{Chabuad} with output $\L$, one computes the permanent via the call
\ttt{perm}$\L=\L$\ttt{//Permanent}.
In \Fig{fig:test:Lambda:Chabaud}(left) we show the additional code \ttt{FactorPerm}$\L$, which factors $\L(\om)$ as a function of $\om$, after reducing all powers $p$ of $\om^p\to \om^{\Mod[p,N]}$ to modulo $N$.
\Fig{fig:test:Lambda:Chabaud}(right) shows and example of creating $\L$ for the zero amplitude $A=0$,  $N=3$ transition 
$\ket{1,2,3}\overset{S_3}{\to}\ket{2,2,2}$, first creating $\L$, then factoring it directly, and subsequently re-factoring it after reducing all powers $p$ of $\om^p$ to modulo $N$, using \ttt{FactorPerm}$\L$.
Note the appearance of the \tit{full} FSR for $S_3$.
We also display $\L$\ttt{Analytic}, as a check that we formed the correct $\L(S_3)$ for this transition.

\begin{thebibliography}{30}%
\makeatletter
\providecommand \@ifxundefined [1]{%
 \@ifx{#1\undefined}
}%
\providecommand \@ifnum [1]{%
 \ifnum #1\expandafter \@firstoftwo
 \else \expandafter \@secondoftwo
 \fi
}%
\providecommand \@ifx [1]{%
 \ifx #1\expandafter \@firstoftwo
 \else \expandafter \@secondoftwo
 \fi
}%
\providecommand \natexlab [1]{#1}%
\providecommand \enquote  [1]{``#1''}%
\providecommand \bibnamefont  [1]{#1}%
\providecommand \bibfnamefont [1]{#1}%
\providecommand \citenamefont [1]{#1}%
\providecommand \href@noop [0]{\@secondoftwo}%
\providecommand \href [0]{\begingroup \@sanitize@url \@href}%
\providecommand \@href[1]{\@@startlink{#1}\@@href}%
\providecommand \@@href[1]{\endgroup#1\@@endlink}%
\providecommand \@sanitize@url [0]{\catcode `\\12\catcode `\$12\catcode
  `\&12\catcode `\#12\catcode `\^12\catcode `\_12\catcode `\%12\relax}%
\providecommand \@@startlink[1]{}%
\providecommand \@@endlink[0]{}%
\providecommand \url  [0]{\begingroup\@sanitize@url \@url }%
\providecommand \@url [1]{\endgroup\@href {#1}{\urlprefix }}%
\providecommand \urlprefix  [0]{URL }%
\providecommand \Eprint [0]{\href }%
\providecommand \doibase [0]{http://dx.doi.org/}%
\providecommand \selectlanguage [0]{\@gobble}%
\providecommand \bibinfo  [0]{\@secondoftwo}%
\providecommand \bibfield  [0]{\@secondoftwo}%
\providecommand \translation [1]{[#1]}%
\providecommand \BibitemOpen [0]{}%
\providecommand \bibitemStop [0]{}%
\providecommand \bibitemNoStop [0]{.\EOS\space}%
\providecommand \EOS [0]{\spacefactor3000\relax}%
\providecommand \BibitemShut  [1]{\csname bibitem#1\endcsname}%
\let\auto@bib@innerbib\@empty
\bibitem [{\citenamefont {Spagnolo}\ \emph {et~al.}(2013)\citenamefont
  {Spagnolo}, \citenamefont {Vitelli}, \citenamefont {Aparo}, \citenamefont
  {Mataloni}, \citenamefont {Sciarrino}, \citenamefont {Crespi}, \citenamefont
  {Ramponi},\ and\ \citenamefont {Osellame}}]{Spagnolo:2013}%
  \BibitemOpen
  \bibfield  {author} {\bibinfo {author} {\bibfnamefont {N.}~\bibnamefont
  {Spagnolo}}, \bibinfo {author} {\bibfnamefont {C.}~\bibnamefont {Vitelli}},
  \bibinfo {author} {\bibfnamefont {L.}~\bibnamefont {Aparo}}, \bibinfo
  {author} {\bibfnamefont {P.}~\bibnamefont {Mataloni}}, \bibinfo {author}
  {\bibfnamefont {F.}~\bibnamefont {Sciarrino}}, \bibinfo {author}
  {\bibfnamefont {A.}~\bibnamefont {Crespi}}, \bibinfo {author} {\bibfnamefont
  {R.}~\bibnamefont {Ramponi}}, \ and\ \bibinfo {author} {\bibfnamefont
  {R.}~\bibnamefont {Osellame}},\ }\bibfield  {title} {\enquote {\bibinfo
  {title} {Three-photon bosonic coalescence in an integrated tritter},}\
  }\href@noop {} {\bibfield  {journal} {\bibinfo  {journal} {Nat. Comm}\
  }\textbf {\bibinfo {volume} {4}},\ \bibinfo {pages} {1606} (\bibinfo {year}
  {2013})}\BibitemShut {NoStop}%
\bibitem [{\citenamefont {Lu}\ \emph {et~al.}(2018)\citenamefont {Lu},
  \citenamefont {Lukens}, \citenamefont {Peters}, \citenamefont {Odele},
  \citenamefont {Leaird}, \citenamefont {Weiner},\ and\ \citenamefont
  {Lougovski}}]{Lu:2018}%
  \BibitemOpen
  \bibfield  {author} {\bibinfo {author} {\bibfnamefont {H.~H.}\ \bibnamefont
  {Lu}}, \bibinfo {author} {\bibfnamefont {J.~M.}\ \bibnamefont {Lukens}},
  \bibinfo {author} {\bibfnamefont {N.~A.}\ \bibnamefont {Peters}}, \bibinfo
  {author} {\bibfnamefont {O.~D.}\ \bibnamefont {Odele}}, \bibinfo {author}
  {\bibfnamefont {D.~E.}\ \bibnamefont {Leaird}}, \bibinfo {author}
  {\bibfnamefont {A.~M.}\ \bibnamefont {Weiner}}, \ and\ \bibinfo {author}
  {\bibfnamefont {P.}\ \bibnamefont {Lougovski}},\ }\bibfield  {title}
  {\enquote {\bibinfo {title} {Electro-optic frequency beam splitters and
  tritters for high-fidelity photonic quantum information processing},}\
  }\href@noop {} {\bibfield  {journal} {\bibinfo  {journal} {Phys. Rev. Lett.}\
  }\textbf {\bibinfo {volume} {120}},\ \bibinfo {pages} {030502} (\bibinfo
  {year} {2018})}\BibitemShut {NoStop}%
\bibitem [{\citenamefont {Suryadi}\ \emph {et~al.}(2025)\citenamefont
  {Suryadi}, \citenamefont {Amadi},\ and\ \citenamefont {Ali}}]{Suryadi:2025}%
  \BibitemOpen
  \bibfield  {author} {\bibinfo {author} {\bibnamefont {Suryadi}}, \bibinfo
  {author} {\bibfnamefont {P.~O.}\ \bibnamefont {Amadi}}, \ and\ \bibinfo
  {author} {\bibfnamefont {N.}~\bibnamefont {Ali}},\ }\bibfield  {title}
  {\enquote {\bibinfo {title} {Three-photon pulse interference in a tritter:
  {A} novel approach for a three-party {Quantum Key Distribution Protocol}},}\
  }\href@noop {} {\bibfield  {journal} {\bibinfo  {journal} {Physics}\ }\textbf
  {\bibinfo {volume} {7}},\ \bibinfo {pages} {14} (\bibinfo {year}
  {2025})}\BibitemShut {NoStop}%
\bibitem [{\citenamefont {Su}\ \emph {et~al.}(2017)\citenamefont {Su},
  \citenamefont {L.}, \citenamefont {Rohde}, \citenamefont {Huang},
  \citenamefont {Wang}, \citenamefont {Li}, \citenamefont {Liu}, \citenamefont
  {Dowling}, \citenamefont {Lu},\ and\ \citenamefont {Pan}}]{Su:2017}%
  \BibitemOpen
  \bibfield  {author} {\bibinfo {author} {\bibfnamefont {Z.~E.}\ \bibnamefont
  {Su}}, \bibinfo {author} {\bibfnamefont {Y.}~\bibnamefont {L.}}, \bibinfo
  {author} {\bibfnamefont {Peter~P.}\ \bibnamefont {Rohde}}, \bibinfo {author}
  {\bibfnamefont {H.~L.}\ \bibnamefont {Huang}}, \bibinfo {author}
  {\bibfnamefont {X.~L.}\ \bibnamefont {Wang}}, \bibinfo {author}
  {\bibfnamefont {L.}~\bibnamefont {Li}}, \bibinfo {author} {\bibfnamefont
  {N.~L.}\ \bibnamefont {Liu}}, \bibinfo {author} {\bibfnamefont {J.~P.}\
  \bibnamefont {Dowling}}, \bibinfo {author} {\bibfnamefont {C.~Y.}\
  \bibnamefont {Lu}}, \ and\ \bibinfo {author} {\bibfnamefont {J.~W.}\
  \bibnamefont {Pan}},\ }\bibfield  {title} {\enquote {\bibinfo {title}
  {Multiphoton interference in quantum {F}ourier transform circuits and
  applications to quantum metrology},}\ }\href@noop {} {\bibfield  {journal}
  {\bibinfo  {journal} {Phys. Rev. Lett.}\ }\textbf {\bibinfo {volume} {119}},\
  \bibinfo {pages} {080502} (\bibinfo {year} {2017})}\BibitemShut {NoStop}%
\bibitem [{\citenamefont {Ou}(1996)}]{Ou:1996}%
  \BibitemOpen
  \bibfield  {author} {\bibinfo {author} {\bibfnamefont {Z.Y.}\ \bibnamefont
  {Ou}},\ }\bibfield  {title} {\enquote {\bibinfo {title} {Quantum
  multiparticle interference due to a single photon},}\ }\href@noop {}
  {\bibfield  {journal} {\bibinfo  {journal} {Quant. and Semiclass Opt.}\
  }\textbf {\bibinfo {volume} {8}},\ \bibinfo {pages} {315} (\bibinfo {year}
  {1996})}\BibitemShut {NoStop}%
\bibitem [{\citenamefont {Ou}(2007)}]{Ou:2007}%
  \BibitemOpen
  \bibfield  {author} {\bibinfo {author} {\bibfnamefont {Z.Y.J.}\ \bibnamefont
  {Ou}},\ }\href@noop {} {\emph {\bibinfo {title} {Multi-photon Quantum
  Interference}}}\ (\bibinfo  {publisher} {Springer-Verlag US},\ \bibinfo
  {address} {New York},\ \bibinfo {year} {2007})\BibitemShut {NoStop}%
\bibitem [{\citenamefont {Ou}(2017)}]{Ou_Book:2017}%
  \BibitemOpen
  \bibfield  {author} {\bibinfo {author} {\bibfnamefont {Z.Y.}\ \bibnamefont
  {Ou}},\ }\href@noop {} {\emph {\bibinfo {title} {Quantum {O}ptics for
  {E}xperimentalists, (Chap. 6.2.2, p162 and 8.3.2, p246)}}}\ (\bibinfo
  {publisher} {World Scientific},\ \bibinfo {address} {Singapore},\ \bibinfo
  {year} {2017})\BibitemShut {NoStop}%
\bibitem [{\citenamefont {Pan}\ \emph {et~al.}(2012)\citenamefont {Pan},
  \citenamefont {Chen}, \citenamefont {Lu},\ and\ \citenamefont
  {Weinfurter}}]{Pan:2012}%
  \BibitemOpen
  \bibfield  {author} {\bibinfo {author} {\bibfnamefont {J.W.}\ \bibnamefont
  {Pan}}, \bibinfo {author} {\bibfnamefont {Z.B.}\ \bibnamefont {Chen}},
  \bibinfo {author} {\bibfnamefont {C.Y.}\ \bibnamefont {Lu}}, \ and\ \bibinfo
  {author} {\bibfnamefont {H.}~\bibnamefont {Weinfurter}},\ }\bibfield  {title}
  {\enquote {\bibinfo {title} {Multiphoton entanglement and interferometry},}\
  }\href@noop {} {\bibfield  {journal} {\bibinfo  {journal} {Revs. Mod. Phys.}\
  }\textbf {\bibinfo {volume} {84}},\ \bibinfo {pages} {777} (\bibinfo {year}
  {2012})}\BibitemShut {NoStop}%
\bibitem [{\citenamefont {Dakna}\ \emph {et~al.}(1997)\citenamefont {Dakna},
  \citenamefont {Anhut}, \citenamefont {Opatrny}, \citenamefont {Kn{\"o}ll},\
  and\ \citenamefont {Welsh}}]{Dakna:1997}%
  \BibitemOpen
  \bibfield  {author} {\bibinfo {author} {\bibfnamefont {M.}~\bibnamefont
  {Dakna}}, \bibinfo {author} {\bibfnamefont {T.}~\bibnamefont {Anhut}},
  \bibinfo {author} {\bibfnamefont {T.}~\bibnamefont {Opatrny}}, \bibinfo
  {author} {\bibfnamefont {L.}~\bibnamefont {Kn{\"o}ll}}, \ and\ \bibinfo
  {author} {\bibfnamefont {D.G.}\ \bibnamefont {Welsh}},\ }\bibfield  {title}
  {\enquote {\bibinfo {title} {Generating schrodinger cat-like states by means
  of conditional measurements of a beam splitter},}\ }\href@noop {} {\bibfield
  {journal} {\bibinfo  {journal} {Phys. Rev. A}\ }\textbf {\bibinfo {volume}
  {55}},\ \bibinfo {pages} {3184} (\bibinfo {year} {1997})}\BibitemShut
  {NoStop}%
\bibitem [{\citenamefont {Carranza}\ and\ \citenamefont
  {Gerry}(2012)}]{Carranza:2012}%
  \BibitemOpen
  \bibfield  {author} {\bibinfo {author} {\bibfnamefont {R.}~\bibnamefont
  {Carranza}}\ and\ \bibinfo {author} {\bibfnamefont {C.~C.}\ \bibnamefont
  {Gerry}},\ }\bibfield  {title} {\enquote {\bibinfo {title} {Photon-subtracted
  two-mode squeezed vacuum states and applications to quantum optical
  interferometry},}\ }\href@noop {} {\bibfield  {journal} {\bibinfo  {journal}
  {J. Opt. Soc. Am. B}\ }\textbf {\bibinfo {volume} {29}},\ \bibinfo {pages}
  {2581} (\bibinfo {year} {2012})}\BibitemShut {NoStop}%
\bibitem [{\citenamefont {Maga{\~n}a-Loaiza}\ \emph {et~al.}(2019)\citenamefont
  {Maga{\~n}a-Loaiza}, \citenamefont {Le{\'o}n-Montiel}, \citenamefont
  {Perez-Leija}, \citenamefont {A.B.}, \citenamefont {U’Ren}, \citenamefont
  {You}, \citenamefont {Busch}, \citenamefont {Lita}, \citenamefont {Nam},
  \citenamefont {Mirin},\ and\ \citenamefont {Gerrits}}]{Magana-Loaiza:2019}%
  \BibitemOpen
  \bibfield  {author} {\bibinfo {author} {\bibfnamefont {O.~S.}\ \bibnamefont
  {Maga{\~n}a-Loaiza}}, \bibinfo {author} {\bibfnamefont {R.d.J.}\ \bibnamefont
  {Le{\'o}n-Montiel}}, \bibinfo {author} {\bibfnamefont {A.}~\bibnamefont
  {Perez-Leija}}, \bibinfo {author} {\bibnamefont {A.B.}}, \bibinfo {author}
  {\bibnamefont {U’Ren}}, \bibinfo {author} {\bibfnamefont {C.}~\bibnamefont
  {You}}, \bibinfo {author} {\bibfnamefont {K.}~\bibnamefont {Busch}}, \bibinfo
  {author} {\bibfnamefont {A.E.}\ \bibnamefont {Lita}}, \bibinfo {author}
  {\bibfnamefont {S.W.}\ \bibnamefont {Nam}}, \bibinfo {author} {\bibfnamefont
  {R.P.}\ \bibnamefont {Mirin}}, \ and\ \bibinfo {author} {\bibfnamefont
  {T.}~\bibnamefont {Gerrits}},\ }\bibfield  {title} {\enquote {\bibinfo
  {title} {Multiphoton quantum-state engineering using conditional
  measurements},}\ }\href {\doibase 10.1038/s41534-019-0195-2} {\bibfield
  {journal} {\bibinfo  {journal} {NPJ Quant. Info.}\ }\textbf {\bibinfo
  {volume} {5}},\ \bibinfo {pages} {80} (\bibinfo {year} {2019})}\BibitemShut
  {NoStop}%
\bibitem [{\citenamefont {Dakna}\ \emph {et~al.}(1998)\citenamefont {Dakna},
  \citenamefont {Kn{\"o}ll},\ and\ \citenamefont {Welsh}}]{Dakna:1998}%
  \BibitemOpen
  \bibfield  {author} {\bibinfo {author} {\bibfnamefont {M.}~\bibnamefont
  {Dakna}}, \bibinfo {author} {\bibfnamefont {L.}~\bibnamefont {Kn{\"o}ll}}, \
  and\ \bibinfo {author} {\bibfnamefont {D.G.}\ \bibnamefont {Welsh}},\
  }\bibfield  {title} {\enquote {\bibinfo {title} {Photon-added state
  preparation via conditional measurement on a beam splitter},}\ }\href@noop {}
  {\bibfield  {journal} {\bibinfo  {journal} {Opt. Comm.}\ }\textbf {\bibinfo
  {volume} {145}},\ \bibinfo {pages} {309} (\bibinfo {year}
  {1998})}\BibitemShut {NoStop}%
\bibitem [{\citenamefont {Lvovsky}\ and\ \citenamefont
  {Mlynek}(2002)}]{Lvovsky:2002}%
  \BibitemOpen
  \bibfield  {author} {\bibinfo {author} {\bibfnamefont {A.I.}\ \bibnamefont
  {Lvovsky}}\ and\ \bibinfo {author} {\bibfnamefont {J.}~\bibnamefont
  {Mlynek}},\ }\bibfield  {title} {\enquote {\bibinfo {title} {Quantum-optical
  catalysis: Generating nonclassical states of light by means of linear
  optics},}\ }\href@noop {} {\bibfield  {journal} {\bibinfo  {journal} {Phys.
  Rev. Lett.}\ }\textbf {\bibinfo {volume} {88}},\ \bibinfo {pages} {250401}
  (\bibinfo {year} {2002})}\BibitemShut {NoStop}%
\bibitem [{\citenamefont {Bartley}\ \emph {et~al.}(2012)\citenamefont
  {Bartley}, \citenamefont {Donati}, \citenamefont {Spring}, \citenamefont
  {Min}, \citenamefont {Barbieri}, \citenamefont {Datta}, \citenamefont
  {Smith},\ and\ \citenamefont {Walmsley}}]{Bartley:2012}%
  \BibitemOpen
  \bibfield  {author} {\bibinfo {author} {\bibfnamefont {T.J.}\ \bibnamefont
  {Bartley}}, \bibinfo {author} {\bibfnamefont {G.}~\bibnamefont {Donati}},
  \bibinfo {author} {\bibfnamefont {J.B.}\ \bibnamefont {Spring}}, \bibinfo
  {author} {\bibfnamefont {X.J.}\ \bibnamefont {Min}}, \bibinfo {author}
  {\bibfnamefont {M.}~\bibnamefont {Barbieri}}, \bibinfo {author}
  {\bibfnamefont {A.}~\bibnamefont {Datta}}, \bibinfo {author} {\bibfnamefont
  {B.J.}\ \bibnamefont {Smith}}, \ and\ \bibinfo {author} {\bibfnamefont
  {I.~A.}\ \bibnamefont {Walmsley}},\ }\bibfield  {title} {\enquote {\bibinfo
  {title} {Multiphoton state engineering by heralded interference between
  single photons and coherent states},}\ }\href@noop {} {\bibfield  {journal}
  {\bibinfo  {journal} {Phys. Rev. A}\ }\textbf {\bibinfo {volume} {86}},\
  \bibinfo {pages} {043820} (\bibinfo {year} {2012})}\BibitemShut {NoStop}%
\bibitem [{\citenamefont {Birrittella}\ \emph {et~al.}(2018)\citenamefont
  {Birrittella}, \citenamefont {El-Baz},\ and\ \citenamefont
  {Gerry}}]{Birrittella:2018}%
  \BibitemOpen
  \bibfield  {author} {\bibinfo {author} {\bibfnamefont {R.~J.}\ \bibnamefont
  {Birrittella}}, \bibinfo {author} {\bibfnamefont {M.}~\bibnamefont {El-Baz}},
  \ and\ \bibinfo {author} {\bibfnamefont {C.~C.}\ \bibnamefont {Gerry}},\
  }\bibfield  {title} {\enquote {\bibinfo {title} {Photon catalysis and quantum
  state engineering},}\ }\href@noop {} {\bibfield  {journal} {\bibinfo
  {journal} {JOSA B}\ }\textbf {\bibinfo {volume} {35}},\ \bibinfo {pages}
  {1514} (\bibinfo {year} {2018})}\BibitemShut {NoStop}%
\bibitem [{\citenamefont {Hong}\ \emph {et~al.}(1987)\citenamefont {Hong},
  \citenamefont {Ou},\ and\ \citenamefont {Mandel}}]{HOM:1987}%
  \BibitemOpen
  \bibfield  {author} {\bibinfo {author} {\bibfnamefont {C.~K.}\ \bibnamefont
  {Hong}}, \bibinfo {author} {\bibfnamefont {Z.~Y.}\ \bibnamefont {Ou}}, \ and\
  \bibinfo {author} {\bibfnamefont {L.}~\bibnamefont {Mandel}},\ }\bibfield
  {title} {\enquote {\bibinfo {title} {Measurement of subpicosecond time
  intervals between two photons by interference},}\ }\href@noop {} {\bibfield
  {journal} {\bibinfo  {journal} {Phys. Rev. Lett.}\ }\textbf {\bibinfo
  {volume} {59}},\ \bibinfo {pages} {2044} (\bibinfo {year}
  {1987})}\BibitemShut {NoStop}%
\bibitem [{\citenamefont {Bouchard}\ \emph {et~al.}(2021)\citenamefont
  {Bouchard}, \citenamefont {Sit}, \citenamefont {Zhang}, \citenamefont
  {Fickler}, \citenamefont {Miatto}, \citenamefont {Yao}, \citenamefont
  {Sciarrino},\ and\ \citenamefont {Karimi}}]{Bouchard:2021}%
  \BibitemOpen
  \bibfield  {author} {\bibinfo {author} {\bibfnamefont {F.}~\bibnamefont
  {Bouchard}}, \bibinfo {author} {\bibfnamefont {A.}~\bibnamefont {Sit}},
  \bibinfo {author} {\bibfnamefont {Y.}~\bibnamefont {Zhang}}, \bibinfo
  {author} {\bibfnamefont {R.}~\bibnamefont {Fickler}}, \bibinfo {author}
  {\bibfnamefont {F.~M.}\ \bibnamefont {Miatto}}, \bibinfo {author}
  {\bibfnamefont {Y.}~\bibnamefont {Yao}}, \bibinfo {author} {\bibfnamefont
  {F.}~\bibnamefont {Sciarrino}}, \ and\ \bibinfo {author} {\bibfnamefont
  {E.}~\bibnamefont {Karimi}},\ }\bibfield  {title} {\enquote {\bibinfo {title}
  {Two-photon interference: The {H}ong-{O}u-{M}andel effect},}\ }\href@noop {}
  {\bibfield  {journal} {\bibinfo  {journal} {Rep. Prog. Phys.}\ }\textbf
  {\bibinfo {volume} {84}},\ \bibinfo {pages} {012402} (\bibinfo {year}
  {2021})}\BibitemShut {NoStop}%
\bibitem [{\citenamefont {Alsing}\ \emph {et~al.}(2022)\citenamefont {Alsing},
  \citenamefont {Birrittella}, \citenamefont {Gerry}, \citenamefont {Mimih},\
  and\ \citenamefont {Knight}}]{eHOM_PRA:2022}%
  \BibitemOpen
  \bibfield  {author} {\bibinfo {author} {\bibfnamefont {P.~M.}\ \bibnamefont
  {Alsing}}, \bibinfo {author} {\bibfnamefont {R.~J.}\ \bibnamefont
  {Birrittella}}, \bibinfo {author} {\bibfnamefont {C.~C.}\ \bibnamefont
  {Gerry}}, \bibinfo {author} {\bibfnamefont {J.}~\bibnamefont {Mimih}}, \ and\
  \bibinfo {author} {\bibfnamefont {P.}~\bibnamefont {Knight}},\ }\bibfield
  {title} {\enquote {\bibinfo {title} {Extending the {H}ong-{O}u-{M}andel
  {E}ffect: the power of nonclassiciality},}\ }\href@noop {} {\bibfield
  {journal} {\bibinfo  {journal} {Phys. Rev. A}\ }\textbf {\bibinfo {volume}
  {105}},\ \bibinfo {pages} {013712} (\bibinfo {year} {2022})}\BibitemShut
  {NoStop}%
\bibitem [{\citenamefont {Alsing}\ and\ \citenamefont
  {Birrittella}(2025)}]{eHOM_PRA:2025}%
  \BibitemOpen
  \bibfield  {author} {\bibinfo {author} {\bibfnamefont {P.~M.}\ \bibnamefont
  {Alsing}}\ and\ \bibinfo {author} {\bibfnamefont {R.~J.}\ \bibnamefont
  {Birrittella}},\ }\bibfield  {title} {\enquote {\bibinfo {title} {Examination
  of the extended {H}ong-{O}u-{M}andel {E}ffect and considerations for
  experimental detection},}\ }\href@noop {} {\bibfield  {journal} {\bibinfo
  {journal} {Phys. Rev. A}\ }\textbf {\bibinfo {volume} {111}},\ \bibinfo
  {pages} {032616} (\bibinfo {year} {2025})}\BibitemShut {NoStop}%
\bibitem [{\citenamefont {Aaronson}\ and\ \citenamefont
  {Arkhipov}(2010)}]{Aaronson:2010}%
  \BibitemOpen
  \bibfield  {author} {\bibinfo {author} {\bibfnamefont {S.}~\bibnamefont
  {Aaronson}}\ and\ \bibinfo {author} {\bibfnamefont {A.}~\bibnamefont
  {Arkhipov}},\ }\bibfield  {title} {\enquote {\bibinfo {title} {{The
  Computational Complexity of Linear Optics}},}\ }\href@noop {} {\  (\bibinfo
  {year} {2010})},\ \Eprint {http://arxiv.org/abs/1011.3245v1 [quant-ph]}
  {arXiv:1011.3245v1 [quant-ph]} \BibitemShut {NoStop}%
\bibitem [{\citenamefont {Scheel}(2004)}]{Scheel:2004}%
  \BibitemOpen
  \bibfield  {author} {\bibinfo {author} {\bibfnamefont {S.}~\bibnamefont
  {Scheel}},\ }\bibfield  {title} {\enquote {\bibinfo {title} {Permanents in
  linear optical networks},}\ }\href@noop {} {\bibfield  {journal} {\bibinfo
  {journal} {arXiv:0406127v1 [quant-ph]}\ } (\bibinfo {year}
  {2004})}\BibitemShut {NoStop}%
\bibitem [{\citenamefont {Scheel}(2008)}]{Scheel:2008}%
  \BibitemOpen
  \bibfield  {author} {\bibinfo {author} {\bibfnamefont {S.}~\bibnamefont
  {Scheel}},\ }\bibfield  {title} {\enquote {\bibinfo {title} {Permanents in
  linear optical networks},}\ }\href@noop {} {\bibfield  {journal} {\bibinfo
  {journal} {Acta Physica Slovaca}\ }\textbf {\bibinfo {volume} {58}},\
  \bibinfo {pages} {675} (\bibinfo {year} {2008})}\BibitemShut {NoStop}%
\bibitem [{\citenamefont {Lim}\ and\ \citenamefont
  {Beige}(2005{\natexlab{a}})}]{Lim:Beige:2005}%
  \BibitemOpen
  \bibfield  {author} {\bibinfo {author} {\bibfnamefont {Y.~L.}\ \bibnamefont
  {Lim}}\ and\ \bibinfo {author} {\bibfnamefont {A.}~\bibnamefont {Beige}},\
  }\bibfield  {title} {\enquote {\bibinfo {title} {Generalized
  {H}ong-{O}u-{M}andel experiments with bosons and fermions},}\ }\href@noop {}
  {\bibfield  {journal} {\bibinfo  {journal} {New J. Phys.}\ }\textbf {\bibinfo
  {volume} {7}},\ \bibinfo {pages} {155} (\bibinfo {year}
  {2005}{\natexlab{a}})}\BibitemShut {NoStop}%
\bibitem [{\citenamefont {Lim}\ and\ \citenamefont
  {Beige}(2005{\natexlab{b}})}]{Lim:Beige_PRA:2005}%
  \BibitemOpen
  \bibfield  {author} {\bibinfo {author} {\bibfnamefont {Y.~L.}\ \bibnamefont
  {Lim}}\ and\ \bibinfo {author} {\bibfnamefont {A.}~\bibnamefont {Beige}},\
  }\bibfield  {title} {\enquote {\bibinfo {title} {Multiphoton entanglement
  through a {B}ell-multiport beam splitter},}\ }\href@noop {} {\bibfield
  {journal} {\bibinfo  {journal} {Phys. Rev. A}\ }\textbf {\bibinfo {volume}
  {71}},\ \bibinfo {pages} {062311} (\bibinfo {year}
  {2005}{\natexlab{b}})}\BibitemShut {NoStop}%
\bibitem [{\citenamefont {Tichy}\ \emph {et~al.}(2010)\citenamefont {Tichy},
  \citenamefont {Tiersch}, \citenamefont {de~Melo}, \citenamefont {Mintert},\
  and\ \citenamefont {Buchleitner}}]{Tichy:2010}%
  \BibitemOpen
  \bibfield  {author} {\bibinfo {author} {\bibfnamefont {M.~C.}\ \bibnamefont
  {Tichy}}, \bibinfo {author} {\bibfnamefont {M.}~\bibnamefont {Tiersch}},
  \bibinfo {author} {\bibfnamefont {F.}~\bibnamefont {de~Melo}}, \bibinfo
  {author} {\bibfnamefont {F.}~\bibnamefont {Mintert}}, \ and\ \bibinfo
  {author} {\bibfnamefont {A.}~\bibnamefont {Buchleitner}},\ }\bibfield
  {title} {\enquote {\bibinfo {title} {Zero-transmission law for multiport beam
  splitters},}\ }\href@noop {} {\bibfield  {journal} {\bibinfo  {journal}
  {Phys. Rev. Lett.}\ }\textbf {\bibinfo {volume} {104}},\ \bibinfo {pages}
  {220405} (\bibinfo {year} {2010})}\BibitemShut {NoStop}%
\bibitem [{\citenamefont {Tichy}\ \emph {et~al.}(2014)\citenamefont {Tichy},
  \citenamefont {Mayer}, \citenamefont {Buchleitner},\ and\ \citenamefont
  {M{\o}lmer}}]{Tichy:2014}%
  \BibitemOpen
  \bibfield  {author} {\bibinfo {author} {\bibfnamefont {M.~C.}\ \bibnamefont
  {Tichy}}, \bibinfo {author} {\bibfnamefont {K.}~\bibnamefont {Mayer}},
  \bibinfo {author} {\bibfnamefont {A.}~\bibnamefont {Buchleitner}}, \ and\
  \bibinfo {author} {\bibfnamefont {K.}~\bibnamefont {M{\o}lmer}},\ }\bibfield
  {title} {\enquote {\bibinfo {title} {Stringent and efficient assessment of
  {B}oson-sampling devices},}\ }\href@noop {} {\bibfield  {journal} {\bibinfo
  {journal} {Phys. Rev. Lett.}\ }\textbf {\bibinfo {volume} {113}},\ \bibinfo
  {pages} {020502} (\bibinfo {year} {2014})}\BibitemShut {NoStop}%
\bibitem [{\citenamefont {Zukowski}\ \emph {et~al.}(1997)\citenamefont
  {Zukowski}, \citenamefont {Zeilinger},\ and\ \citenamefont
  {Horne}}]{Zukowski:Zeilinger:Horne:1997}%
  \BibitemOpen
  \bibfield  {author} {\bibinfo {author} {\bibfnamefont {M.}~\bibnamefont
  {Zukowski}}, \bibinfo {author} {\bibfnamefont {A.}~\bibnamefont {Zeilinger}},
  \ and\ \bibinfo {author} {\bibfnamefont {M.~A.}\ \bibnamefont {Horne}},\
  }\bibfield  {title} {\enquote {\bibinfo {title} {Realizable
  higher-dimensional two-particle entanglements via multiport beam
  splitters},}\ }\href@noop {} {\bibfield  {journal} {\bibinfo  {journal}
  {Phys. Rev. A}\ }\textbf {\bibinfo {volume} {55}},\ \bibinfo {pages} {2564}
  (\bibinfo {year} {1997})}\BibitemShut {NoStop}%
\bibitem [{\citenamefont {Campos}(2000)}]{Campos:2000}%
  \BibitemOpen
  \bibfield  {author} {\bibinfo {author} {\bibfnamefont {R.}~\bibnamefont
  {Campos}},\ }\bibfield  {title} {\enquote {\bibinfo {title} {Three-photon
  {H}ong-{O}u-{M}andel interference at a multiport mixer},}\ }\href@noop {}
  {\bibfield  {journal} {\bibinfo  {journal} {Phys. Rev. A}\ }\textbf {\bibinfo
  {volume} {62}},\ \bibinfo {pages} {013809} (\bibinfo {year}
  {2000})}\BibitemShut {NoStop}%
\bibitem [{\citenamefont {Chabaud}\ \emph {et~al.}(2022)\citenamefont
  {Chabaud}, \citenamefont {Deshpande},\ and\ \citenamefont
  {Mehraban}}]{Chabaud:2022}%
  \BibitemOpen
  \bibfield  {author} {\bibinfo {author} {\bibfnamefont {U.}~\bibnamefont
  {Chabaud}}, \bibinfo {author} {\bibfnamefont {A.}~\bibnamefont {Deshpande}},
  \ and\ \bibinfo {author} {\bibfnamefont {S.}~\bibnamefont {Mehraban}},\
  }\bibfield  {title} {\enquote {\bibinfo {title} {Quantum-inspired permanent
  identities},}\ }\href@noop {} {\bibfield  {journal} {\bibinfo  {journal}
  {Quantum}\ }\textbf {\bibinfo {volume} {6}},\ \bibinfo {pages} {877}
  (\bibinfo {year} {2022})},\ \Eprint {http://arxiv.org/abs/2208.00327v1
  [quant-ph]} {arXiv:2208.00327v1 [quant-ph]} \BibitemShut {NoStop}%
\bibitem [{\citenamefont {M.~Jerdee}\ and\ \citenamefont
  {Newman}(2022)}]{Jerdee:2022}%
  \BibitemOpen
  \bibfield  {author} {\bibinfo {author} {\bibfnamefont {A.~Kirkley}\
  \bibnamefont {M.~Jerdee}}\ and\ \bibinfo {author} {\bibfnamefont {M.~E.~J.}\
  \bibnamefont {Newman}},\ }\bibfield  {title} {\enquote {\bibinfo {title}
  {Improved estimates for the number of non-negative integer matrices with
  given row and column sums},}\ }\href {\doibase
  {doi.org/10.1098/rspa.2023.0470}} {\bibfield  {journal} {\bibinfo  {journal}
  {Proc. Royal Soc. A}\ }\textbf {\bibinfo {volume} {480}},\ \bibinfo {pages}
  {2282} (\bibinfo {year} {2022})},\ \Eprint {http://arxiv.org/abs/2209.14869v1
  [stat.{CO}]} {arXiv:2209.14869v1 [stat.{CO}]} \BibitemShut {NoStop}%
\end{thebibliography}
\end{document}